\documentclass[
aps,
prd,
superscriptaddress,
twocolumn,
notitlepage,
nofootinbib,
floatfix,
reprint,
preprintnumbers
]{revtex4-1}
\usepackage{comment}
\usepackage[english]{babel}
\usepackage{graphicx}
\usepackage[T1]{fontenc}
\usepackage[utf8]{inputenc}   
\usepackage{epsf,amsmath,bbold,amsfonts,stmaryrd}
\usepackage{mathrsfs}
\usepackage{appendix}
\usepackage{amssymb}
\usepackage{float}
\usepackage{color} 
\usepackage[normalem]{ulem}
\usepackage[colorlinks]{hyperref}
\hypersetup{pageanchor=false,citecolor=blue,
urlcolor=blue,linkcolor=blue}
\usepackage{indentfirst}

\def\bea{\begin{eqnarray}}
\def\eea{\end{eqnarray}}
\def\ba{\begin{array}}
\def\ea{\end{array}}
\def\bc{\begin{center}}
\def\ec{\end{center}}
\def\bl{\begin{flushleft}}
\def\el{\end{flushleft}}
\def\br{\begin{flushright}}
\def\er{\end{flushright}}
\def\bi{\begin{itemize}}
\def\ei{\end{itemize}}

\def\tb{\begin{tabular}}
\def\et{\end{tabular}}
\newtheorem{question}{Question}
\def\bq{\begin{question}}
\def\eq{\end{question}}
\newtheorem{definition}{Def}
\def\bd{\begin{definition}}
\def\ed{\end{definition}}
\newtheorem{answer}{Answer}
\def\ban{\begin{answer}}
\def\ean{\end{answer}}
\newtheorem{possibleanswer}{Possible answer}
\def\bpa{\begin{possibleanswer}\normalfont}
\def\epa{\end{possibleanswer}}
\newtheorem{theorem}{Theorem}
\def\bth{\begin{theorem}}
\def\eth{\end{theorem}}

\newcommand{\be}[1]{\begin{equation} \centering \label{#1}}
\newcommand{\ee}{\end{equation}}
\newcommand{\beq}{\begin{equation}}
\newcommand{\beqn}{\begin{eqnarray}}

\begin{document}
\title{Breaking the axion abundance–isocurvature relation: \\ 
Scale invariance and high-scale inflation}
\author{
Mariaveronica De Angelis}
 	\email{mdeangel@ucm.es}
\affiliation{Departamento de Física Teórica, Facultad de Ciencias Físicas, Universidad Complutense de Madrid, 28040 Madrid, Spain}
\author{Javier Rubio}
 	\email{javier.rubio@ucm.es}
    \affiliation{Departamento de Física Teórica, Facultad de Ciencias Físicas, Universidad Complutense de Madrid, 28040 Madrid, Spain}
\affiliation{Instituto de Física de Partículas y del Cosmos (IPARCOS), Universidad Complutense de Madrid, 28040 Madrid, Spain}
\begin{abstract}
A QCD axion accounting for dark matter through pre-inflationary misalignment is strongly constrained by CMB isocurvature bounds, which place the classic axion window in tension with high-scale inflation. The tension arises because the same decay constant controls both the relic abundance and the primordial axion fluctuations. We show how this link can be broken in a broad class of scale-invariant models, without introducing any dimensionful parameter into the classical action. Rather than imposing a fixed Peccei-Quinn scale, the axion's kinetic normalisation is set by a dynamical field ratio. The resulting field-space metric controls the abundance through its value today and the primordial fluctuations through its value at horizon crossing. A sufficiently large metric during inflation suppresses isocurvature, allowing high-scale inflation in the classic axion window with an untuned misalignment angle and sub-Planckian decay constants. The exact scale and Peccei-Quinn symmetries also enforce a non-turning inflationary attractor, conserving the super-Hubble curvature perturbation and preserving the single-field adiabatic spectrum. We realise the mechanism in the Higgs-Dilaton model by deforming the Peccei-Quinn kinetic term. The same deformation turns the inflaton sector into an $\alpha$-attractor pinned near the Starobinsky point. This explicit realisation has a super-Planckian present decay constant, which places it in the anthropic window, now reopened at high inflationary scales, with an ultralight QCD axion constrained by black-hole superradiance.
\end{abstract}
\preprint{}
\preprint{}
\maketitle
\section{Introduction}

The QCD axion \cite{Weinberg:1977ma,Wilczek:1977pj} remains one of the most compelling dark-matter candidates \cite{Visinelli:2009zm,Wantz:2009it,Sikivie:2006ni}. It resolves the strong-CP problem through the Peccei-Quinn (PQ) mechanism~\cite{Peccei:1977hh,Peccei:2006as} and naturally reproduces the observed relic density via the so-called misalignment mechanism~\cite{Preskill:1982cy,Abbott:1982af,Dine:1982ah}, where the axion is initially displaced from the minimum of its QCD potential and begins to oscillate once its mass becomes comparable to the Hubble rate. These oscillations behave as cold dark matter. For an initial misalignment angle $\varphi_{\rm ini}$ of order unity, the observed abundance requires a decay constant $f\sim10^{12}\,$GeV, in the classic axion window \cite{Wantz:2009it,Borsanyi:2016ksw,Marsh:2015xka}.

This picture depends on the timing of PQ symmetry breaking. If it occurs before or during inflation, and the symmetry is not subsequently restored, the axion is homogenised across the observable Universe at a value $\varphi_{\rm ini}$ inherited from inflation. A light axion nevertheless acquires quantum fluctuations, whose phase amplitude is proportional to $H_*/f$, with $H_*$ being the Hubble rate at horizon crossing. These fluctuations later become spatial variations in the axion abundance, producing a dark-matter isocurvature mode \cite{Axenides:1983hj,Seckel:1985tj,Linde:1985yf,Lyth:1989pb,Turner:1990uz}. Planck bounds its contribution to the CMB power at the few-percent level \cite{Planck:2018jri}, creating a tension between the classic axion window and high-scale inflation when axions account for all dark matter \cite{Fox:2004kb,Beltran:2006sq,Hertzberg:2008wr}.

The origin of this tension is that the same decay constant controls both the relic abundance and the primordial fluctuations. Raising $f$ suppresses the phase fluctuation, but maintaining the observed abundance then requires a smaller $\varphi_{\rm ini}$. The fractional density perturbation is consequently enhanced, partly compensating for the suppression. The product $f\varphi_{\rm ini}$, which controls isocurvature, therefore grows too slowly to accommodate high-scale inflation over the usual range of decay constants \cite{Wantz:2009it,Marsh:2015xka}. A small misalignment angle opens the anthropic axion window, with $f\gg10^{12}\,$GeV, but does not by itself remove the conflict \cite{Preskill:1982cy,Linde:1987bx,Tegmark:2005dy}.

Standard alternatives carry costs. PQ breaking after inflation removes the inflationary isocurvature bound but produces topological defects. Their evolution depends on the domain-wall number $N_{\rm DW}$, which counts the degenerate minima of the axion potential within one period of the phase. For $N_{\rm DW}>1$, stable walls are cosmologically unacceptable \cite{Sikivie:1982qv,Vilenkin:1982ks} unless the degeneracy is lifted \cite{Hiramatsu:2012sc}; for $N_{\rm DW}=1$, the defects decay but contribute to the abundance~\cite{Hiramatsu:2012gg,Kawasaki:2014sqa,Gorghetto:2020qws,Buschmann:2021sdq}. Alternatively, lowering the inflationary scale enough to satisfy the bound leaves the primordial tensor modes unobservable.

Within the pre-inflationary scenario, recent approaches suppress axion fluctuations by introducing explicit PQ breaking or enhancing the effective decay constant during inflation~\cite{Kearney:2016vqw,Kobayashi:2016qld,Ballesteros:2021bee,Graham:2025iwx,Chakraborty:2025lyp,Kawasaki:2026jen}. A larger PQ scale during inflation, first proposed in Ref.~\cite{Linde:1991km}, has been realised in many constructions \cite{Folkerts:2013tua,Chun:2014xva,Fairbairn:2014zta,Allali:2022yvx}, including Higgs inflation \cite{Nakayama:2015pba}. Its weak point is the subsequent relaxation of that scale, which can restore the PQ symmetry non-thermally \cite{Kawasaki:2013iha,Harigaya:2015hha}. An intermediate vacuum reached before inflation ends has recently been proposed to avoid this problem \cite{Lyu:2026rwo}.

Field-space geometry provides another way to obtain a larger effective decay constant during inflation. The angular metric sets the canonical normalisation of the axion phase, so a larger metric suppresses the phase fluctuation associated with a given canonical fluctuation. Closest to our work is Ref.~\cite{Tadepalli:2026mdc}, where a hyperbolic metric suppresses the observable angular fluctuation of a spectator PQ scalar with a quartic potential, without explicit symmetry breaking. For extra-dimensional axions, a radion coupled to the inflaton plays a similar role~\cite{Choi:2026tup}. A different possibility uses a non-compact axion field space, allowing inflationary fluctuations to populate many degenerate QCD vacua and evade the isocurvature constraint, at the cost of a domain-wall network and residual strong CP violation~\cite{Karananas:2025uhy}.

In this paper, we study the geometric separation of abundance and isocurvature in a class of scale-invariant models. The axion metric depends on a dynamical, dimensionless field ratio, parametrised by $\Theta$. Its value at horizon crossing sets the primordial fluctuations, while its value today sets the decay constant entering the relic abundance. A metric much larger during inflation therefore suppresses the fluctuations without changing the late-time abundance. Here, the axion field space remains compact and the axion potential unchanged; the suppression arises entirely from kinetic geometry. Unlike in Ref.~\cite{Tadepalli:2026mdc}, the PQ scalar is not a spectator, since the field ratio controlling its angular metric is also the one that drives inflation.

Scale invariance allows this hierarchy to arise without introducing any dimensionful parameter into the classical action. It does not fix the hierarchy: a concrete model must provide the required kinetic function and evolution of the field ratio. Within the general class, a sufficiently large metric at horizon crossing makes high-scale inflation compatible with the classic axion window, an untuned misalignment angle, and sub-Planckian decay constants. We derive the required hierarchy before choosing a realisation.

Our analysis applies to models containing a real scalar and a complex PQ scalar coupled to gravity, with kinetic and potential functions depending on their ratio. Scale invariance and chiral PQ symmetry are exact symmetries of the classical action. The radial expectation value breaks both spontaneously, generating the Planck and Peccei-Quinn scales, a recurring feature of classically scale-invariant constructions~\cite{Wetterich:1987fm,Shaposhnikov:2008xi,Ferreira:2016wem,Casas:2017wjh,Casas:2018fum,Kubo:2018kho,Rubio:2020zht,Kubo:2020fdd,Garces:2026cbj,DeAngelis:2026mwx}. An identification of the PQ scalar with the dilaton has been considered in such models~\cite{Tokareva:2017nng}, in a scenario where the axion does not account for dark matter; here, it does. Only the confining dynamics breaks the chiral symmetry explicitly, generating the axion potential long after inflation. Our inflationary analysis concerns the classical background and its tree-level fluctuations.

The two symmetries also simplify the inflationary dynamics. In the Einstein frame, they shift the dilaton and the axion phase, on which neither the metric nor the potential depends. Along the trajectories of interest, the associated charge densities are diluted by the expansion, leaving the background motion along the single field-space direction orthogonal to both symmetry directions. The covariant turning rate therefore vanishes on the resulting attractor. As a consequence, entropy perturbations do not source the curvature perturbation during inflation: the latter is conserved on super-Hubble scales, the axion isocurvature mode remains uncorrelated with it, and the adiabatic power spectrum and bispectrum keep their single-field form. Although the shift symmetry forbids a scalar-sector potential for the phase, the variation of its metric gives the canonically normalised entropy mode a geometric mass. This mass is slow-roll suppressed at CMB horizon crossing in the realisation considered below.

We realise the mechanism in the Higgs-Dilaton scenario \cite{Blas:2011ac,GarciaBellido:2011de,Garcia-Bellido:2012npk,Bezrukov:2012hx,Rubio:2014wta,Karananas:2016kyt,Casas:2018fum,Piani:2022gon,Piani:2023aof}, where the radial PQ field and the Higgs field couple non-minimally to gravity \cite{Bezrukov:2007ep}. In the original model, the axion metric is smaller during inflation than it is today, reversing the required hierarchy. We therefore multiply the kinetic term of the complex PQ scalar by $\Theta^{-n}$, with $n>0$ the deformation exponent. This makes the metric larger during inflation while preserving both shift symmetries and leaving the scalar potential unchanged.

The same deformation has a useful consequence for the inflaton sector. It couples the kinetic terms of the inflaton and the dilaton, and imposing the vanishing-charge attractor gives an effective inflaton kinetic function governed by a second-order pole throughout the observable regime. Its residue is essentially independent of both $n$ and the non-minimal coupling $\xi_\chi$ of the radial PQ field. The model therefore behaves as an $\alpha$-attractor \cite{Kallosh:2013hoa,Kallosh:2013yoa,Artymowski:2016pjz}, with predictions $n_s\simeq1-2/N_*$ and $r\simeq12/N_*^2$ pinned near the Starobinsky point \cite{Starobinsky:1980te}, where $N_*$ is the number of $e$-folds between horizon crossing and the end of inflation. This is in contrast with the standard Higgs-Dilaton model, whose predictions drift away from this point as $\xi_\chi$ grows. The isocurvature suppression, instead, grows steeply with $n$, which can therefore be raised without appreciably changing the inflationary observables.

With the tensor-to-scalar ratio near its Starobinsky value, the measured scalar amplitude fixes $H_*\simeq1.5\times10^{13}\,$GeV for $N_*=55$. The isocurvature bound then constrains $n$ and $\xi_\chi$ jointly: a larger coupling weakens the suppression and requires a larger exponent. At the reference coupling $\xi_\chi=5\times10^{-5}$, an exponent of four provides the required inflationary suppression. We also estimate the interaction scales associated with the deformation and find them well above the inflationary Hubble rate.

The same coupling fixes the late-time axion metric, making the present decay constant super-Planckian. The realisation therefore sits in the anthropic window, with an ultralight QCD axion and $\varphi_{\rm ini}\simeq2.4\times10^{-6}$ at the reference parameters. The small misalignment angle remains, but the inflationary isocurvature constraint can now be satisfied at high $H_*$. Since the axion mass grows with $\xi_\chi$, black-hole superradiance supplies a complementary constraint, restricting the coupling to $\xi_\chi\lesssim5.5\times10^{-4}$ under the assumptions of Ref.~\cite{Aswathi:2025nxa}. Reaching the classic window would require a different embedding that separates the late-time decay constant from the coupling controlling the inflationary geometry. This possibility is allowed by the general framework but is not realised by the Higgs-Dilaton model studied here.

The suppression derived during inflation must also survive the subsequent evolution. After inflation, oscillations of $\Theta$ modulate the axion metric and can amplify fluctuations. We discuss this possibility but leave a quantitative treatment of reheating for future work. The late-time isocurvature bounds quoted here are therefore conditional on the absence of significant post-inflationary amplification. We also discuss the axion-quality problem associated with the explicit embedding.

The paper is organised as follows. Section~\ref{sec:scaleinv} introduces the scale-invariant models and their Einstein-frame formulation, together with the elements of covariant perturbation theory we need. Section~\ref{sec:currents} derives the conserved currents and background evolution and shows that the turning rate vanishes on the inflationary attractor. Section~\ref{se:ps} derives the adiabatic and isocurvature spectra and non-Gaussianity. Section~\ref{sec:DM} derives the relic abundance and isocurvature bound. Section~\ref{sec:HD} specialises to the deformed Higgs-Dilaton model, its inflationary predictions, and axion constraints. We conclude in Section~\ref{sec:conclusions}; the appendices detail the field-space geometry, the covariant perturbation equations, and the effective-theory cutoff.

\section{Scale-invariant models}\label{sec:scaleinv}

We consider a real scalar $h$ and a complex PQ scalar $X$
coupled to gravity, with the axion identified with the phase
of $X$. We impose global scale invariance and chiral PQ
symmetry on the classical action. These symmetries constrain
the field dependence of the potential and kinetic terms,
while allowing the axion normalisation to vary with the
ratio of the scalar fields.

We restrict ourselves to an action linear in the Ricci scalar
and quadratic in first derivatives of the fields, invariant
under the global scale transformations
\begin{equation}
g_{\mu\nu}\mapsto c_1^{-2}g_{\mu\nu}\,,
\qquad
h\mapsto c_1h\,,
\qquad
X\mapsto c_1X\,,
\end{equation}
and the chiral transformations
\begin{equation}
X\mapsto e^{i c_2}X\,,
\qquad
X^\dagger\mapsto e^{-i c_2}X^\dagger\,,
\end{equation}
with constant $c_1>0$ and real $c_2$. In polar variables
$X=\chi e^{i\varphi}/\sqrt{2}$, the chiral symmetry shifts
the phase $\varphi$ while leaving $h$ and $\chi$ unchanged.

To express the scale symmetry explicitly, we introduce
$\rho=\rho(h,\chi)$, homogeneous of degree one in its
arguments, and $\Theta=\Theta(h/\chi)$, unchanged under
scale transformations. For the class considered here,
the Lagrangian reads
\begin{equation}
\frac{\mathcal{L}}{\sqrt{-g}}
=\frac{\rho^2F(\Theta)}{2}R
-\rho^4u(\Theta)
-\frac{\rho^2}{2}
(Y^\mu)^T\cdot G(\Theta)\cdot Y_\mu\,,
\label{eqn:hdm_lagrangianL2}
\end{equation}
where
$Y_\mu=(\partial_\mu\Theta,\partial_\mu\rho/\rho,
\partial_\mu\varphi)^T$ and
\begin{equation}
G(\Theta)=
\begin{pmatrix}
G_1(\Theta) & G_2(\Theta) & 0\\
G_2(\Theta) & G_3(\Theta) & 0\\
0 & 0 & G_4(\Theta)
\end{pmatrix}.
\end{equation}
Here $F$, $G_1$, $G_2$, $G_3$, $G_4$, and $u$ are
dimensionless functions of $\Theta$. The powers of $\rho$
supply the required mass dimensions, so no dimensionful
parameter appears in the action. In these variables,
the Lagrangian is manifestly invariant under
\begin{equation}
g_{\mu\nu}\mapsto c_1^{-2}g_{\mu\nu}\,,
\qquad
\rho\mapsto c_1\rho\,,
\qquad
\Theta\mapsto\Theta\,,
\end{equation}
and
\begin{equation}
\varphi\mapsto\varphi+c_2\,.
\end{equation}
Assuming the prefactor of the Ricci scalar to be positive-definite ($\rho^2 F(\Theta) > 0$), we can perform a Weyl rescaling of the metric $g_{\mu\nu}\to M_P^2/(\rho^2F(\Theta))\, g_{\mu\nu}$ together  with a field redefinition \cite{Blas:2011ac,Karananas:2016kyt,Casas:2018fum}
\begin{equation}
\Phi=\ln\left(\frac{\rho}{M_P}\right)  +  \int^{\Theta}
 \frac{3\, F'(\Theta)+G_2(\Theta)
}{6 F(\Theta)+G_3(\Theta)}\, d\Theta   \,,
\end{equation}
 with the primes denoting derivatives with respect to $\Theta$. The Lagrangian \eqref{eqn:hdm_lagrangianL2}  is then reduced to the form
\begin{equation}\label{eq:sigmamodel}
\frac{\mathcal{L}}{\sqrt{-g}}=\frac{M_P^2}{2}R-
\frac{1}{2} g^{\mu\nu} K_{IJ}\,\partial_{\mu}\phi^I
\partial_{\nu}\phi^J- U\,,
\end{equation}
where we have adopted a compact notation with  $\phi^I=(\Theta,\Phi,\varphi)$ denoting collectively the three fields in the model, $K_{IJ}=
{M_P^2}\,{\rm diag}(K_\Theta, K_\Phi, K_\varphi)$ a $\Theta$-dependent diagonal field metric with entries
\begin{equation}
K_\Theta=\dfrac{G_1(\Theta)}{F(\Theta)}+ \dfrac{3}{2} \left(\dfrac{F'(\Theta)}{F(\Theta)}\right)^2- \frac{(3F'(\Theta)+G_2(\Theta))^2}{6 F^2(\Theta)+F(\Theta) G_3(\Theta)}\,,  \label{eq:Kdiag}
\end{equation}
\begin{equation}
K_\Phi\equiv 6+\dfrac{G_3(\Theta)}{F(\Theta)}  \hspace{10mm}
K_\varphi\equiv \dfrac{G_4(\Theta)}{F(\Theta)}\,,
\end{equation}
and
\begin{equation}
U(\Theta)\equiv \frac{M_P^4\,  u(\Theta)}{F^2(\Theta)}\,.
\end{equation} 
In these variables, the scale symmetry acts as
$\Phi\mapsto\Phi+\ln c_1$, while the PQ symmetry shifts
$\varphi$. Neither transformation acts on the Einstein-frame
metric.

Given the Lagrangian density \eqref{eq:sigmamodel}, it is still possible to perform additional field redefinitions to modify the precise structure of the theory-defining functions $K_\Theta$, $K_\Phi$ and $K_\varphi$. Such redefinitions correspond to changes of coordinates on the three-dimensional field-space manifold with
metric $K_{IJ}$.\footnote{The associated curvature in configuration space is, however, generically non-vanishing, meaning that not all the 
kinetic terms in the action can be brought into a canonical form.} Using this metric, we can define covariant derivatives
\begin{equation}
\nabla_J A^I = \partial_J A^I + \Gamma^I_{\>\> JK} A^K\,,
\end{equation}
and directional derivatives along the trajectory
\begin{equation}
{\cal D}_t A^I \equiv \dot{\phi}^J \nabla_J A^I = \dot{A}^I + \Gamma^I_{\>\> JK} A^J \dot{\phi}^K\,,
\end{equation}
for any field-space vector $A^I$. 

For later use, we only need a few general results of covariant multifield perturbation theory; the complete derivation and our conventions are collected in App.~\ref{sec:cov_multif}.

Defining the adiabatic velocity and the unit vector tangent to the background trajectory as
\begin{equation}
\dot\sigma^2=K_{IJ}\dot\phi^I\dot\phi^J,
\qquad
\hat\sigma^I=\frac{\dot\phi^I}{\dot\sigma},
\end{equation}
the perturbations can be decomposed into components that are parallel and orthogonal to the trajectory. Their coupling is controlled by the covariant turning rate
\begin{equation}\label{eq:turning}
\omega^I\equiv \mathcal{D}_t\hat\sigma^I
=-\frac{1}{\dot\sigma}U_{,K}\hat s^{IK},
\qquad
\hat s^{IK}\equiv K^{IK}-\hat\sigma^I\hat\sigma^K .
\end{equation}
In particular, the comoving curvature perturbation \(\mathcal R_c=HQ_\sigma/\dot\sigma\) obeys, on super-Hubble scales,
\begin{equation}
\dot{\mathcal R}_c=
\frac{2H}{\dot\sigma}\omega_I Q_s^I
+\mathcal O(k^2/a^2).
\end{equation}
Entropy perturbations can therefore affect the adiabatic sector only
if the background trajectory turns in field space. We now show that
the two exact shift symmetries of the scale-invariant theory forbid
precisely this possibility on the inflationary attractor.

\section{Conserved currents}\label{sec:currents}

The independence of the Einstein-frame action from $\Phi$ and $\varphi$ implies the existence of two conserved currents,
\begin{equation}
    J_\Phi^\mu=-M_P ^2\, K_\Phi\, \partial^\mu \Phi\,, \label{eq:noetherphi}\qquad     J_\varphi^\mu=-M_P ^2\, K_\varphi\, \partial^\mu \varphi\,,
\end{equation}
satisfying
\begin{equation}
\nabla_\mu J_\Phi^\mu=0\,,\hspace{15mm} \nabla_\mu J_\varphi^\mu=0 \,.
\end{equation}
Let us consider the evolution of the $\Phi$ and $\varphi$ fields in an expanding Universe. Assuming a flat Friedmann-Lemaitre-Robertson-Walker metric $ds^2=-dt^2+a^2(t)\delta_{ij}dx^i d x^j$ with scale factor $a(t)$ and initial conditions $(\Phi_i,\dot \Phi_i)$ and $(\varphi_i,\dot \varphi_i)$, the above conservation equations reduce to 
\begin{equation}
    a^3 K_\Phi \dot{\Phi}=\mathrm{const.},
    \qquad
    a^3 K_\varphi \dot{\varphi}=\mathrm{const.}.
\end{equation}
In the diagonal parametrization of Eq.~\eqref{eq:sigmamodel}, these expressions admit formal solutions
\begin{eqnarray}
\Phi&=&\Phi_i + \dot \Phi_i K_{\Phi}(t_i)a^3(t_i) \int_{t_i}^t \frac{dt'}{K_{\Phi}(t')a^3(t')}\,,\\
\varphi&=&\varphi_i + \dot \varphi_iK_{\varphi}(t_i)a^3(t_i)\int_{t_i}^t \frac{dt'}{K_{\varphi}(t')a^3(t')}\,.
\end{eqnarray}
Along the inflationary trajectories of interest, for which
$a^3K_\Phi$ and $a^3K_\varphi$ grow with the expansion, this implies
\begin{equation}
 \Phi\rightarrow \;\;\makebox{constant}\,, \hspace{10mm} \varphi\rightarrow \;\;\makebox{constant}\,,
\end{equation}
as $t\rightarrow\infty$ without requiring any special fine-tuning of the initial conditions. The simultaneous breaking of scale and chiral symmetries is not controlled by a potential, nor does it imply the existence of a massive (radial) degree of freedom, as is usually found in the standard axion formulation. 

However, the relevant statement is more general than the freezing of any
particular field coordinate. Indeed, the two shift symmetries define two
independent Killing directions in field space. Once their conserved
charge densities have been diluted, the background velocity covector
has vanishing projection along both of them. The potential gradient
has the same property, since $U$ is invariant under both symmetries.
In the three-dimensional field space considered here, the subspace
orthogonal to these two independent symmetry directions is
one-dimensional. Hence, the velocity covector and the potential
gradient are necessarily parallel on the zero-charge attractor,
\begin{equation}
    K_{IJ}\dot{\phi}^J \propto U_{,I}.
\end{equation}
Since the entropic projector annihilates the background velocity,
$\hat s^{IJ}K_{JK}\dot\phi^K=0$, it therefore also annihilates
$U_{,I}$, and the turning rate \eqref{eq:turning} vanishes,
\begin{equation}
\omega^I=0\,,  \hspace{10mm} \omega^2 =0 \,.
\label{eq:turning_rate}
\end{equation}
This coordinate-independent statement will be central to the perturbation analysis below. In particular, it remains valid in the deformed Higgs--Dilaton realisation of Sec.~\ref{sec:HD}, where the field-space metric is non-diagonal and the dilaton coordinate itself need not remain constant. The component-by-component check in the undeformed case is given in App.~\ref{appendix:multifield}.

\section{Power spectra}\label{se:ps}

On the zero-turning attractor, the adiabatic and entropic
sectors evolve independently, and the curvature perturbation
is conserved on super-Hubble scales. The entropy modes,
however, remain sensitive to the field-space geometry.
Although $U=U(\Theta)$ is independent of $\Phi$ and
$\varphi$,\footnote{A potential that breaks one of the two symmetries
would introduce an additional contribution. An explicit
example is worked out in App.~\ref{appendix:multifield}.}
the variation of their kinetic normalisations along the
inflationary trajectory gives the canonically normalised
entropy perturbations effective masses
\begin{eqnarray}
{\cal M}_{\Phi\Phi}
&=& \beta_{,\sigma}U_{,\sigma}
-\left(\beta_{,\sigma}^2+\beta_{,\sigma\sigma}\right)
\dot{\sigma}^2\,,
\\
{\cal M}_{\varphi\varphi}
&=& \tilde{\gamma}_{,\sigma}U_{,\sigma}
-\left(\tilde{\gamma}_{,\sigma}^2
+\tilde{\gamma}_{,\sigma\sigma}\right)
\dot{\sigma}^2\,,
\label{eq:m_phiphi}
\end{eqnarray}
where
\begin{equation}
\beta\equiv\frac12\ln K_{\Phi\Phi}\,,
\qquad
\tilde{\gamma}\equiv\frac12\ln K_{\varphi\varphi}\,,
\end{equation}
with $K_{\Phi\Phi}=M_P^2K_\Phi$ and
$K_{\varphi\varphi}=M_P^2K_\varphi$ the diagonal entries
of $K_{IJ}$ in \eqref{eq:sigmamodel}.

To express the perturbation equations in terms of ordinary
time derivatives, we introduce the canonically normalised
entropy perturbations $\delta s_I=\sqrt{K_{II}}\,Q_s^I$
(no sum over $I$), with $I=\Phi,\varphi$, and define
\begin{equation}\label{eq:normalised_pert}
{\cal R}_c=\frac{H}{\dot\sigma}Q_\sigma
=\frac{Q_\sigma}{\sqrt{2\epsilon}\,M_P}\,,
\qquad
{\cal S}_I=\frac{H}{\dot\sigma}\delta s_I
=\frac{\delta s_I}{\sqrt{2\epsilon}\,M_P}\,.
\end{equation}
Here $Q_\sigma$ and $Q_s^I$ are the projections defined
in App.~\ref{sec:cov_multif}. In this basis, the connection
terms are absorbed into the effective masses above, and
the mode equations become
\begin{equation}\label{eq:curvcons}
\ddot{\cal R}_c+(3+\epsilon_2)H\dot{\cal R}_c
+\frac{k^2}{a^2}{\cal R}_c=0\,,
\end{equation}
\begin{equation}\label{eq:ddotSvarphi}
\ddot{\cal S}_I+(3+\epsilon_2)H\dot{\cal S}_I
+\left(\frac{k^2}{a^2}
+{\cal M}_{II}+\Delta_\epsilon\right){\cal S}_I=0\,,
\end{equation}
with no sum over $I$, and
\begin{equation}
\Delta_\epsilon
=\frac{\epsilon_2}{2}
\left(3-\epsilon+\frac{\epsilon_2}{2}+\epsilon_3\right)H^2\,.
\end{equation}
The derivation is given in App.~\ref{appendix:multifield}.
The curvature equation has the standard single-field form,
giving the super-Hubble power spectrum
\begin{equation}
{\cal P}_{\cal R}(k)
={\cal P}_{{\cal R}*}
\left(\frac{k}{k_*}\right)^{n_s-1}\,,
\end{equation}
where $k_*=a_*H_*$ is the pivot scale and, at leading order
in slow roll,
\begin{equation}
{\cal P}_{{\cal R}*}
\simeq\frac{H_*^2}{8\pi^2\epsilon_*M_P^2}\,,
\qquad
n_s-1=-2\epsilon_*-\epsilon_{2*}\,.
\end{equation}

\subsection{Non-Gaussianities}

Despite the multi-field nature of the dynamics, the curvature perturbation is Gaussian to the accuracy of any foreseeable measurement, and its bispectrum is that of single-field inflation. The reason is the symmetry structure that enforces \eqref{eq:turning_rate} in Sec.~\ref{sec:currents}, the scale and chiral transformations acting as shift symmetries of the full action and not merely of the potential.

This is most transparent in the $\delta N$ formalism~\cite{Sasaki:1995aw,Lyth:2005fi}. It relates the curvature perturbation on a final uniform-density hypersurface to the variation of the number of $e$-folds $N$ elapsed from an initial flat hypersurface during inflation, here expanded in the covariant perturbations ${\cal Q}^I$ of App.~\ref{appendix:multifield}~\cite{Gong:2011uw,Elliston:2012ab}. The local non-Gaussianity parameter reads
\begin{equation}
    f_{\rm NL} = \frac{5}{6} \frac{N^{,J} N^{,K} (\nabla_J \nabla_K N)}{(N_{,I} N^{,I})^2} \;,
\end{equation}
where $N_{,I} \equiv \partial N / \partial \phi^I$ is the derivative of the number of $e$-folds with respect to the initial field values. A constant shift of either $\Phi$ or $\varphi$ maps any solution of \eqref{eq:sigmamodel} onto a solution with the same expansion history, so that
\begin{equation}
N_{,\Phi} = N_{,\varphi} = 0\,,
\end{equation}
identically, with no appeal to slow roll and irrespective of the off-diagonal structure that the deformation of Sec.~\ref{sec:HD} induces in the field-space metric.\footnote{In the slow-roll approximation the same conclusion follows from $N_{,I} =M_P^{-2}\,U\,U_{,I}/(U^{,J}U_{,J})$ together with $U_{,\Phi} = U_{,\varphi} = 0$, the number of $e$-folds reducing to the one-dimensional integral $N \simeq -\int_{\Theta_*}^{\Theta_{\rm end}} K_\Theta\,(U/U_{,\Theta})\,{\rm d}\Theta$ over the initial value of $\Theta$ alone.} All cross-terms and pure entropic contributions in the multi-field expansion therefore vanish, and the expression reduces to its single-field form along the canonically normalised adiabatic field $\sigma$, for which the covariant Hessian is an ordinary second derivative,
\begin{equation}
    f_{\rm NL} = \frac{5}{6} \frac{N_{,\sigma\sigma}}{(N_{,\sigma})^2} \,.
\end{equation}
At leading order in slow roll $N_{,\sigma} \simeq -(\sqrt{2\epsilon} M_P)^{-1}$ and $N_{,\sigma\sigma} \simeq \epsilon_2/(4\epsilon M_P^2)$, so that $f_{\rm NL}=(5/12)\,\epsilon_2$. This is the contribution of the super-horizon evolution of Gaussian field fluctuations. In addition, the intrinsic non-Gaussianity of the modes at horizon crossing adds $(5/6)\,\epsilon$, and the two combine into the single-field squeezed-limit result~\cite{Maldacena:2002vr}
\begin{equation}
    f_{\rm NL} \simeq \frac{5}{12}\left(2\epsilon+\epsilon_2\right)= \frac{5}{12} (1 - n_s) \,,
\end{equation}
the split between them being conventional and only the sum physically meaningful. The adiabatic bispectrum therefore satisfies the single-field consistency relation
and is slow-roll suppressed.

Axion isocurvature, however, is non-Gaussian because the axion energy density depends quadratically on the phase, rather than because of inflationary dynamics (we will return to this in Sec.~\ref{sec:DM}).

\section{Axion dark matter and isocurvature}\label{sec:DM}
The absence of entropy-to-curvature transfer during inflation
does not remove the axion fluctuation. Once the axion potential
becomes relevant, that fluctuation produces spatial variations
in the misalignment abundance.

We derive here the two constraints that any realisation of the mechanism must satisfy, and combine them into a single statement about the field-space geometry. The first is the relic abundance. The axion is produced by misalignment, and matching the observed dark matter density ties the metric at the present epoch, $K_\varphi(\Theta_0)$, to the initial misalignment angle. The second is the isocurvature bound. The inflationary fluctuation of the phase is controlled by the metric at horizon crossing, $K_\varphi(\Theta_*)$, and the CMB limits its amplitude. The misalignment angle enters both and cancels between them, so that the pair collapses to a requirement on the geometry alone. Supplying that hierarchy is what a concrete model has to do.

The phase field $\varphi$ remains classically displaced and inherits a shift symmetry at high energies. At low temperatures, this symmetry is mildly broken by non-perturbative effects (e.g. instantons), generating a periodic potential for the canonically normalised field
\begin{equation}\label{eq: Adef}
A \equiv f_0\,\varphi\,, \qquad f_0 \equiv M_P \sqrt{K_\varphi(\Theta_0)}\,,
\end{equation}
where $\Theta_0$ denotes the present value of $\Theta$ and $\Theta_*$ its value at horizon exit of the CMB pivot scale; $f_0$ is the present decay constant, and $f_*\equiv M_P\sqrt{K_\varphi(\Theta_*)}$ its value at horizon crossing. Both $f_0$ and $f_*$ depend on the fields only through the ratio $\Theta$, since scale invariance removes the overall scale, i.e.~the dilaton, from $K_\varphi$. The hierarchy $f_*\gg f_0$ is therefore not the displacement of a radial field but the same function evaluated at two values of the ratio.
The corresponding energy density is
\begin{equation}
\rho_A = \frac12 \dot{A}^2 + m_A^2(T)\,f_0^2 \left(1-\cos\frac{A}{f_0}\right)\,,\label{eq:energy_density}
\end{equation}
which, for small displacements, reduces to the harmonic form $\rho_A \simeq \frac12 (\dot{A}^2 + m_A^2 A^2)$. 

We use two assumptions throughout. First, we assume that the PQ symmetry is broken before inflation and is not subsequently restored, giving a homogeneous $\varphi_{\rm ini}$ across the observable Universe, perturbed by the fluctuations below. Topological defects then contribute no relic abundance, which comes solely from misalignment. Second, we set $N_{\rm DW}=1$. This domain-wall number counts degenerate instanton-potential minima within $\varphi\to\varphi+2\pi$ and is fixed by the PQ-current anomaly with respect to the confining gauge group. General $N_{\rm DW}$ replaces $A/f_0$ by $N_{\rm DW}A/f_0$ in \eqref{eq:energy_density} and multiplies the zero-temperature mass \eqref{eq:m0_chi} by $N_{\rm DW}$. The final condition depends on this sector only through $m_0^{5/6}$ or $m_0^{1/2}$, depending on the regime, so this choice does not affect our conclusions. Pre-inflationary symmetry breaking also avoids the domain-wall
problem for $N_{\rm DW}>1$.

We assume a temperature-dependent axion mass\footnote{The temperature dependence comes from non-perturbative instanton effects in a confining gauge sector \cite{Abbott:1982af,Dine:1982ah,Pisarski:1980md,PhysRevD.33.889,Wantz:2009it}.} of the form
\begin{equation}
m_A(T)=
\begin{cases}
m_0\left(\dfrac{\Lambda}{T}\right)^p\,, & T>\Lambda \,, \\
m_0\,, & T\leq \Lambda \,,
\end{cases}
\label{eq:m_A}
\end{equation}
where $\Lambda$ is the non-perturbative scale at which the potential becomes effective \cite{Visinelli:2009kt,Marsh:2015xka}, and $p$ the temperature exponent. The confining sector has two independent quantities: the crossover temperature $\Lambda$, where the mass saturates, and the zero-temperature topological susceptibility $\chi_{\rm top}$, which sets its magnitude,\footnote{Both are Einstein-frame quantities and introduce no scale by hand. The potential matters only from the onset of oscillations, when $\Theta=\Theta_0$, so $\Lambda$ and $\chi_{\rm top}$ are measured constants of the confining sector.}
\begin{equation}
m_0=\frac{\sqrt{\chi_{\rm top}}}{f_0}\,.
\label{eq:m0_chi}
\end{equation}
Thus $m_0\propto f_0^{-1}$ is not independent. We define the dimensionless ratio
\begin{equation}
\zeta\equiv\frac{\sqrt{\chi_{\rm top}}}{\Lambda^{2}}=\left(\frac{\chi_{\rm top}^{1/4}}{\Lambda}\right)^{2},
\label{eq:zeta}
\end{equation}
a dimensionless number $\zeta\simeq0.5$ for pure $SU(3)$ gauge theory~\cite{Ce:2015qha,Boyd:1996bx}, while in QCD the light quarks suppress the susceptibility and $\chi_{\rm top}^{1/4}\simeq75.5\,$MeV~\cite{GrillidiCortona:2015jxo} against $\Lambda\simeq0.155\,$GeV, giving $\zeta\simeq0.24$.
The homogeneous mode obeys
\begin{equation}
\ddot A + 3H\dot A + m_A^2(T)\,A = 0\,,
\end{equation}
in the small-angle regime (i.e. $A/f_0 \ll 1$), and starts oscillating when $3H(T_{\rm osc}) \simeq m_A(T_{\rm osc}).$
Assuming radiation domination,
\begin{equation}
H(T)=\sqrt{\frac{\pi^2 g_*(T)}{90}}\frac{T^2}{M_P}\,,
\end{equation}
implies that
\begin{align}
T_{\rm osc}^{\,p+2}&=\frac{m_0\,\Lambda^p\,M_P}{3\sqrt{\pi^2 g_*(T_{\rm osc})/90}}\,,
\label{eq:Tosc}\\
\left(\frac{T_{\rm osc}}{\Lambda}\right)^{p+2}&=\frac{\zeta\,M_P}{3\sqrt{\pi^2 g_*(T_{\rm osc})/90}\;f_0}\,,
\label{eq:Tosc_ratio}
\end{align}
for $T_{\rm osc}>\Lambda$. The second form follows on inserting \eqref{eq:m0_chi} and \eqref{eq:zeta}. Apart from the order-one factor $\zeta$, the onset of oscillations, measured in units of the crossover temperature, is fixed by the decay constant alone,
\begin{equation}
f_0\simeq0.22\,\zeta\,M_P\left(\frac{20}{g_*(\Lambda)}\right)^{1/2},
\label{eq:f0_cross}
\end{equation}
irrespective of the temperature exponent $p$. The relativistic degrees of freedom entering here are those at $T_{\rm osc}=\Lambda\simeq 0.155\,$GeV, and the equation is normalised accordingly. For QCD, $g_*(\Lambda)\simeq20$ gives the crossover at $f_0\simeq0.05\,M_P\simeq1.3\times10^{17}\,$GeV \cite{Borsanyi:2016ksw}. Below it, the temperature-dependent branch covers the classic window and most of the anthropic one. Above it lies the constant-mass branch realised in Sec.~\ref{sec:HD}.

At the onset of oscillations, the energy density is
\begin{equation}
\rho_{A,{\rm osc}}\simeq \frac12\,m_A^2(T_{\rm osc})\,f_0^2\,\varphi_{\rm ini}^2\,,\label{eq:energy_density_osc}
\end{equation}
with $\varphi_{\rm ini}$ the initial classical misalignment. Using conservation of the comoving number density, $n_A/s={\rm const}$ (valid for adiabatic expansion\footnote{The axion is a classical field and does not exchange energy with radiation. Comoving number conservation additionally requires the axion mass to vary adiabatically, $\dot{m_A}/m_A =p\,H\ll m_A$ which holds after the onset of oscillations.}), with entropy density
\begin{equation}
s(T)=\frac{2\pi^2}{45}g_{*s}(T)T^3\,,
\end{equation}
the present-day energy density is
\begin{equation}
\rho_{A,0}=\frac12\,m_0\,m_A(T_{\rm osc})\,f_0^2\,\varphi_{\rm ini}^2\,\frac{g_{*s,0}}{g_{*s}(T_{\rm osc})}\left(\frac{T_0}{T_{\rm osc}}\right)^3.
\end{equation}

\paragraph{Temperature-dependent regime.} The relic abundance follows,
\begin{equation}
\begin{aligned}
\Omega_A h^2=\frac{\rho_{A,0}}{\rho_{c,0}/h^2}=
&\frac{f_0^2\,\varphi_{\rm ini}^2\,T_0^3}{2(\rho_{c,0}/h^2)}\frac{g_{*s,0}}{g_{*s}(T_{\rm osc})}
\left(3\sqrt{\frac{\pi^2 g_*(T_{\rm osc})}{90}}\right)^{\frac{p+3}{p+2}} \\
& \times M_P^{-\frac{p+3}{p+2}}m_0^{\frac{p+1}{p+2}}\Lambda^{-\frac{p}{p+2}}.
\end{aligned}
\end{equation}
Taking the reference value $p=4$~\cite{Borsanyi:2016ksw} together with the standard values $T_0=2.35\times10^{-13}\,$GeV, $g_{*s,0}=3.91$ and $\rho_{c,0}/h^2=8.10\times10^{-47}\,$GeV$^4$, one finds
\begin{equation}
\begin{aligned}
\Omega_A h^2 \simeq&0.20 \left(\frac{f_0}{10^{12}\,{\rm GeV}}\right)^2\, \varphi_{\rm ini}^2\,\left(\frac{m_0}{10^{-5}\,{\rm eV}}\right)^{5/6}\\
&\times \left(\frac{\Lambda}{0.1\,{\rm GeV}}\right)^{-2/3}\left(\frac{g_*(T_{\rm osc})}{61.75}\right)^{7/12}\left(\frac{61.75}{g_{*s}(T_{\rm osc})}\right)\,,
\label{eq:Omega_general}
\end{aligned}
\end{equation}
valid when $T_{\rm osc}>\Lambda$. Matching the observed dark matter abundance $\Omega_{\rm DM} h^2 \simeq 0.12$ then gives
\begin{equation}\label{eq:f0_pred}
\begin{aligned}
f_0 \simeq7.8 \times 10^{11}\,{\rm GeV}\varphi_{\rm ini}^{-1}\left(\frac{m_0}{10^{-5}\,{\rm eV}}\right)^{-5/12}\left(\frac{\Lambda}{0.1\,{\rm GeV}}\right)^{1/3}\\
\times\left(\frac{g_*(T_{\rm osc})}{61.75}\right)^{-7/24}\left(\frac{g_{*s}(T_{\rm osc})}{61.75}\right)^{1/2}\,.
\end{aligned}
\end{equation}
Substituting the definition \eqref{eq: Adef} into \eqref{eq:f0_pred} fixes
\begin{equation}
\begin{aligned}
K_\varphi(\Theta_0) \simeq1.0 \times 10^{-13}&\varphi_{\rm ini}^{-2}\left(\frac{m_0}{10^{-5}\,{\rm eV}}\right)^{-5/6}
\left(\frac{\Lambda}{0.1\,{\rm GeV}}\right)^{2/3}\\
&\times\left(\frac{g_*(T_{\rm osc})}{61.75}\right)^{-7/12}\left(\frac{g_{*s}(T_{\rm osc})}{61.75}\right)\,.\label{eq:K_class}
\end{aligned}
\end{equation}
Equations \eqref{eq:Omega_general}--\eqref{eq:K_class} require $T_{\rm osc}>\Lambda$, comfortably satisfied by the present estimates with $f_0\sim10^{12}\,$GeV through \eqref{eq:Tosc_ratio}.

\paragraph{Constant-mass regime.} For $f_0$ above \eqref{eq:f0_cross}, the right-hand side of \eqref{eq:Tosc_ratio} is below unity for every $p$. The mass has saturated before oscillations, $m_A(T_{\rm osc})=m_0$, and $p$ drops out. Equation \eqref{eq:Tosc} becomes $T_{\rm osc}=[m_0M_P/3]^{1/2}(90/\pi^2g_*)^{1/4}$, and \eqref{eq:Omega_general} becomes
\begin{equation}
\begin{aligned}
\Omega_Ah^2\simeq2.7\times10^{-3}&\left(\frac{f_0}{10^{12}\,{\rm GeV}}\right)^2\varphi_{\rm ini}^2\left(\frac{m_0}{10^{-5}\,{\rm eV}}\right)^{1/2}\\
&\times\left(\frac{g_*(T_{\rm osc})}{86.25}\right)^{3/4}\left(\frac{86.25}{g_{*s}(T_{\rm osc})}\right) ,
\label{eq:Omega_cold}
\end{aligned}
\end{equation}
and correspondingly
\begin{equation}
\begin{aligned}
K_\varphi(\Theta_0)\simeq7.6\times10^{-12}\,&\varphi_{\rm ini}^{-2}\left(\frac{m_0}{10^{-5}\,{\rm eV}}\right)^{-1/2}\\
&\times\left(\frac{g_*(T_{\rm osc})}{86.25}\right)^{-3/4}\left(\frac{g_{*s}(T_{\rm osc})}{86.25}\right)
\label{eq:K_class_cold}
\end{aligned}
\end{equation}
in place of \eqref{eq:K_class}. Here $m_0=10^{-5}\,$eV gives $T_{\rm osc}\simeq50\,$GeV and $g_*=g_{*s}=86.25$, whereas the normalisation $61.75$ in \eqref{eq:Omega_general}--\eqref{eq:K_class} corresponds to $T_{\rm osc}\simeq1\,$GeV above the QCD transition.

In this regime, neither the crossover temperature $\Lambda$ nor the exponent $p$ appears in \eqref{eq:Omega_cold} and the only property of the confining sector that survives is the zero-temperature susceptibility. 

\paragraph{The two regimes compared.} Eliminating $m_0$ through \eqref{eq:m0_chi} gives
\begin{equation}
\Omega_Ah^2\;\propto\;\varphi_{\rm ini}^2\times
\begin{cases}
\chi_{\rm top}^{5/12}\,\Lambda^{-2/3}\,f_0^{7/6}, & T_{\rm osc}>\Lambda ,\\[2pt]
\chi_{\rm top}^{1/4}\,f_0^{3/2}, & T_{\rm osc}\leq\Lambda ,
\end{cases}
\label{eq:Omega_both}
\end{equation}
with the familiar QCD-axion scaling in the upper line. There, $f_0$ is the only field-space quantity, fixed by the observed abundance as
\begin{equation}
\begin{aligned}
f_0\simeq6.5\times10^{11}\,{\rm GeV}\;&\varphi_{\rm ini}^{-12/7}\left(\frac{\chi_{\rm top}^{1/4}}{0.1\,{\rm GeV}}\right)^{-10/7}\\
&\times\left(\frac{\Lambda}{0.1\,{\rm GeV}}\right)^{4/7} ,
\label{eq:f0_eliminated}
\end{aligned}
\end{equation}
leaving $(\chi_{\rm top},\Lambda)$ and $\varphi_{\rm ini}$ as the only free quantities. For QCD, with the values quoted after \eqref{eq:zeta}, this returns $f_0\simeq1.2\times10^{12}\,$GeV, $m_0\simeq4.6\,\mu$eV and $T_{\rm osc}\simeq1\,$GeV at $\varphi_{\rm ini}\sim1$, the standard QCD axion, of which the fiducial values quoted above are a round-number approximation. The constant-mass regime is necessarily the anthropic one. Already at the crossover \eqref{eq:f0_cross} the observed abundance requires $\varphi_{\rm ini}\simeq9\times10^{-4}$ for a QCD-like sector, and smaller values beyond it.

The classical abundance, therefore, depends on the late-time metric $K_\varphi(\Theta_0)$ through $f_0$, the initial angle $\varphi_{\rm ini}$, and the confining sector, but neither on $H_*$ nor on $K_\varphi(\Theta_*)$.

\paragraph{Isocurvature.} 
We now derive the isocurvature constraint. Since the instanton potential that generates the axion mass switches on only once the Universe has cooled to $T\sim\Lambda$, long after inflation, the chiral symmetry is exact during inflation and $U_{,\varphi}=U_{,\varphi\varphi}=0$. 
Consequently, the effective isocurvature mass $\mathcal{M}_{\varphi\varphi}$ arises only from the variation of the field-space metric along the trajectory, which is slow-roll suppressed around CMB horizon crossing but can become sizeable towards the end of inflation, as verified for the realisation in Sec.\,\ref{sec:axion_sector}. Moreover, on the attractor of Sec.~\ref{sec:currents}, the homogeneous phase is frozen and its fluctuation evolves independently of the curvature perturbation. The phase is therefore a light spectator at horizon crossing. For such a light mode and a slowly varying background, the phase fluctuation is
\begin{equation}
\delta\varphi_* =  \frac{H_*}{2\pi M_P\sqrt{K_\varphi(\Theta_*)}}.
\label{eq:deltaa_star}
\end{equation}
The canonically normalised fluctuation $M_P\sqrt{K_\varphi(\Theta_*)}\,\delta\varphi_*$ retains the light-field value $H_*/2\pi$, so \eqref{eq:deltaa_star} suppresses the angle through a large metric, without making the axion heavy. Between horizon exit and the end of slow roll, the phase has no potential and super-Hubble gradients are negligible. Its perturbation obeys the linearised chiral-current conservation law associated with
Eq.~\eqref{eq:noetherphi}, while the decaying mode is suppressed by
the expansion. The phase perturbation therefore approaches a constant
super-Hubble mode, so that $\varphi_{\rm ini}$ and $\delta\varphi$
retain their horizon-crossing values throughout the slow-roll stage.

Extending this result to the onset of axion oscillations requires that
the subsequent evolution does not significantly amplify the
super-Hubble phase perturbation. This need not hold after inflation, since oscillations of $\Theta$ periodically modulate $K_\varphi$ and can
induce parametric amplification. We discuss this caveat in
Sec.~\ref{sec:axion_sector}. It concerns the statistics of the mode and its correlation with the curvature perturbation as well as its amplitude.

Once $H\sim m_A(T)$, oscillations produce cold dark matter by misalignment. The quadratic amplitude dependence in \eqref{eq:energy_density_osc} gives
\begin{equation}
    \frac{\delta\rho_A}{\rho_A}\simeq 2\,\frac{\delta \varphi_{\rm ini}}{\varphi_{\rm ini}}\equiv S_A\,.
    \label{eq: drho_over_rho_axion}
\end{equation}
Then, if axions constitute a fraction
\begin{equation}
    F_A \equiv \frac{\rho_A}{\rho_{\rm DM}}\,,
\end{equation}
of the total dark matter, while the remaining component is purely
adiabatic, the total CDM entropy perturbation is the mixture
\begin{equation}
    S_{\rm DM}=F_A\,S_A+(1-F_{A})\,S_{\rm c}\,,
    \label{eq: S_ad}
\end{equation}
where $S_X\equiv\delta_X-\tfrac34\delta_\gamma$ is the entropy perturbation of the species $X$, with $\delta_X$ its density contrast and $\delta_\gamma$ that of radiation. Here $c$ denotes the remaining non-axion component. Adiabatic here means $\delta_c=\tfrac34\delta_\gamma$, so its entropy perturbation vanishes identically. In the axion contribution the inherited adiabatic part cancels in the same way, leaving only the misalignment-induced perturbation \eqref{eq: drho_over_rho_axion}. The isocurvature signal is thus carried by the axion alone, diluted by its fraction $F_A$. Moreover, since the turning rate vanishes, the mode is uncorrelated with the curvature perturbation during inflation, $\langle S_{\rm DM}\mathcal R\rangle=0$.

Combining \eqref{eq: drho_over_rho_axion}, \eqref{eq: S_ad} and \eqref{eq:deltaa_star} gives the axion isocurvature power spectrum
\begin{equation}
    \mathcal{P}_S^{(A)}(k) = F_A^2 \left(\frac{H_*}{\pi M_P\sqrt{K_{\varphi}(\Theta_*)}\varphi_{\rm ini}}\right)^2.
    \label{eq:PS_axion}
\end{equation}
The CMB constrains the isocurvature fraction
\begin{equation}
    \beta_{\rm iso}(k_\ast)\equiv\frac{\mathcal P_S^{(A)}(k_\ast)}{A_s+\mathcal P_S^{(A)}(k_\ast)} ,
\end{equation}
which for $\mathcal P_S^{(A)}\ll A_s$ implies
\begin{equation}
    \mathcal P_S^{(A)}(k_\ast)\lesssim \beta_{\rm iso}A_s .
\end{equation}
Combining this with \eqref{eq:PS_axion} bounds the axion fraction in terms of the inflationary scale
\begin{equation}
F_A \lesssim \pi\sqrt{\beta_{\rm iso}A_s}\;\frac{f_*\,\varphi_{\rm ini}}{H_\ast}\simeq2.8\times10^{-5}\,\frac{f_*\,\varphi_{\rm ini}}{H_\ast}\,,
\label{eq:F_bound_iso}
\end{equation}
using $A_s\simeq2.1\times10^{-9}$ and $\beta_{\rm iso}\lesssim0.039$ at $k=0.05\,$Mpc$^{-1}$\,\cite{Planck:2018jri}.
CMB bounds on isocurvature therefore translate into limits on $F_{A}$ and $H_{\ast}$. Since the axion potential is periodic, the homogeneous displacement\footnote{Since the background phase is frozen during inflation, its value at horizon exit coincides with the initial misalignment $\varphi_{\rm ini}$ entering the relic abundance \eqref{eq:K_class} or \eqref{eq:K_class_cold}, according to the regime.} cannot exceed the field-space period, $\varphi_{\rm ini}\lesssim \mathcal{O}(1)$ (for a displacement close to the top of the potential, anharmonic corrections enhance the abundance, and we neglect them here), so the right-hand side of \eqref{eq:F_bound_iso} is also the largest fraction the axion can supply.
Equation \eqref{eq:PS_axion} and the bounds drawn from it are statements at the pivot scale. However, the spectrum is not exactly scale invariant, since $K_\varphi$ is evaluated at each mode's own horizon exit. To first order in slow roll one finds, as shown in App.~\ref{app:mass_der},
\begin{equation}
n_{\rm iso}-1=\frac{2}{3}\frac{\mathcal{M}_{\varphi\varphi}(\Theta_*)}{H_*^2}-2\epsilon_*\,.
\label{eq:n_iso}
\end{equation}
The geometric mass and the variation of the axion metric
describe the same dynamics and must not be counted as
independent sources of suppression. A constant super-Hubble
phase perturbation corresponds to a canonically normalised
entropy mode evolving as $\delta s_\varphi\propto\sqrt{K_\varphi}$. The tilt is blue wherever $K_\varphi$ is decreasing at horizon crossing, as for other axions whose effective decay constant decreases during inflation~\cite{Kasuya:2009up,Chung:2017uzc,Tadepalli:2026mdc}, which is the case in the realisation of Sec.~\ref{sec:axion_sector}. The constraint at CMB scales then says nothing about the small-scale behaviour, examined there. In quoting $\beta_{\rm iso}$ we use the bound derived for an uncorrelated cold dark matter isocurvature mode with a scale-invariant spectrum, which is adequate as long as $|n_{\rm iso}-1|\ll1$.

The isocurvature non-Gaussianity anticipated in Sec.~\ref{se:ps} follows from the second-order relation $S_A=2\,\delta\varphi_{\rm ini}/\varphi_{\rm ini}+(\delta\varphi_{\rm ini}/\varphi_{\rm ini})^2$ \cite{Kawasaki:2008sn,Kawasaki:2008pa}. Since \eqref{eq:PS_axion} bounds the product $F_A\,\delta\varphi_*/\varphi_{\rm ini}$, that weight is at most $(4F_A)^{-1}\sqrt{\beta_{\rm iso}A_s}$, which is $2\times10^{-6}$ for an axion making up the whole of the dark matter, as assumed in the joint condition below and realised in Sec.~\ref{sec:axion_sector}. The linearisation of \eqref{eq: drho_over_rho_axion} is therefore self-consistent and the isocurvature bispectrum unobservable, although both degrade as $F_A^{-1}$ for a subdominant axion. 

\paragraph{The joint condition.} The two constraints can now be combined into a single condition. The observed dark-matter abundance fixes the product $K_\varphi(\Theta_0)\varphi_{\rm ini}^2$ through \eqref{eq:K_class} in the temperature-dependent regime and through \eqref{eq:K_class_cold} in the constant-mass one, while the isocurvature bound (taking $F_A=1$) requires
\begin{equation}
    K_{\varphi}(\Theta_*)\,\varphi_{\rm ini}^2\gtrsim\frac{H_*^2}{\pi^2\,\beta_{\rm iso}\,A_s\,M_P^2} \,.
    \label{eq:iso_constraint}
\end{equation}
Equivalently, \eqref{eq:iso_constraint} is a bound on the decay constant at horizon crossing alone,
\begin{equation}
    f_*\gtrsim\frac{H_*}{\pi\,\varphi_{\rm ini}\sqrt{\beta_{\rm iso}A_s}}\simeq3.5\times10^{4}\,\frac{H_*}{\varphi_{\rm ini}} \,,
    \label{eq:fstar_bound}
\end{equation}
which involves no assumptions about the confining sector or the relic-abundance regime.
Dividing \eqref{eq:iso_constraint} by \eqref{eq:K_class} and by \eqref{eq:K_class_cold} yields, respectively, 
\begin{align}
\frac{K_{\varphi}(\Theta_*)}{K_{\varphi}(\Theta_0)}&\gtrsim1.2\times10^{22}\left(\frac{m_0}{10^{-5}{\rm eV}}\right)^{5/6}\left(\frac{\Lambda}{0.1{\rm GeV}}\right)^{-\frac{2}{3}}\left(\frac{H_*}{M_P}\right)^2,
\label{eq:joint_num}\\
\frac{K_{\varphi}(\Theta_*)}{K_{\varphi}(\Theta_0)}&\gtrsim1.6\times10^{20}\left(\frac{m_0}{10^{-5}\,{\rm eV}}\right)^{1/2}\left(\frac{H_*}{M_P}\right)^2\,,
\label{eq:joint_num_cold}
\end{align}
up to the inverse $g_*$ factors of \eqref{eq:K_class} and \eqref{eq:K_class_cold}. The angle cancels in both bounds. For externally specified $m_0$, their right-hand sides are fixed by $m_0$ and $\Lambda$ in the temperature-dependent regime and by $m_0$ alone in the constant-mass regime. Imposing \eqref{eq:m0_chi} with \eqref{eq: Adef} instead retains the dependence $m_0^{5/6}\propto K_\varphi(\Theta_0)^{-5/12}$ or $m_0^{1/2}\propto K_\varphi(\Theta_0)^{-1/4}$. The bounds then constrain $K_\varphi(\Theta_*)K_\varphi(\Theta_0)^{-7/12}$ and $K_\varphi(\Theta_*)K_\varphi(\Theta_0)^{-3/4}$, with the abundance exponents of \eqref{eq:Omega_both}. This form applies to Sec.~\ref{sec:axion_sector}, where the non-minimal coupling fixes $K_\varphi(\Theta_0)$ and the residual dependence appears in \eqref{eq:joint_num_QCD}.

\paragraph{Comparison with the standard scenario.} 
With a flat field-space metric and fixed dimensionful decay constant $f$, the observed abundance at $f\gg10^{12}$~GeV requires $\varphi_{\rm ini}\ll1$, the \lq\lq anthropic axion window\rq\rq~\cite{Wantz:2009it}. The same $f$ controls axion fluctuations, giving an isocurvature signal $H_*/(f\,\varphi_{\rm ini})$ enhanced by the small angle \cite{Wantz:2009it}. For the QCD-axion scaling in \eqref{eq:Omega_both}, fixed abundance fixes $\varphi_{\rm ini}^2f^{7/6}$, so $f\varphi_{\rm ini}\propto f^{5/12}$. Raising $f$ thus relaxes the bound too slowly to make this window viable at high-scale inflation. In the scale-invariant setup, instead, the two roles are carried by $f_*$ and $f_0$, which the evolution of $\Theta$ can make very different. The enhancement of isocurvature by a small $\varphi_{\rm ini}$ is unchanged, but the geometry
can compensate for it. The decoupling, therefore, has two distinct consequences. It allows the small misalignment angles of the anthropic window to coexist with high-scale inflation, and it permits the classic window with an angle of order unity if the
required metric hierarchy can be realised.

For high-scale inflation $H_*\sim10^{13}\,$GeV $\simeq4\times10^{-6}M_P$, the temperature-dependent condition requires a metric ratio $\gtrsim2\times10^{11}$,
while for low-scale inflation $H_*\sim10^9\,$GeV the bound
relaxes to $\gtrsim2\times10^3$. In the constant-mass regime \eqref{eq:joint_num_cold} gives $\gtrsim2.8\times10^{9}$ and $\gtrsim2.8\times10^1$ at the same two scales for $m_0=10^{-5}\,$eV, and less for a lighter axion.

In terms of physical scales, for $\varphi_{\rm ini}\sim\mathcal{O}(1)$ the observed abundance fixes $f_0\simeq7.8\times10^{11}\,$GeV through \eqref{eq:f0_pred} at the fiducial values, or $f_0\simeq1.2\times10^{12}\,$GeV through \eqref{eq:f0_eliminated} for a QCD confining sector, i.e.\ the classic axion window in either case. The isocurvature bound \eqref{eq:fstar_bound} at $H_*\sim10^{13}\,$GeV demands in turn $f_*\gtrsim0.15\,M_P$. Both scales are compatible with sub-Planckian values, and the required hierarchy is realised without tuning the misalignment angle. Within the general framework, high-scale inflation and a decay constant in the classic axion window are compatible as soon as the field-space metric relaxes by the factor \eqref{eq:joint_num} between horizon crossing and the present epoch ($8\times10^{10}$ rather than $2\times10^{11}$ if the QCD values are used). It is the geometry, and not the misalignment angle or the inflationary scale, that carries the burden. We refer to this configuration, a decay constant in the classic window with $\varphi_{\rm ini}\sim\mathcal{O}(1)$ at $H_*\sim10^{13}\,$GeV, as the classic-window benchmark. Which of the two situations is realised is decided by $f_0$ through \eqref{eq:Tosc_ratio}: the benchmark lies in the temperature-dependent regime, whereas the concrete realisation of Sec.~\ref{sec:HD}, with its super-Planckian $f_0$, lies in the constant-mass one and is anthropic. 

\section{The Higgs-Dilaton scenario}\label{sec:HD}
We now ask whether the metric hierarchy required in
Sec.~\ref{sec:DM} can be realised in the Higgs-Dilaton
scenario \cite{Blas:2011ac,GarciaBellido:2011de,Garcia-Bellido:2012npk,Bezrukov:2012hx,Rubio:2014wta,Karananas:2016kyt,Casas:2017wjh,Casas:2018fum,Piani:2022gon,Piani:2023aof}, and how obtaining it affects inflation.

We consider a scale-invariant extension of the Standard Model with Lagrangian density 
\begin{eqnarray}
\label{eqn:hdm_lagrangianHD}
\frac{\mathcal{L}}{\sqrt{g}}&=& M^2(H,X) R-V(H,X) \\ 
&-& g^{\mu\nu} (D_\mu H)^\dagger(D_\nu H)-g^{\mu\nu} (\partial_\mu X)^\dagger(\partial_\nu X)\nonumber \,,
\end{eqnarray}
with 
\begin{align}
M^2(H,X)&=\xi_h\, H^\dagger H +\xi_\chi X^\dagger X \\
V(H,X)&=\lambda \left(H^\dagger H-\alpha\,X^\dagger X \right)^2\,.
\end{align}
In the unitary gauge, the non-minimal coupling and the kinetic sector are conveniently expressed through the two quadratic invariants
\begin{equation}
    \begin{aligned}
        B&= \xi_h h^2 + \xi_\chi \chi^2 \equiv 2M^2\,,\\
        A&= (1+6\xi_h)h^2 + (1+6\xi_\chi)\chi^2\equiv \bar{\rho}^2\,,
        \label{eq:A}
    \end{aligned}
\end{equation}
in terms of which the dimensionless combination controlling the dynamics reads
\begin{equation}
    \Theta=\gamma^2 \frac{A}{B}\,,
    \qquad
    \gamma^2\equiv\frac{\xi_\chi}{1+6\xi_\chi}\,,
    \label{eq:Theta_AB}
\end{equation}
consistently with the field redefinition \eqref{eq:Thetadef} of App.~\ref{app:HD}, where we review how a Weyl rescaling brings the model to the Einstein-frame form \eqref{eq:sigmamodel} with kinetic functions \eqref{Ktheta}--\eqref{eq:Utheta}. At late times the dilaton dominates the field content, $h \ll \chi$, so that $A/B \rightarrow (1+6\xi_{\chi})/\xi_{\chi}$ and $\Theta \rightarrow \gamma^2(1+6\xi_{\chi})/\xi_{\chi}=1$. We therefore identify $\Theta_0$ with the present epoch.
 
For the axion kinetic function \eqref{eq:Kvarphi}, the geometric suppression of Sec.~\ref{sec:DM} requires $K_\varphi(\Theta_*)/K_\varphi(\Theta_0)\gg1$, quantified by \eqref{eq:joint_num} or \eqref{eq:joint_num_cold}. The Higgs-Dilaton model
by itself fails this requirement, since $\Theta_*\ll\Theta_0=1$,
\begin{equation}
    \frac{K_\varphi^{\rm HD}(\Theta_*)}{K_\varphi^{\rm HD}(\Theta_0)}=\frac{\kappa\,\Theta_*+c}{\kappa+c}\simeq\Theta_*\ll 1\,,
    \label{eq:HDratio}
\end{equation}
i.e. the angular metric is smaller during inflation than today, reversing the required hierarchy, as is generic in non-minimally coupled setups~\cite{Rigouzzo:2025hza,Rigouzzo:2025ycb,Karananas:2025uhy}.
In regimes with $f_*\gg f_0$, post-inflationary resonance and non-thermal PQ restoration have been reported in several constructions~\cite{Kawasaki:2013iha,Harigaya:2015hha,Kearney:2016vqw,Kobayashi:2016qld,Ballesteros:2021bee,Graham:2025iwx,Lyu:2026rwo,Kofman:1995fi,Tkachev:1995md}. We discuss their applicability in Sec.~\ref{sec:axion_sector}.

To restore the hierarchy, we deform the angular kinetic sector by a factor $\Theta^{-n}$ with $n>0$, to be implemented at the level of the Jordan-frame action below. Since $\Theta$ is a dimensionless ratio of the fields that depends on $X$ only through $|X|^2$, the deformation preserves both scale invariance and the chiral symmetry, while the scalar potential \eqref{eq:Utheta} is left untouched.
By construction it then satisfies both requirements of Sec.~\ref{sec:DM}, namely $K_\varphi^{\rm new}(\Theta_0=1)=(\kappa+c)/(\kappa c) =K_\varphi^{\rm HD}(\Theta_0)$, so the late-time value (and with it $f_0$ and the relic abundance) is unchanged, while $K_\varphi^{\rm new}(\Theta_*)\gg K_\varphi^{\rm new}(\Theta_0)$ during inflation, suppressing the isocurvature amplitude.
 
In terms of $A$ and $B$, the deformation is realised through
\begin{equation}
    \frac{1}{\Theta^{n}}=\frac{1}{\gamma^{2n}}\left(\frac{B}{A}\right)^{n}\,,
\end{equation}
where the ratio $B/A=\gamma^2/\Theta$ is much larger during inflation
($\Theta_*\ll1$) than its late-time value $\gamma^2$ at $\Theta_0=1$.

The Jordan-frame action then reads
\begin{equation}
    \begin{aligned}
        \frac{\mathcal{L}}{\sqrt{-g}} &= M^2(H,X) R - (D_\mu H)^\dagger D^\mu H \\
        &- \frac{1}{\gamma^{2n}} \left( \frac{B(H,X)}{A(H,X)} \right)^{n} (\partial_\mu X)^\dagger \partial^\mu X-V(H,X),
    \end{aligned}
\end{equation}
which is still invariant under $X \to e^{ic_2}X$. In terms of $h$, $\chi$ and $\varphi$ it becomes
\begin{widetext}
\begin{equation}
\frac{\mathcal{L}}{\sqrt{-g}}=\frac{\xi_h h^2 + \xi_\chi\chi^2}{2}\,R- \frac{1}{2}(\partial h)^2- \frac{1}{2\gamma^{2n}}\!\left(\frac{\xi_h h^2+\xi_\chi\chi^2}{(1+6\xi_h)h^2+(1+6\xi_\chi)\chi^2}\right)^{\!n}\!\!\left[(\partial\chi)^2 + \chi^2(\partial\varphi)^2\right]- \frac{\lambda}{4}(h^2-\alpha\chi^2)^2.\label{eq:Jmod}
\end{equation}
\end{widetext}
By performing the Weyl rescaling
\begin{equation}
    \tilde g_{\mu\nu}=\Omega^2 g_{\mu\nu}\,,
    \qquad
    \Omega^2 \equiv \frac{B}{M_P^2} = \frac{\xi_h h^2+\xi_\chi\chi^2}{M_P^2}\,,
    \label{eq:Weyl}
\end{equation}
the non-minimal coupling $(B/2)\,R$ turns into the Einstein-Hilbert term $(M_P^2/2)\,\tilde R$, and the kinetic terms of \eqref{eq:Jmod} combine into an Einstein-frame field-space metric $\mathcal{G}_{AB}$ for $\phi^A=(h,\chi,\varphi)$, derived in Apps.~\ref{app:HD} and~\ref{app:kinetic}. The deformation multiplies only the explicit $\chi$ and $\varphi$ kinetic terms by $\Theta^{-n}$, so it modifies only the $\chi\chi$ and $\varphi\varphi$ entries of $\mathcal{G}_{AB}$. The gradient term generated by the Weyl rescaling is that of the Higgs-Dilaton model, and the axion direction remains block-diagonal, $\mathcal{G}_{h\varphi}=\mathcal{G}_{\chi\varphi}=0$. The Higgs-Dilaton metric is recovered for $n=0$ and, at late times, at $\Theta_0=1$.

We now pass to the variables $(\Theta,\Phi,\varphi)$. Moving along $\Phi$ at fixed $\Theta$ is a uniform rescaling $h\to\lambda h$, $\chi\to\lambda\chi$ (the dilaton direction), while moving along $\Theta$ at fixed $\Phi$ changes the ratio $h/\chi$ at constant scale $\bar\rho$ (the inflaton direction), along which we write $h=\bar\rho\,f_1(\Theta)$ and $\chi=\bar\rho\,f_2(\Theta)$. The kinetic coefficients $K_\Theta^{\rm new}$, $K_\Phi^{\rm new}$ and $K_{\Theta\Phi}$ are the projections of $\mathcal{G}_{AB}$ on these two directions; the change of variables, the functions $f_{1,2}$ and the projections are worked out in App.~\ref{app:kinetic}. Here and below the superscript \lq\lq new\rq\rq\ distinguishes the kinetic coefficients of the modified model from their HD counterparts. The variables $(\Theta, \Phi)$ diagonalise the HD sector (App.\,\ref{app:kinetic}), so $K_{\Theta\Phi}$, which vanishes identically for $n=0$, carries no superscript.

$K_\varphi$ requires no projection
\begin{equation}
    M_P^2 K_\varphi^{\rm new}= \mathcal{G}_{\varphi\varphi}= \frac{\Theta^{-n}\chi^2 M_P^2}{B}\,,
    \label{eq:Kphi_from_G}
\end{equation}
and using \eqref{eq:Kvarphi} we can rewrite it as
\begin{equation}
    K_\varphi^{\rm new}(\Theta)= \Theta^{-n}\,K_\varphi^{\rm HD}(\Theta)\,.
    \label{eq:Kphi_new}
\end{equation}
The dilaton coefficient reads
\begin{align}
&K_\Phi^{\rm new}(\Theta)= \Theta\bigl[1+\bigl(\Theta^{-n}-1\bigr)f_2^2\bigr]\,,\label{eq:KPhi_new}\\
f_2^2(\Theta)&=\frac{\xi_h(\Theta-\Theta_{\rm bd})}{(\xi_h-\xi_\chi)\,\Theta}\,,\quad\Theta_{\rm bd}\equiv\frac{\gamma^2(1+6\xi_h)}{\xi_h}\,,\nonumber
\end{align}
where $\Theta_{\rm bd}$ is the value of $\Theta$ at which $\chi\to0$ ($f_2(\Theta_{\rm bd})=0$); it will play the role of the inflationary boundary in Sec.~\ref{subsec:inflation_obs}. We recover $K_\Phi^{\rm HD}=\Theta$ for $n=0$. The correction is the projection of the modified $\chi$ kinetic term onto the $\Phi$ direction. 

The inflaton coefficient reads
\begin{equation}
K_\Theta^{\rm new}(\Theta)= K_\Theta^{\rm HD}(\Theta)+ \bigl(\Theta^{-n}-1\bigr)\frac{\Theta\,f_2^{\prime\,2}}{\gamma^2}\,,
\label{eq:KTheta_new}
\end{equation}
where $K_\Theta^{\rm HD}$ is given in \eqref{Ktheta} and the correction is the projection of the modified explicit $\chi$ kinetic term onto the $\Theta$ direction.
Since the HD sector is diagonal in these variables, the off-diagonal coefficient comes entirely from the $\Theta^{-n}$ factor in the $\chi\chi$ entry of $\mathcal{G}_{AB}$ and reads
\begin{equation}
K_{\Theta\Phi}(\Theta)= \bigl(\Theta^{-n}-1\bigr)\frac{\gamma\,(1+6\xi_h)}{2\Theta\,(\xi_h-\xi_\chi)}\,.
\label{eq:KTP_box}
\end{equation}
For $\xi_h\gg\xi_\chi$ the off-diagonal term is not a small perturbation and has significant dynamical consequences, as we show below. The complete $(\Theta,\Phi,\varphi)$ kinetic Lagrangian is
\begin{align}
\frac{\mathcal{L}}{M_P^2}= -\frac{1}{2}K_\Theta^{\rm new}(\partial\Theta)^2& -K_{\Theta\Phi}\,\partial\Theta\,\partial\Phi- \frac{1}{2}K_\Phi^{\rm new}(\partial\Phi)^2\nonumber\\
&- \frac{1}{2}K_\varphi^{\rm new}(\partial\varphi)^2- \frac{U(\Theta)}{M_P^2}\,.
\label{eq:Lkin_full}
\end{align}
The mixed kinetic term does not take the model outside
the class of Sec.~\ref{sec:scaleinv}: a shift of the dilaton
coordinate by a function of $\Theta$ restores the diagonal
form.\footnote{The redefinition $\Phi'=\Phi+\int(K_{\Theta\Phi}/K_\Phi^{\rm new})\,d\Theta$ brings the metric to the diagonal form, with
$K_\Theta\to\widetilde K_\Theta$ and $K_\Phi\to K_\Phi^{\rm new}$. The attractor then
corresponds to $\dot\Phi'=0$.} We retain the present coordinates to make the
relation to the original Higgs-Dilaton model explicit.

\begin{figure}
    \centering
    \includegraphics[width=1\linewidth]{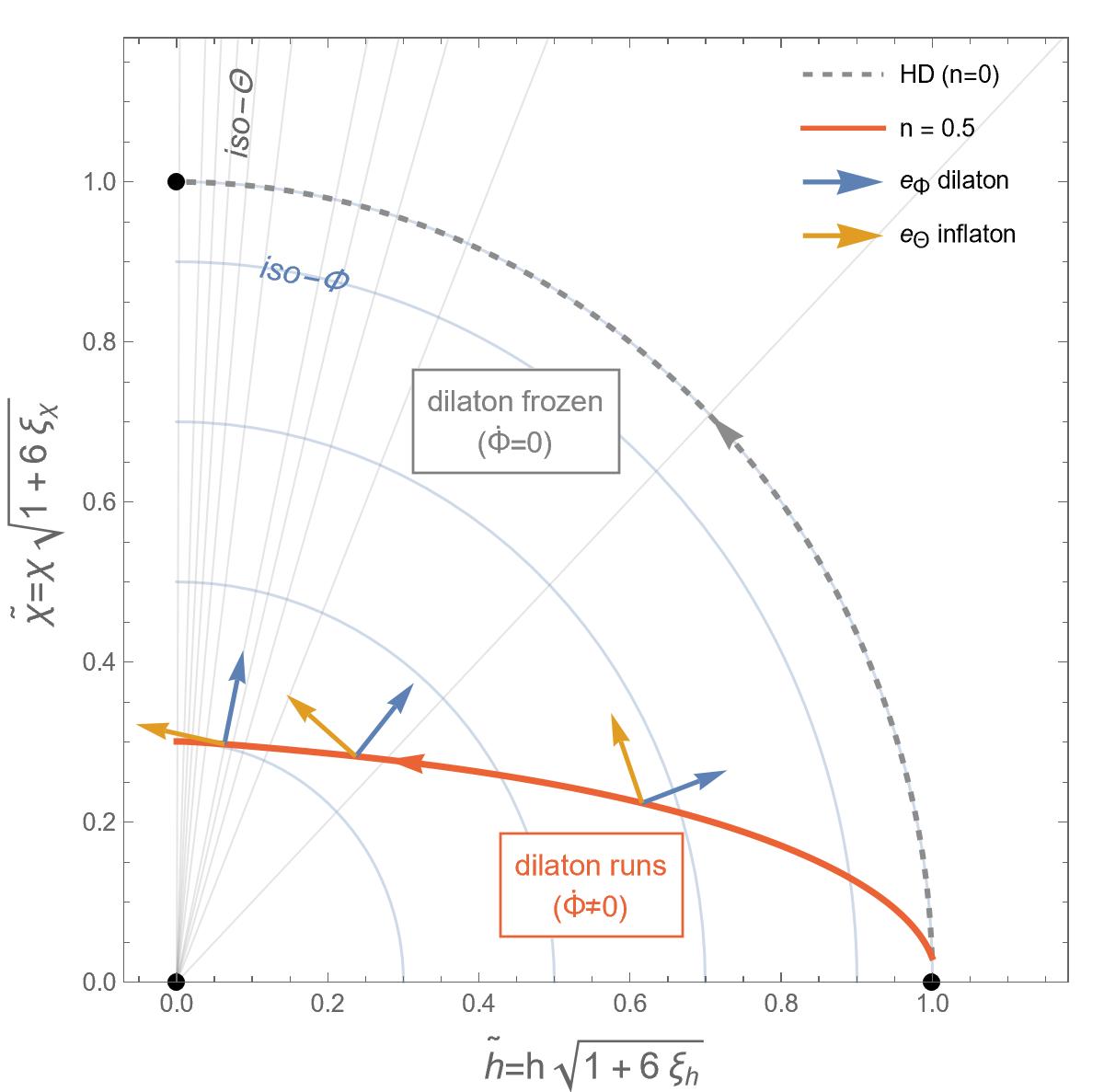}
    \caption{Field-space trajectories in the $(h,\chi)$ plane. The axes are the rescaled variable $(\tilde{h},\tilde{\chi})$ in units of $\bar{\rho}$, for which lines of constant $\bar{\rho}$ are circular arcs. For $n=0$ (HD) the dilaton is frozen and the trajectory follows an arc of constant $\bar\rho$; for $n>0$ the off-diagonal coefficient $K_{\Theta\Phi}$ drives $\dot{\bar\rho}<0$, eq.~\eqref{eq:rhodot}, and the trajectory spirals inward.} 
    \label{fig:field_space}
\end{figure}

The two-field kinetic energy in the $(\Theta,\Phi)$ sector
defines the total squared field velocity
\begin{equation}
\dot\sigma^2 \equiv M_P^2\bigl(K_\Theta^{\rm new}\dot\Theta^2+ 2K_{\Theta\Phi}\dot\Theta\dot\Phi+ K_\Phi^{\rm new}\dot\Phi^2\bigr)\,.
\label{eq:sigmadot}
\end{equation}
We stress that the $\varphi$ sector decouples from the inflationary dynamics; $K_\varphi^{\rm new}$ enters only the isocurvature sector. 
Since the kinetic coefficients in~\eqref{eq:Lkin_full} depend only on $\Theta$ and the potential is independent of $\Phi$, the Lagrangian possesses the shift symmetry $\Phi\to\Phi+c$. Hence, the associated Noether current \eqref{eq:noetherphi} generalises to
\begin{equation}
J^\mu_\Phi \equiv -M_P^2\bigl(K_\Phi^{\rm new}\,\partial^\mu\Phi + K_{\Theta\Phi}\,\partial^\mu\Theta\bigr)\,,
\label{eq:Noether_Phi}
\end{equation}
and satisfies $\nabla_\mu J^\mu_\Phi = 0$. In flat FLRW this reads
\begin{equation}
\frac{d}{dt}\!\left[a^3 J^0_\Phi\right] = 0\,,
\end{equation}
so the canonical momentum $J^0_\Phi=M_P^2\bigl(K_\Phi^{\rm new}\dot\Phi+K_{\Theta\Phi}\dot\Theta\bigr)$ redshifts as $a^{-3}$ and is rapidly diluted during inflation.
On the resulting attractor
\begin{equation}
\dot\Phi = -\frac{K_{\Theta\Phi}}{K_\Phi^{\rm new}}\,\dot\Theta.
\label{eq:attractor}
\end{equation}
For $n=0$, the vanishing off-diagonal coefficient makes dilution of $J^0_\Phi$ enforce $\dot\Phi=0$, the standard frozen-dilaton HD attractor. For $n>0$, dilution instead sets the full combination in~\eqref{eq:Noether_Phi} to zero, carrying the dilaton along with the inflaton through~\eqref{eq:attractor}.
Using \eqref{eq:Thetadef}, the attractor~\eqref{eq:attractor} translates directly into a rate of change of the overall scale,
\begin{equation}
\frac{\dot{\bar\rho}}{\bar\rho}= -\frac{\gamma\,K_{\Theta\Phi}}{K_\Phi^{\rm new}}\,\dot\Theta\,.
\label{eq:rhodot}
\end{equation}
In the original variables, the field velocity decomposes along the two basis vectors,
\begin{equation}
(\dot h,\dot\chi)=\dot\Theta\,e_\Theta+\dot\Phi\,e_\Phi\,,
\label{eq:velocity_decomp}
\end{equation}
combining inflaton motion along $e_\Theta$ with scaling motion along $e_\Phi$. For $n>0$, $K_{\Theta\Phi}>0$ in the inflationary window $\Theta_{\rm bd}<\Theta<1$ (where $\Theta^{-n}>1$), while $\dot\Theta>0$ as the inflaton rolls toward $\Theta_0=1$. Equation~\eqref{eq:rhodot} then gives $\dot{\bar\rho}<0$. The scale therefore shrinks monotonically and the trajectory spirals inward (Fig.~\ref{fig:field_space}). 
Substituting~\eqref{eq:attractor} into the field-space kinetic energy \eqref{eq:sigmadot} defines the effective inflaton kinetic function,
\begin{equation}
\dot\sigma^2 = M_P^2\,\widetilde K_\Theta\,\dot\Theta^2\,, \quad \widetilde K_\Theta(\Theta) \equiv K_\Theta^{\rm new} - \frac{K_{\Theta\Phi}^2}{K_\Phi^{\rm new}}.
\label{eq:Keff_def}
\end{equation}
Thus $\widetilde K_\Theta$ governs the inflaton dynamics. Equations~\eqref{eq:KPhi_new}, \eqref{eq:KTheta_new} and~\eqref{eq:KTP_box} give
\begin{equation}
\widetilde K_\Theta(\Theta) = K_\Theta^{\rm HD}(\Theta)+ \frac{\bigl(\Theta^{-n}-1\bigr)\,\dfrac{\Theta}{\gamma^2}\,f_2^{\prime\,2}}{1+\bigl(\Theta^{-n}-1\bigr)\,f_2^2}\,.
\label{eq:Keff_closed}
\end{equation}
Along the attractor \eqref{eq:attractor}, the $(\Theta,\Phi)$ sector reduces to the single adiabatic field $\sigma$ defined above, and the turning rate
continues to vanish. Although $\dot{\phi}^I=\dot{\Theta}\,\delta^I_{\Theta}$ no longer holds when $K_{\Theta\Phi}\neq0$, the covariant argument of Sec.~\ref{sec:currents} does not rely on it, as anticipated there. Hence $\omega^I=0$ also in the deformed theory, without invoking slow roll. The curvature perturbation is therefore still conserved and the equations of Sec.~\ref{se:ps} apply with $K_{\Theta}\rightarrow\widetilde K_\Theta$, the only change in the decomposition of App.~\ref{appendix:multifield} being that $Q_s^\Phi$ no longer coincides with $Q^\Phi$.
Since the axion block is unchanged, the $\varphi$ equation \eqref{Qvarphi} and \eqref{eq:ddotSvarphi} carry over unchanged. However, what the reduction does not remove is the entropic mode of the $(\Theta,\Phi)$ sector. It is light, obeys the same bound as the axion one, and acquires de Sitter fluctuations of its own. These are harmless. Equation\,\eqref{eq:turning_rate} forbids their conversion into curvature perturbations, and $\Phi$ is the Goldstone mode of the exact dilatation symmetry, with no potential and with the Einstein-frame masses of the matter fields depending on the ratio $\Theta$ alone, so that a displacement of $\Phi$ is not observable~\cite{GarciaBellido:2011de}.

\subsection{Pole structure and inflationary observables}\label{subsec:inflation_obs}

\begin{figure}
    \centering
    \includegraphics[width=1\linewidth]{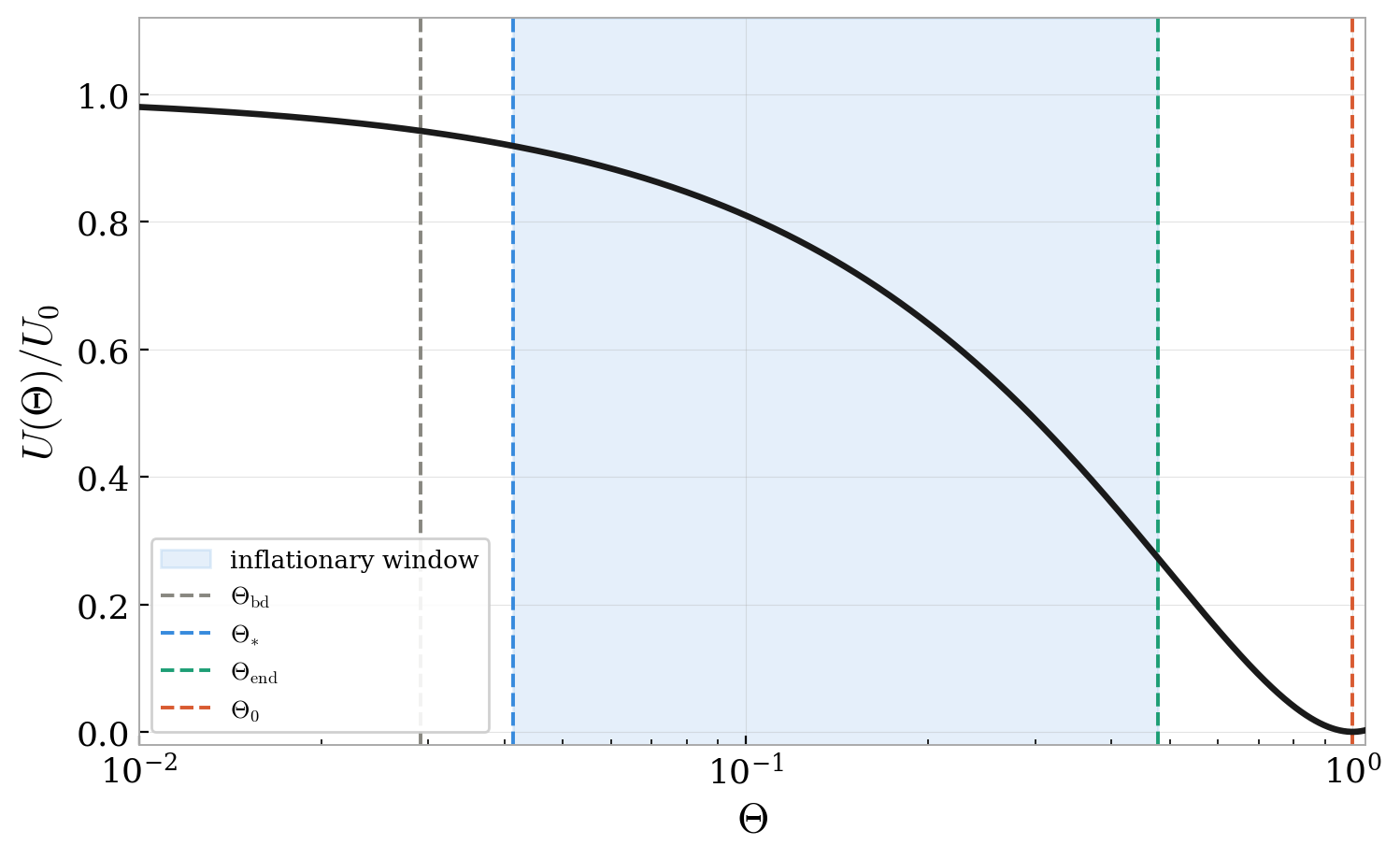}
   \caption{Scale-invariant normalised potential (black); the shaded region is the observable inflationary window (defined below). This is computed with the effective kinetic function \eqref{eq:Keff_closed} for $\xi_h=10^3$, $\xi_\chi=5\times10^{-3}$, $n=2$ and $N_*=55$. We get $\Theta_{\rm bd}=0.029$, $\Theta_*=0.041$, $\Theta_{\rm end}=0.478$. The pivot scale sits just above the pole at $\Theta_{\rm bd}$; the minimum lies at $\Theta_0=1$.}
    \label{fig:potential_theta}
\end{figure}

The observables follow from the pole structure of $\widetilde K_\Theta$ in \eqref{eq:Keff_closed}. In pole inflation, the pole order and residue at the field-space boundary determine the spectral index and tensor-to-scalar ratio~\cite{Galante:2014ifa}. Here the pole lies at $\Theta_{\rm bd}$, fixed geometrically by $\xi_\chi$ and $\xi_h$. 
As illustrated in Fig.~\ref{fig:potential_theta}, the pivot scale $\Theta_*$ sits close to this pole on the scale set by the potential, whose minimum is at $\Theta_0=1$, so the entire observable window
$\Theta_*\lesssim\Theta\lesssim\Theta_{\rm end}$ is controlled by the pole structure. 
In terms of the couplings, the boundary introduced in \eqref{eq:KPhi_new} reads
\begin{equation}
    \Theta_{\rm bd} = \frac{\gamma^2(1+6\xi_h)}{\xi_h}= \frac{\xi_\chi(1+6\xi_h)}{\xi_h(1+6\xi_\chi)}\,.
    \label{eq:Thetabd}
\end{equation}
Defining the distance to the boundary as $\delta\equiv\Theta-\Theta_{\rm bd}$, the effective kinetic function \eqref{eq:Keff_closed} has two regimes, worked out in App.~\ref{app:pole}. In a thin shell $\delta\lesssim\delta_c$ around the pole, with $\delta_c$ given in \eqref{eq:crossover}, the dilaton subtraction in \eqref{eq:Keff_closed} is negligible and $\widetilde K_\Theta$ keeps the simple pole of the HD model, with residue enhanced by $\Theta_{\rm bd}^{-n}$. Outside it, the subtraction dominates and the two regimes match at the crossover.

For the parameters of interest, this shell is extremely thin -- e.g.\ $\delta_c\sim2\times10^{-5}$ for $n=2$ and $\xi_\chi=5\times10^{-3}$ -- while the pivot scale sits at $(\Theta_*-\Theta_{\rm bd})\sim10^{-2}$, some hundred times further out. The entire observable window $\Theta_*\lesssim\Theta\lesssim\Theta_{\rm end}$ therefore lies in the regime $\delta\gg\delta_c$, where the dilaton subtraction dominates over unity in the denominator of \eqref{eq:Keff_closed}. There, the correction in~\eqref{eq:Keff_closed} saturates to
\begin{equation}
    \widetilde K_\Theta - K_\Theta^{\rm HD}\xrightarrow{\;\delta\gg\delta_c\;}\frac{\Theta_{\rm bd}^2}{4\,\gamma^2\,\Theta\,(\Theta-\Theta_{\rm bd})^2}\equiv\frac{a\, \Theta_{\rm bd}}{\Theta(\Theta-\Theta_{\rm bd})^2}\,.
    \label{eq:saturated}
\end{equation}
Adding $K_{\Theta}^{\rm HD}\simeq a/(\Theta(\Theta-\Theta_{\rm bd}))$, which follows from \eqref{Ktheta} and the identity $c=-\kappa\Theta_{\rm bd}$ up to a term suppressed by $1/\xi_h$, cancels the factors of $\Theta$ between the two contributions and leaves
\begin{equation}
    \widetilde K_\Theta \simeq \frac{a}{(\Theta-\Theta_{\rm bd})^2}
    \label{eq:Ktil_pole}
\end{equation}
a second-order pole at $\Theta_{\rm bd}$, with residue
\begin{equation}
    a = \frac{\Theta_{\rm bd}}{4\gamma^2}=\frac{1+6\xi_h}{4\,\xi_h}\;\xrightarrow{\;\xi_h\gg1\;}\;\frac{3}{2}\,,
    \label{eq:residue_universal}
\end{equation}
independent of both $n$ and $\xi_\chi$. With this constant residue, \eqref{eq:Ktil_pole} holds across the observable window, not just near the boundary. Since $\widetilde K_\Theta$ depends only on $\delta=\Theta-\Theta_{\rm bd}$, so do $\varepsilon_{\rm SR}$ and $dN/d\Theta$, up to the slowly varying factor $(1-\Theta)$. The pivot therefore lies at an essentially fixed \emph{distance} from the pole. Hence, increasing $\xi_\chi$ moves the pole and observable window together, raising $\Theta_*$ without changing that distance.

\begin{figure}
    \centering
    \includegraphics[width=1\linewidth]{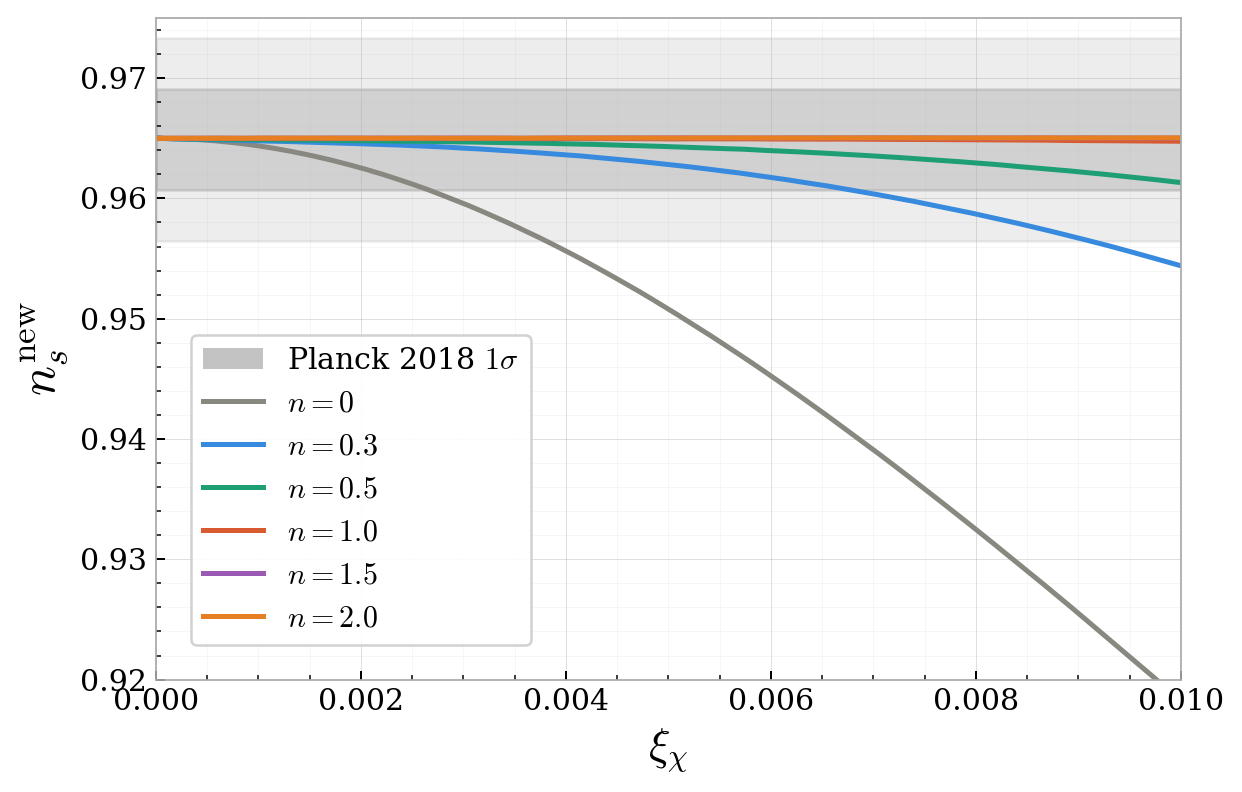}
    \caption{Spectral index $n_s$ vs\ $\xi_\chi$ for several values of $n$, from numerical slow-roll integration with the effective inflaton kinetic function $\widetilde K_\Theta$. Grey bands show the Planck\,2018 $1\sigma$/$2\sigma$ regions~\cite{Planck:2018jri}. For $n\gtrsim1$ the curves collapse onto \eqref{eq:obs_staro},
    essentially independent of $n$ and $\xi_\chi$; the $n=0$ (HD) curve drifts to lower $n_s$ as $\xi_\chi$ grows.}
    \label{fig:ns_numerical}
\end{figure}

\begin{figure}
    \centering
    \includegraphics[width=1\linewidth]{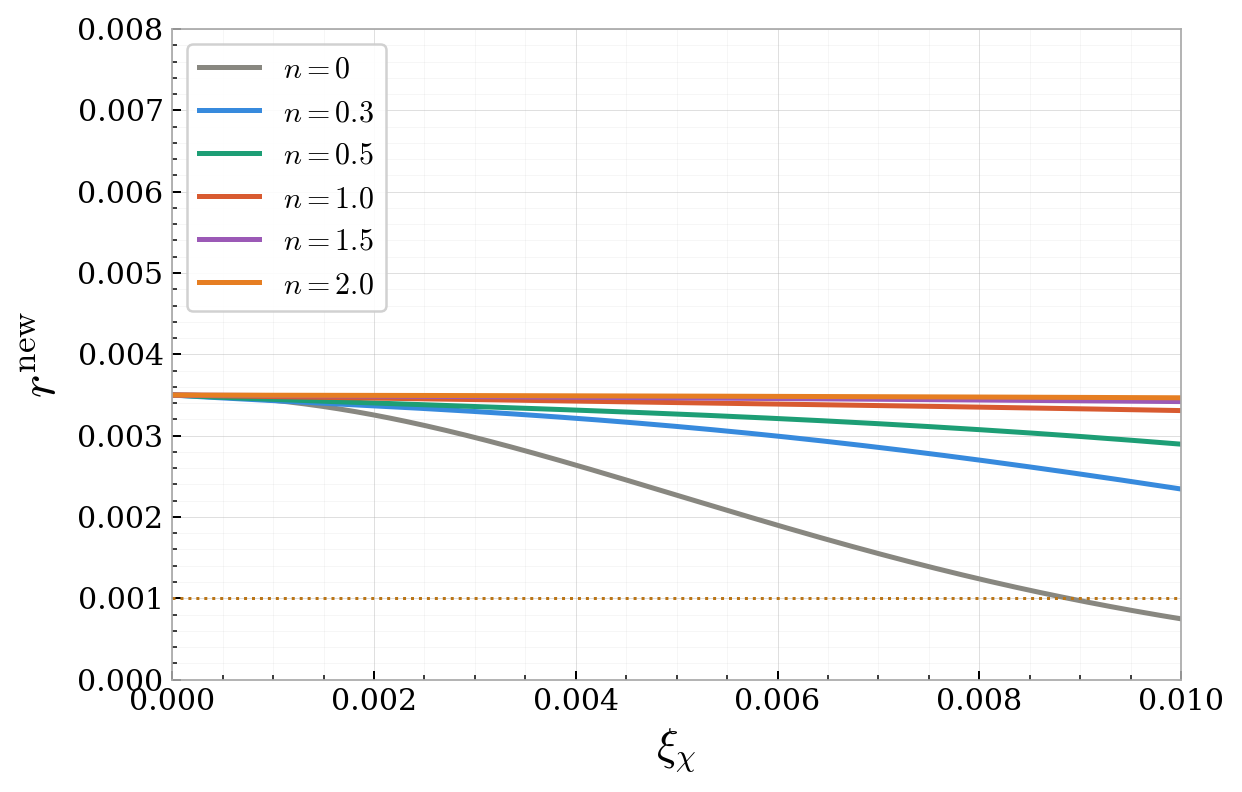}
    \caption{Tensor-to-scalar ratio $r$ vs\ $\xi_\chi$ for several values of $n$. The dotted line is the projected LiteBIRD sensitivity $r\sim10^{-3}$~\cite{LiteBIRD:2022cnt}, the current bound is $r<0.036$ from BICEP/Keck+Planck~\cite{BICEP:2021xfz}. For $n\gtrsim1$ the curves stay at the plateau of \eqref{eq:obs_staro} across the whole range, while the $n=0$ (HD) curve falls toward $r\sim10^{-3}$ at large $\xi_\chi$.}
    \label{fig:r_vs_xic}
\end{figure}

The same pole structure also explains the behaviour of the undeformed model \cite{Karananas:2016kyt,Casas:2017wjh,Casas:2018fum,Karananas:2020qkp}. Written in the same way, $K_\Theta^{\rm HD}\simeq a/[\Theta(\Theta-\Theta_{\rm bd})]$ carries two simple poles, at the origin and at $\Theta_{\rm bd}$, which merge into a second-order pole only in the limit $\Theta_{\rm bd}\to0$. The Higgs--Dilaton model therefore sits at the attractor prediction while $\Theta_{\rm bd}\ll\Theta_*$ and drifts away from it once the two become comparable, which is what the $n=0$ curves of Figs.~\ref{fig:ns_numerical} and~\ref{fig:r_vs_xic} show. In contrast, the deformation removes that condition. By \eqref{eq:Ktil_pole}, the kinetic function exhibits an effective second-order pole at $\Theta_{\rm bd}$ throughout the observable inflationary window, for any $\xi_\chi$ in the regime considered.

\begin{figure}
    \centering
    \includegraphics[width=1\linewidth]{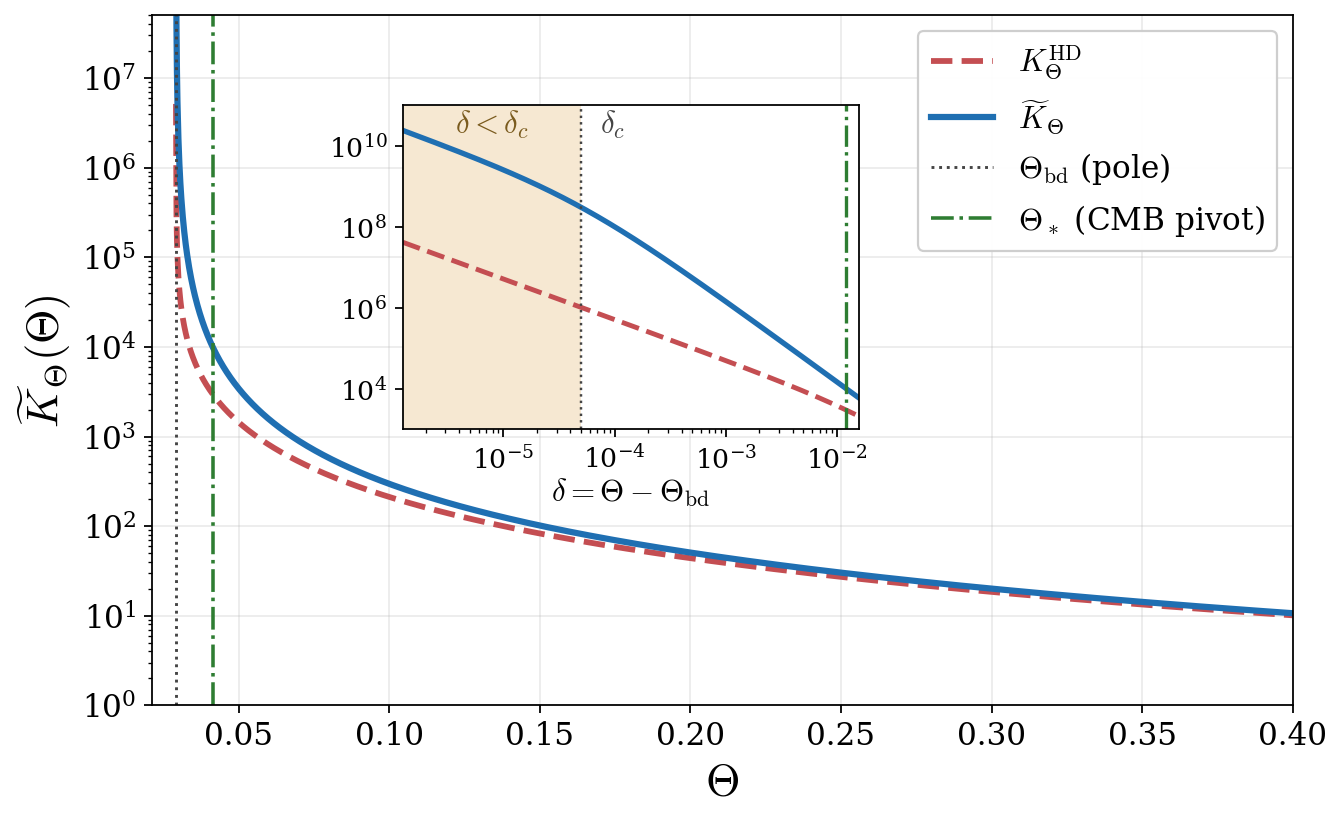}
    \caption{Effective inflaton kinetic function \eqref{eq:Keff_closed} (solid) compared with the HD baseline $K_\Theta^{\rm HD}$ (dashed), for the same parameters as Fig.\,\ref{fig:potential_theta}. The dilaton subtraction promotes the simple pole at $\Theta_{\rm bd}$ to a second-order pole, enhancing $\widetilde K_\Theta$ by $\sim3\times$ at the pivot scale $\Theta_*$. The inset shows that the simple-pole regime survives only in the thin shell $\delta<\delta_c$ (beige shaded), while the entire observable window lies in the second-order regime.}
    \label{fig:pole_offdiag}
\end{figure}

At fixed distance from the pole, the coupling changes only the relative contributions to \eqref{eq:Ktil_pole}. Dividing \eqref{eq:saturated} by \eqref{eq:Ktil_pole}, the deformation contributes a fraction $\Theta_{\rm bd}/\Theta$ of $\widetilde K_\Theta$, explaining the wide separation of the solid and dashed curves in Fig.~\ref{fig:pole_offdiag} and their near-coincidence at $\xi_\chi=5\times10^{-5}$.
An order-two pole with fixed residue defines an $\alpha$-attractor \cite{Kallosh:2013hoa,Kallosh:2013yoa,Galante:2014ifa,Artymowski:2016pjz}. The residue \eqref{eq:residue_universal} corresponds to the Starobinsky point $\alpha=1$, and protects the inflationary predictions against variations of $n$ and $\xi_\chi$.

With \eqref{eq:Keff_closed} the slow-roll parameters and $e$-fold rate read
\begin{align}
    \varepsilon_{\rm SR}(\Theta)    &= \frac{2}{\widetilde K_\Theta(\Theta)\,(1-\Theta)^2}\,,
    \label{eq:epsSR}\\
    \epsilon_2(\Theta)    &= \frac{2}{\widetilde K_\Theta(\Theta)\,(1-\Theta)}\frac{d\ln\varepsilon_{\rm SR}}{d\Theta}\,,\\
    \frac{dN}{d\Theta}&= \frac{\widetilde K_\Theta(\Theta)\,(1-\Theta)}{2}\,.
    \label{eq:dNdT}
\end{align}
We find $\Theta_{\rm end}$ from $\varepsilon_{\rm SR}(\Theta_{\rm end})=1$, determine $\Theta_*$ by integrating~\eqref{eq:dNdT} backward $N_*=55$ $e$-folds\footnote{A representative value for high-scale inflation with efficient reheating. Under a change of $N_*$, the pinned predictions below shift according to $n_s\simeq 1-2/N_*$ and $r\simeq 12/N_*^2$.}, and evaluate $n_s-1=-2\varepsilon_{\rm SR}-\epsilon_2$ and $r=16\,\varepsilon_{\rm SR}$ there, with $\epsilon_2$ the same quantity as in Sec.~\ref{se:ps}. Explicitly,
\begin{align}
    n_s &= 1 - \frac{8}{\widetilde K_\Theta(\Theta_*)\,(1-\Theta_*)^2}+ \frac{2\,\widetilde K_\Theta'(\Theta_*)}{[\widetilde K_\Theta(\Theta_*)]^2\,(1-\Theta_*)}\,,
    \label{eq:ns_formula}\\
    r &= 16\,\varepsilon_{\rm SR}(\Theta_*)\,.
    \label{eq:r_formula}
\end{align}
Fig.~\ref{fig:ns_numerical} and Fig.~\ref{fig:r_vs_xic} show $n_s$ and $r$ as functions of $\xi_\chi$ for several $n$.
Accordingly, the predictions lock onto the $\alpha$-attractor values
\begin{equation}
    n_s\simeq0.965\,,\qquad r\simeq3.5\times10^{-3}\,,
    \label{eq:obs_staro}
\end{equation}
to within a few parts in $10^{3}$ in $n_s$ and a few percent in $r$, for all $n\gtrsim1$ and across the entire HD-allowed range $\xi_\chi\lesssim10^{-2}$. Equation \eqref{eq:obs_staro} is what \eqref{eq:ns_formula} and \eqref{eq:r_formula} return at $N_*=55$; the familiar $\alpha$-attractor expressions $n_s\simeq1-2/N_*$ and $r\simeq12/N_*^2$ are their leading order in $1/N_*$ and give $0.964$ and $4.0\times10^{-3}$. 

\begin{figure}
    \centering
    \includegraphics[width=1\linewidth]{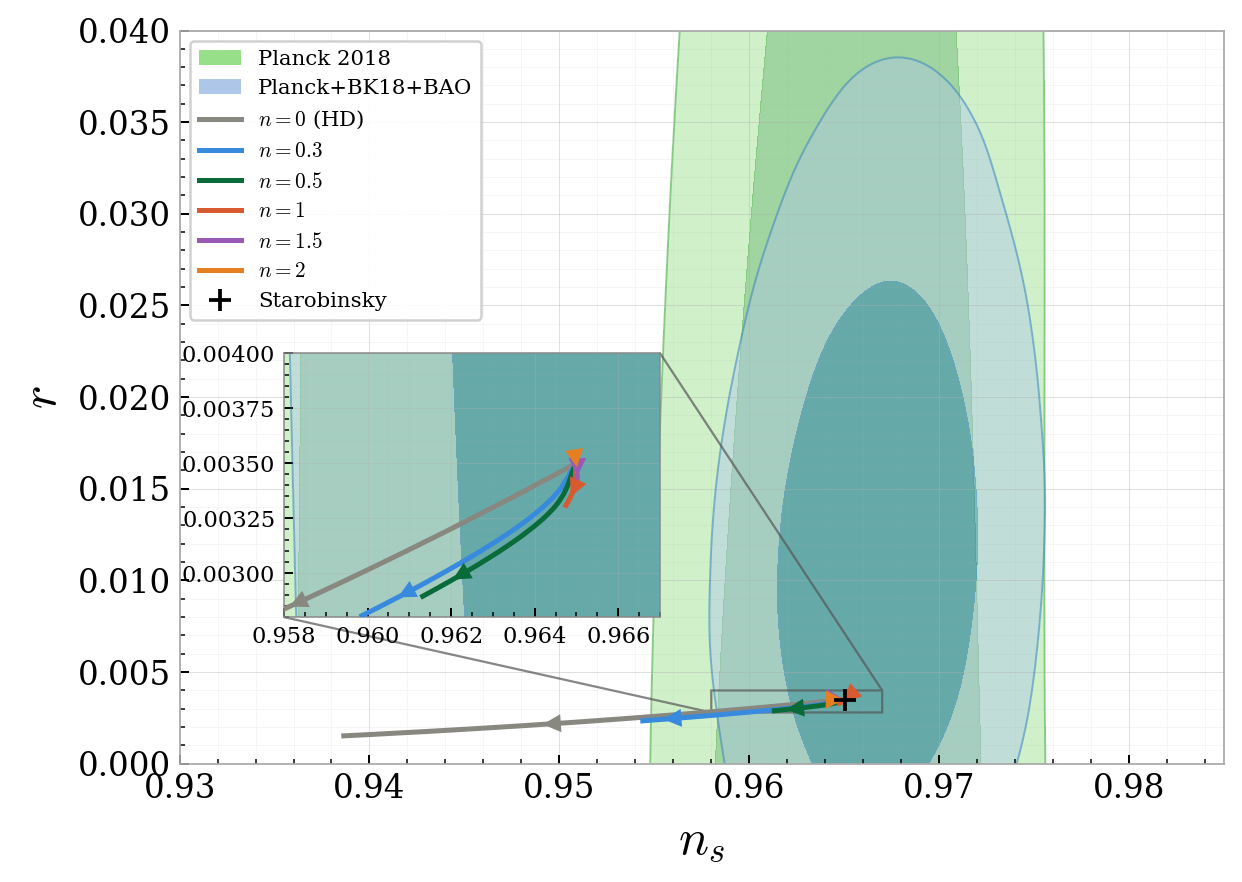}
    \caption{Joint trajectories in the $(n_s,r)$ plane for several values of $n$ as $\xi_\chi$ varies from $0$ to $0.01$; arrows indicate increasing $\xi_\chi$. The $n=0$ curve is the HD baseline. The cross marks \eqref{eq:obs_staro}, on which every $n\gtrsim1$ curve stays. Filled contours show the Planck\,2018 and Planck+BK18+BAO $68\%$ and $95\%$  regions~\cite{Planck:2018jri,BICEP:2021xfz}.}
    \label{fig:ns_r_plane}
\end{figure}

The robustness follows from the dilaton subtraction in \eqref{eq:Keff_def}. The denominator $1+(\Theta^{-n}-1)f_2^2$ bounds the correction, which in $K_\Theta^{\rm new}$ alone would grow linearly in $(\Theta^{-n}-1)$. At the pivot, $\Theta_*\ll1$ and $(\Theta^{-n}-1)f_2^2\gg1$ already for $n\gtrsim1$, so the correction saturates. Thus Figs.~\ref{fig:ns_numerical}--\ref{fig:ns_r_plane}, drawn for $n\leq2$, also represent the larger exponents selected by the isocurvature bound in Sec.~\ref{sec:axion_sector}.

Figure~\ref{fig:ns_r_plane} shows the $(n_s,r)$ trajectories. The modification is distinguishable from HD at moderate $\xi_\chi$, where HD drifts toward lower $n_s$, but different $n>0$ values are nearly degenerate. All $n>0$ predictions lie within the Planck+BK18+BAO $68\%$ region. We note that CMB analyses including ACT DR6 data prefer a somewhat larger spectral index than Planck alone, placing the Starobinsky point at $N_*=55$ up to $\sim3\sigma$ tension with those combinations, the one with CMB lensing and DESI BAO giving $n_s=0.974\pm0.003$ \cite{AtacamaCosmologyTelescope:2025blo,AtacamaCosmologyTelescope:2025nti}. Because \eqref{eq:obs_staro} is pinned, this makes the scenario sharply falsifiable. The only quantity the model leaves free here is $N_*$, and a larger $N_*$, which requires a shorter reheating stage or one with an equation of state stiffer than radiation \cite{Haque:2025uri}, raises $n_s$ through $n_s\simeq1-2/N_*$ without touching the attractor structure.~\footnote{An alternative possibility is to modify the end of
inflation \cite{Cembranos:2026cat}.}

\subsection{Axion sector} \label{sec:axion_sector}

\begin{figure}
    \centering
    \includegraphics[width=1.05\linewidth]{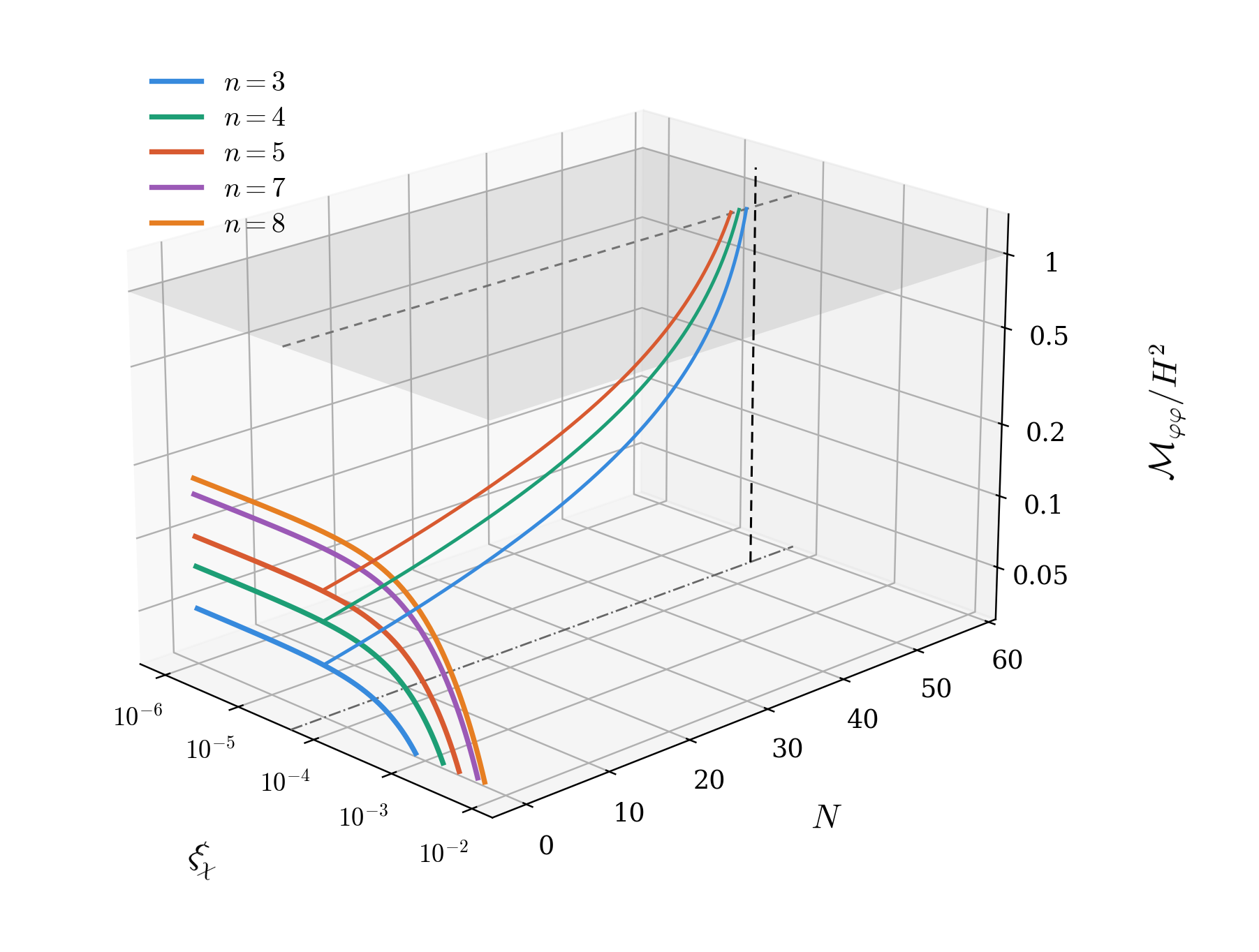}
    \caption{Effective isocurvature mass $\mathcal{M}_{\varphi\varphi}/H^2$ of the axion phase. Thick lines show its value at horizon crossing, in the $\mathcal{M}_{\varphi\varphi}/H^2$--$\xi_\chi$ plane evaluated at the CMB pivot scale, for $n=3,\ldots,8$. Thin lines show its evolution along the trajectory, in the $\mathcal{M}_{\varphi\varphi}/H^2$--$N$ plane at $\xi_\chi=5\times10^{-5}$ (dash-dotted line), for $n=3,4,5$, with $N$ the number of $e$-folds after horizon crossing. The gray surface indicates $\mathcal{M}_{\varphi\varphi}/H^2=1$. The end of inflation is shown by the black dashed line. We take $\xi_h=10^3$.}
    \label{fig:Mphiphi}
\end{figure}

Since the deformation preserves the shift symmetry, the phase remains a Goldstone boson of the scalar sector, becoming a pseudo-Goldstone boson only through the QCD anomaly, and its isocurvature mass \eqref{eq:m_phiphi} comes from the variation of its metric alone (Sec.~\ref{sec:DM}). On the slow-roll attractor,
\begin{equation}
\frac{|\mathcal{M}_{\varphi\varphi}(\Theta_*)|}{H^2(\Theta_*)}\lesssim 2\times10^{-1},
\label{eq:Mphiphi_bound}
\end{equation}
for all $0<n\leq8$ and $\xi_\chi\lesssim10^{-2}$ (Fig.~\ref{fig:Mphiphi}). The phase is light at horizon crossing, as anticipated in Sec.~\ref{sec:DM}, but the ratio reaches order unity in the last few $e$-folds, relevant to the small-scale spectrum below. This does not affect \eqref{eq:deltaa_star}, which uses the horizon-crossing value bounded by \eqref{eq:Mphiphi_bound}. The horizon-crossing value becomes negative above $\xi_\chi\simeq4\times10^{-3}$ for $n=3$, $7\times10^{-3}$ for $n=4$ and $9\times10^{-3}$ for $n=5$, but remains positive for $n\geq6$. These sign changes lie outside the isocurvature-allowed region derived below, within which the smallest value, $6\times10^{-3}$ at $n=7$ and $\xi_\chi=10^{-2}$, is still positive. A negative value does not by itself imply growth of the phase perturbation, since the geometric mass governs the canonically normalised entropy mode. The phase retains a constant super-Hubble solution.

\begin{figure}
    \centering
    \includegraphics[width=1\linewidth]{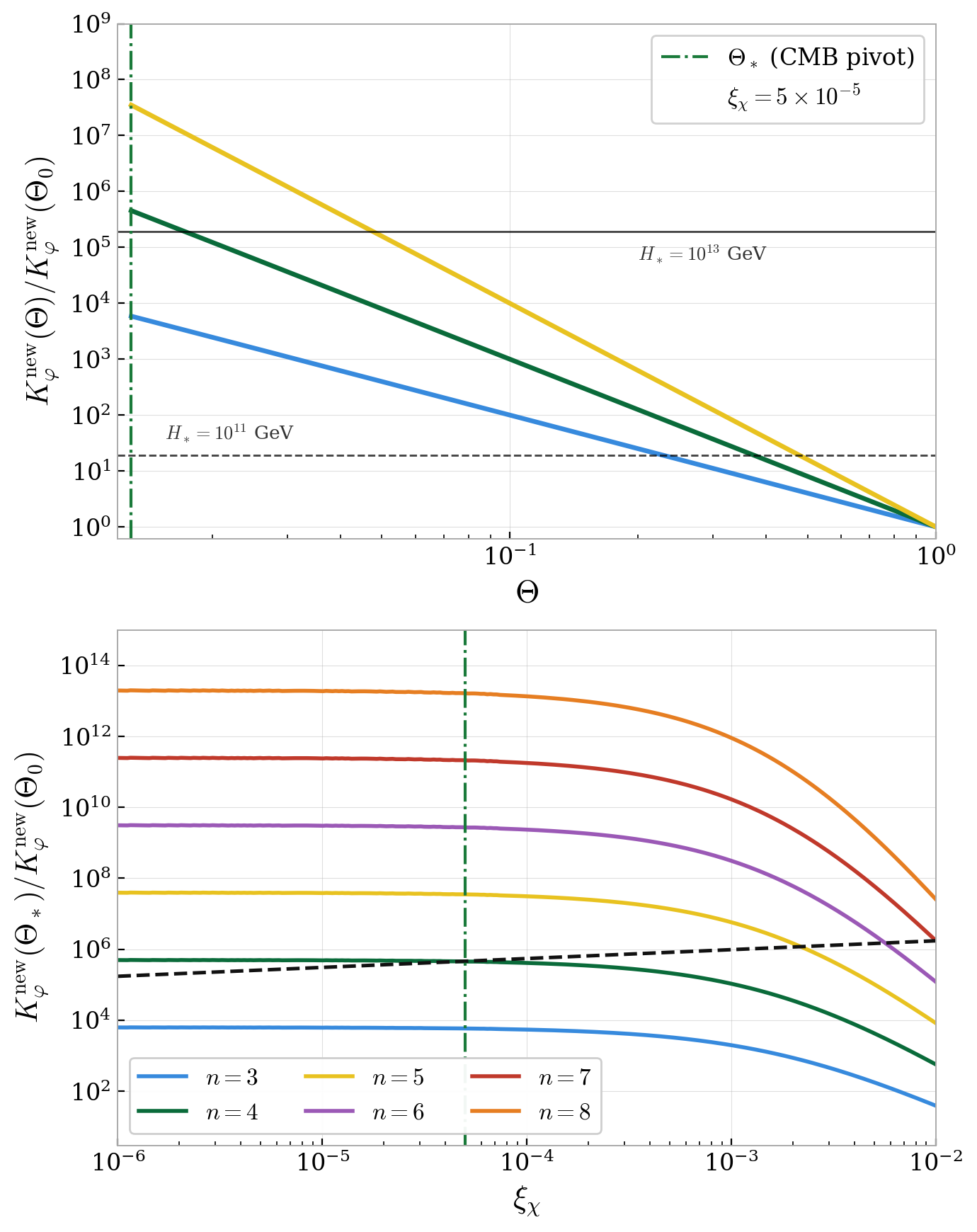}
    \caption{Geometric suppression of the axion fluctuations. \textbf{Top}: $K_\varphi^{\rm new}(\Theta)/K_\varphi^{\rm new}(\Theta_0)$ at $\xi_\chi=5\times10^{-5}$, for $n=3,4,5$. The horizontal lines show the requirement \eqref{eq:joint_num_QCD} at $H_*=10^{13}$ and $10^{11}\,$GeV. \textbf{Bottom}: the same ratio evaluated at $\Theta_*$ as a function of $\xi_\chi$, for $n=3,\dots,8$. The dashed line is the requirement \eqref{eq:joint_num_QCD} at the realised scale and the dash-dotted vertical line marks the slice shown above. A given $(n,\xi_\chi)$ is allowed where its curve lies above the requirement. Computed with $N_*=55$, $\beta_{\rm iso}=0.039$ and $\chi_{\rm top}^{1/4}=75.5\,$MeV.}
    \label{fig:kphi_suppression}
\end{figure}
The amplitude of the axion isocurvature mode is controlled by the field-space metric at horizon crossing relative to its present value, through the ratio that appears on the left-hand side of \eqref{eq:joint_num} and \eqref{eq:joint_num_cold}, here $\Theta_0=1$. 
Explicitly, 
\begin{equation}
\frac{K_\varphi^{\rm new}(\Theta_*)}{K_\varphi^{\rm new}(\Theta_0)}= \Theta_*^{-n}\frac{K_\varphi^{\rm HD}(\Theta_*)}{K_\varphi^{\rm HD}(\Theta_0)}=\Theta_*^{-n}\,\frac{\Theta_*-\Theta_{\rm bd}}{1-\Theta_{\rm bd}}\simeq \Theta_*^{-(n-1)}\,.
    \label{eq:suppression_scaling}
\end{equation}
The second equality is exact, and follows from the identity $c=-\kappa\Theta_{\rm bd}$ of Sec.~\ref{subsec:inflation_obs}. The third holds while $\Theta_{\rm bd}\ll\Theta_*$, and fails at the upper end of the allowed range. Since the Higgs-Dilaton factor \eqref{eq:HDratio} works the wrong way, one power of the exponent is spent undoing it. Each further unit of $n$ then raises the suppression by $\Theta_*^{-1}\sim\mathcal{O}(10^{2})$ (Fig.\,\ref{fig:kphi_suppression}, top), so a higher inflationary scale, which demands a larger ratio, requires a larger $n$.

The required suppression is also fixed: the inflationary observables determine $H_*$ through the measured scalar amplitude, while $\xi_\chi$ fixes the late-time metric, the misalignment regime, and the axion mass. Because the inflaton sector is an $\alpha$-attractor with $r$ pinned at \eqref{eq:obs_staro}, the measured scalar amplitude $A_s\simeq2.1\times10^{-9}$ fixes the Hubble rate through $H_*^2/M_P^2=(\pi^2/2)A_s\,r$ to
\begin{equation}
H_*\simeq6\times10^{-6}\,M_P\simeq1.5\times10^{13}\,\mathrm{GeV}\,,
    \label{eq:Hstar_fixed}
\end{equation}
essentially independently of $n$ and $\xi_\chi$. This realisation therefore predicts \textit{high-scale} inflation; lower-scale isocurvature conditions apply to the general framework of Secs.~\ref{sec:scaleinv}--\ref{sec:DM}, where $H_*$ remains free.

\paragraph{The realised axion parameters.}
The late-time value of the angular metric follows from \eqref{eq:Kvarphi} at $\Theta_0=1$. For $\xi_h\gg1$, which gives $\kappa\simeq-1/6$, and $\xi_h\gg\xi_\chi$, which gives $c\simeq\gamma^2$, it reduces to
\begin{equation}
    K_{\varphi}(\Theta_0)=\frac{1}{c}+\frac{1}{\kappa}\simeq\frac{1}{\xi_{\chi}}.
\end{equation}
Together with Eq.~\eqref{eq: Adef}, this gives
\begin{equation}
    f_0\simeq \frac{M_P}{\sqrt{\xi_{\chi}}}\simeq 141.4 M_P,
    \label{eq:f_0}
\end{equation}
for $\xi_{\chi}=5\times 10^{-5}$. The value at horizon crossing, $f_*$, is larger still, and unlike $f_0$ it depends on the exponent through \eqref{eq:suppression_scaling} ($f_*\simeq1\times10^{5}M_P$ at the minimal exponent $n=4$ selected below, rising to $6\times10^{7}M_P$ at $n=7$).
The super-Planckian $f_0$ fixes the misalignment regime through \eqref{eq:Tosc_ratio}. Its predicted value exceeds the crossover \eqref{eq:f0_cross}, $f_0\simeq0.05\,M_P$, by more than three orders of magnitude, making the right-hand side of \eqref{eq:Tosc_ratio} smaller than unity. No solution with $T_{\rm osc}>\Lambda$ exists for any temperature exponent $p$; the constant-mass branch applies, with onset at $3H=m_0$. This conclusion is sector-independent, because the crossover scales as $\zeta$, while any confining gauge theory has $\chi_{\rm top}\lesssim\Lambda^4$, hence $\zeta\lesssim1$ and a crossover below $f_0\simeq0.2\,M_P$. Therefore, \eqref{eq:Omega_cold} gives the abundance rather than \eqref{eq:Omega_general}.
Equation~\eqref{eq:m0_chi} fixes the axion mass from $f_0$ and $\chi_{\rm top}$, leaving only the choice of confining sector. For QCD, $\chi_{\rm top}^{1/4}\simeq75.5\,$MeV is known~\cite{GrillidiCortona:2015jxo}, so the model predicts
\begin{equation}
m_0=\frac{\sqrt{\chi_{\rm top}}}{f_0}\simeq1.7\times10^{-14}\,{\rm eV}\,,
\label{eq:m0_pred}
\end{equation}
an ultralight QCD axion. Oscillations then begin at $T_{\rm osc}\simeq3.5\,$MeV and the observed abundance requires a small misalignment angle, $\varphi_{\rm ini}\simeq2.4\times10^{-6}$.\footnote{If instead the axion couples to a hidden confining sector, $\chi_{\rm top}$ is a free parameter. The value $m_0=10^{-5}\,$eV used in Sec.~\ref{sec:DM} corresponds, at the $f_0$ fixed by the geometry, to $\chi_{\rm top}^{1/4}=\sqrt{m_0f_0}\simeq1.8\,$TeV and gives $\varphi_{\rm ini}\simeq1.9\times10^{-8}$, $T_{\rm osc}\simeq51\,$GeV. Being heavier, such an axion needs more suppression, and by \eqref{eq:nmin} the minimal exponent rises to $n>6$.} The axion is also harmless for nucleosynthesis. It is frozen, behaving as a constant energy density, until $T_{\rm osc}$, where $\rho_A/\rho_{\rm rad}\simeq2\times10^{-7}$. From then on the ratio grows only as $1/T$.

\begin{figure}
    \centering
    \includegraphics[width=1\linewidth]{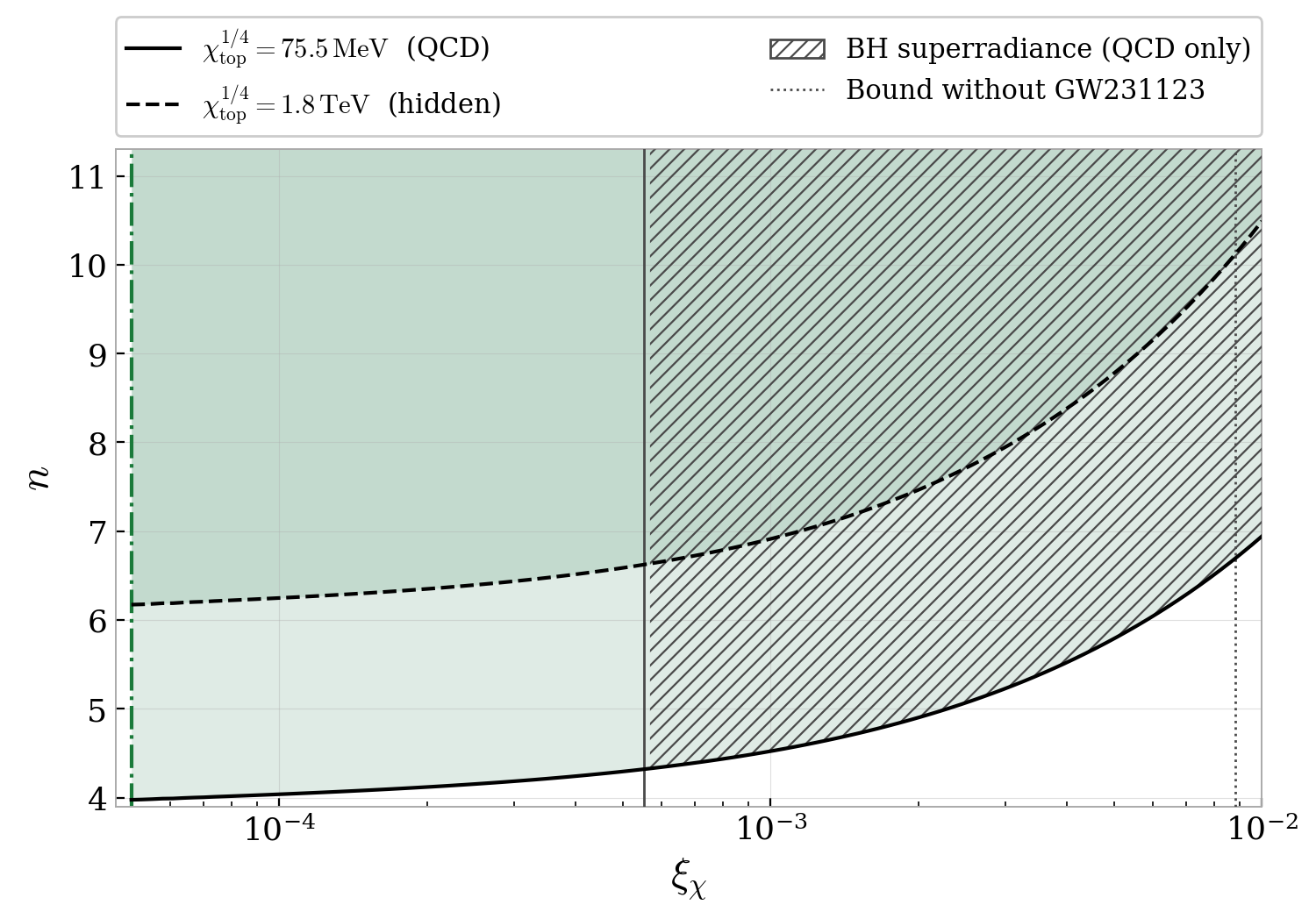}
    \caption{Plot of $n$ vs $\xi_{\chi}$ at the inflationary scale \eqref{eq:Hstar_fixed}, with the region allowed by isocurvature shaded (light: QCD only; dark: both sectors). The solid line represents a QCD confining sector; the dashed one is a hidden sector with $\chi_{\rm top}^{1/4}=1.8\,$TeV. The dash-dotted line marks the reference coupling $\xi_\chi=5\times10^{-5}$. The hatched region applies to the QCD axion only, its mass lies in the range $0.55$--$11\times10^{-13}\,$eV disfavoured by BH superradiance~\cite{Aswathi:2025nxa}. That bound uses the spins of GW231123, observed in the fourth LIGO-Virgo-KAGRA observing run~\cite{LIGOScientific:2025rsn}, and of GW190517 from GWTC-2.1~\cite{LIGOScientific:2021usb}, at 90\% C.L. and for BH ages above $10^5\,$yr. The dotted line is the bound from GW190517 alone, consistent with the earlier GWTC-2 analysis in~\cite{Ng:2020ruv}. The hidden-sector axion, with $m_0\sim10^{-5}$--$10^{-4}\,$eV, is unaffected.}
    \label{fig:nmin}
\end{figure}

An axion this light is constrained by black-hole superradiance. When the Compton wavelength of a boson is comparable to the size of a spinning black hole, a bound cloud grows exponentially from a single quantum and spins the hole down~\cite{Ng:2020ruv,Aswathi:2025nxa}. The bound therefore tests only the existence of the particle, not its cosmological abundance. The spins of GW231123~\cite{LIGOScientific:2025rsn} and GW190517~\cite{LIGOScientific:2021usb} disfavour $5.5\times10^{-14}\,{\rm eV}\lesssim m\lesssim1.1\times10^{-12}\,$eV at 90\% C.L., for black holes older than $10^5\,$yr~\cite{Aswathi:2025nxa}. The bound holds for $f\gtrsim10^{15}\,$GeV, below which self-interactions halt the growth of the cloud, and $f_0$ exceeds this by more than four orders of magnitude. The mass \eqref{eq:m0_pred} lies a factor of three below the disfavoured range. Since the mass grows with the coupling, however, the bound limits $\xi_\chi$ (Fig.~\ref{fig:nmin}).

The joint constraint of Sec.~\ref{sec:DM} takes the constant-mass form \eqref{eq:joint_num_cold}, not \eqref{eq:joint_num}. It scales as $m_0^{1/2}$, with $m_0=\sqrt{\chi_{\rm top}\,\xi_\chi}/M_P$ from \eqref{eq:f_0} and \eqref{eq:m0_chi}, retaining a dependence on $\xi_\chi$. Normalising to \eqref{eq:m0_pred} while leaving the confining sector open,
\begin{align}
\frac{K_\varphi(\Theta_*)}{K_\varphi(\Theta_0)}&\gtrsim\mathcal{K}\,,\nonumber\\
\mathcal{K}\simeq1.1\times10^{16}&\left(\frac{H_*}{M_P}\right)^{2}\left[\frac{\chi_{\rm top}}{\chi_{\rm top}^{\rm QCD}}\,\frac{\xi_\chi}{5\times10^{-5}}\right]^{1/4}\nonumber\\
&\times\left[\frac{g_*(T_{\rm osc})}{10.75}\right]^{-1/4}\,,
\label{eq:joint_num_QCD}
\end{align}
both bracketed factors being unity for QCD. A heavier axion oscillates earlier, where $g_*$ is larger, and the last factor then relaxes the requirement slightly. At the predicted scale \eqref{eq:Hstar_fixed} and the reference coupling this is $\mathcal{K}\simeq4\times10^{5}$, far below the $\sim2\times10^{11}$ of Sec.~\ref{sec:DM}, because the super-Planckian $f_0$ forces an exceptionally light axion. Equation \eqref{eq:joint_num_QCD} is the dashed curve in the lower panel of Fig.~\ref{fig:kphi_suppression}.

Comparing the achieved suppression \eqref{eq:suppression_scaling} with the requirement \eqref{eq:joint_num_QCD} gives the minimal exponent
\begin{equation}
n\geq\frac{1}{\ln(1/\Theta_*)}\,\ln\!\left[\frac{(1-\Theta_{\rm bd})\,\mathcal{K}}{\Theta_*-\Theta_{\rm bd}}\right],
\label{eq:nmin}
\end{equation}
which, at the reference coupling, where $\Theta_*\simeq1.3\times10^{-2}$, reads
\begin{equation}
    n\;\gtrsim\;4\,.
    \label{eq:n_bound}
\end{equation}
The confining sector enters \eqref{eq:nmin} only through $\chi_{\rm top}^{1/4}$ inside the logarithm, giving a mild dependence. The coupling also sets $\Theta_{\rm bd}$ and $\Theta_*$. Their essentially fixed separation (Sec.~\ref{subsec:inflation_obs}) leaves the logarithm varying only through $\mathcal{K}\propto\xi_\chi^{1/4}$ over the range of interest. Increasing $\Theta_*$ reduces $\ln(1/\Theta_*)$, weakening suppression per unit $n$ and raising the minimum exponent from $4$ to $7$ across the Higgs-Dilaton range of $\xi_\chi$. This limits $\xi_\chi$, as Fig.~\ref{fig:nmin} shows for QCD and the heavier hidden-sector axion of Sec.~\ref{sec:DM}, $\chi_{\rm top}^{1/4}=1.8\,$TeV, at the predicted inflationary scale.
The monotonic dependence of \eqref{eq:nmin} on $\xi_\chi$ gives an upper coupling limit at fixed $n$ (Fig.\,\ref{fig:nmin}), namely $\xi_\chi\lesssim 5 \times10^{-5}$ for $n=4$, $\xi_\chi\lesssim2\times10^{-3}$ for $n=5$, and $\xi_\chi\lesssim6\times10^{-3}$ for $n=6$. As far as isocurvature is concerned, the whole HD-allowed range is covered from $n=7$ on. For the QCD axion, superradiance cuts this range. The mass enters the disfavoured interval at $\xi_\chi\simeq5.5\times10^{-4}$ (Fig.~\ref{fig:nmin}), which leaves $\xi_\chi\lesssim5.5\times10^{-4}$, covered from $n=5$. This edge relies on GW231123; using GW190517 alone, the disfavoured region would instead start only at $\xi_\chi\simeq9\times10^{-3}$. The hidden-sector axion, with $m_0\sim10^{-5}$--$10^{-4}\,$eV, is far heavier than the masses probed by stellar-mass black holes and is unaffected.
The $\Theta^{-n}$ model thus makes a sharp joint prediction, high-scale inflation together with a small non-minimal coupling.

The realisation lies deep in the anthropic window, with the small angle found above tied to the super-Planckian $f_0$. This is structural, since $\xi_\chi$ fixes both $f_0$ and the pole \eqref{eq:Thetabd}, $\Theta_{\rm bd}\simeq6\xi_\chi/(1+6\xi_\chi)$ for $\xi_h\gg1$. Lowering $f_0$ raises $\xi_\chi$, moving the pole and, at essentially fixed separation, $\Theta_*$ outward. The suppression in \eqref{eq:suppression_scaling} then collapses toward the wrong-direction regime \eqref{eq:HDratio}. This embedding therefore cannot lower $f_0$ below a few $M_P$. Realising the benchmark of Sec.~\ref{sec:DM}, whose freely chosen late-time metric allows $\varphi_{\rm ini}\sim\mathcal{O}(1)$, would require decoupling the expectation value setting $f_0$ from the ratio driving inflation. This limitation belongs to the embedding; geometric suppression makes its small-misalignment corner viable.

In models where the Peccei-Quinn scale is set by the displacement of a massive radial mode, a large hierarchy $f_* \gg f_0$ has been argued to be problematic for two reasons. The displaced radial field can affect the inflationary dynamics, and its later oscillations can induce non-thermal symmetry restoration via parametric resonance~\cite{Kawasaki:2013iha,Harigaya:2015hha}. Neither argument applies in the same way in the present construction.
The first concern does not arise here, because the hierarchy is not obtained by displacing a spectator field but comes from the inflaton itself, through $f=M_P\sqrt{K_\varphi(\Theta)}$ with $K_\varphi\propto\Theta^{-n}$ and $\Theta_*\ll1$ during inflation.  During inflation the Higgs is large and rolling, while the radial Peccei--Quinn field $\chi$, aligned with the flat dilaton direction, is not displaced, following only the slow drift of the overall scale in \eqref{eq:rhodot}. Since $\Theta$ measures the ratio $h/\chi$, the evolution from $\Theta_*$ to $\Theta_0=1$ is the Higgs rolling down. The decay constant is therefore large during inflation because the Higgs is large, not because $\chi$ has been displaced. There is accordingly no additional field whose energy density or couplings could disturb the trajectory.
The second argument, however, does have a direct analogue. Its severity depends on how much of the hierarchy is left after inflation. For a radial mode in a quartic potential, lattice studies find non-thermal restoration once the field relaxes from more than about $10^4$ times its final value \cite{Kawasaki:2013iha}, which has motivated reaching an intermediate vacuum before inflation ends \cite{Lyu:2026rwo}. Here this happens without further ingredients. $K_\varphi^{\rm new}$ relaxes adiabatically during slow roll, from $\Theta_*$ to $\Theta_{\rm end}$, so that only $f_{\rm end}/f_0\simeq3$--$15$ remains for $n=4$--$8$. Nor can a radial PQ mode pass through the origin, since $\chi$ vanishes only at $\Theta_{\rm bd}\ll\Theta_{\rm end}$. After inflation, oscillations about $\Theta_0=1$ periodically modulate the steep inverse-power function $K_\varphi$ and the axion's canonical normalisation. The resulting oscillating coefficient in the mode equation drives exponential growth within resonance bands, as in geometric and kinetic preheating after multi-field $\alpha$-attractor inflation~\cite{Krajewski:2018moi,Iarygina:2018kee,Adshead:2023nhk} and in the production of phase modes in non-minimally coupled models~\cite{Ema:2016dny,DeCross:2015uza}. Its significance depends on competition between resonance and the redshifting oscillation amplitude; assessing it requires the coupled evolution of $\Theta$ and $\delta\varphi$ through reheating \cite{Garcia-Bellido:2008ycs,Repond:2016sol,Sfakianakis:2018lzf,Rubio:2019ypq,Barman:2025lvk}, left for future work.

The super-Planckian $f_0$ and $f_*$ raise distinct issues. Derivative interactions remain controlled by the cutoffs derived in App.~\ref{app:cutoff}. In particular, $\Lambda_3\simeq7.9\,M_P$ at $n=4$ and it never falls below $3.9\,M_P$, remaining six orders of magnitude above \eqref{eq:Hstar_fixed}, so the $\Theta^{-n}$ deformation introduces no new scale below $M_P$ and the effective description remains weakly coupled throughout inflation. This concerns the operators introduced by the deformation. The Higgs-Dilaton sector carries, in addition, the background-dependent strong-coupling scale of Higgs inflation \cite{Rubio:2018ogq,Bezrukov:2014ipa,Bezrukov:2017dyv}, which during inflation lies well above $H_*$~\cite{Bezrukov:2010jz,Bezrukov:2012hx}. The axion's excursion also remains far below $M_P$ throughout its history, although super-Planckian decay constants sit in tension with the axion weak gravity conjecture~\cite{Arkani-Hamed:2006emk}, which we do not evaluate here. However, the periodicity $2\pi f_0$ exceeds the cutoff, posing an axion-quality problem \cite{Kamionkowski:1992mf,Holman:1992us} that we do not solve. It is aggravated because the Peccei-Quinn radial field is the dilaton, $X=\chi\,e^{i\varphi}/\sqrt2$ in \eqref{eq:Jmod}. Scale-invariant chiral-symmetry-breaking operators take the form $\lambda_k\,\chi^{4-k}X^k$, leaving $\chi^4$ for every $k$ and eliminating the usual suppression by operator dimension. Like the anthropic placement, this is a property of the embedding, since the benchmark mechanism of Sec.~\ref{sec:DM} uses sub-Planckian decay constants.

\paragraph{Small-scale isocurvature.}
As $\Theta$ grows during inflation, \eqref{eq:Kphi_new} decreases and modes leaving later acquire larger $\delta\varphi$ by \eqref{eq:deltaa_star}, giving the blue spectrum described at the pivot by \eqref{eq:n_iso}. Note that the phase and the instanton potential share the period $2\pi$. Fluctuations reaching that scale would populate different minima when the potential switches on, producing domain walls rather than ordinary isocurvature.

Extrapolating \eqref{eq:n_iso} to those scales will not settle the question. At $n=4$ and $\xi_\chi=5\times10^{-5}$, $\mathcal{M}_{\varphi\varphi}/H^2$ grows from $0.074$ at the pivot, where \eqref{eq:n_iso} gives $n_{\rm iso}-1\simeq0.05$, to order unity well before inflation ends, and by then $\epsilon$ is no longer small either. The tilt is a pivot-scale statement. We therefore evaluate \eqref{eq:deltaa_star} mode by mode. At the same $n$ and $\xi_\chi$, $\delta\varphi$ grows from $1.0\times10^{-11}$ at $\Theta_*$ to $1.2\times10^{-9}$ by the end of inflation. For $n=7$ the endpoint is smaller, $3.7\times10^{-10}$, because the fluctuation starts lower. Either way $\delta\varphi$ stays some ten orders of magnitude below $2\pi$. Every patch relaxes to the same minimum, no domain walls form, and the extra power at small scales is ordinary isocurvature.
The only way defects could still be produced is through the parametric amplification discussed above, since a growth of $\delta\varphi$ to values of order the periodicity is equivalent to a restoration of the Peccei-Quinn symmetry after inflation.

\section{Conclusions}\label{sec:conclusions}

We have studied a geometric mechanism for suppressing axion isocurvature in a broad class of scale-invariant models. The axion metric $K_\varphi(\Theta)$ depends on an evolving dimensionless field ratio: its present value fixes the misalignment abundance, while its value at horizon crossing controls the primordial phase fluctuations. A larger metric during inflation therefore separates abundance from isocurvature without lowering the inflationary scale $H_*$, modifying the QCD sector or adding degrees of freedom. At the level of the general framework, the required hierarchy permits high-scale inflation in the classic axion window with an untuned misalignment angle and sub-Planckian decay constants.

The exact scale and chiral Peccei-Quinn symmetries also constrain the inflationary dynamics. They forbid a scalar-sector axion potential and force the turning rate to vanish on the zero-charge attractor. The curvature perturbation is consequently conserved on super-Hubble scales, entropy perturbations do not feed the adiabatic sector, and the adiabatic spectrum and bispectrum retain their single-field form. The changing kinetic normalisation nevertheless gives the canonical entropy mode a geometric mass, small at CMB horizon crossing in our realisation.

We implemented the required metric hierarchy in the Higgs-Dilaton scenario through a $\Theta^{-n}$ deformation of the Peccei-Quinn kinetic term, preserving both symmetries. The deformation has an unexpected payoff in the inflaton sector. Imposing the attractor relation gives an effective kinetic function governed by a second-order pole throughout the observable regime, with universal residue $a\to3/2$. The inflationary predictions therefore remain near the Starobinsky point, with $n_s\simeq0.965$ and $r\simeq3.5\times10^{-3}$ at $N_*=55$, essentially independently of $n$ for $n\gtrsim1$ and across the Higgs-Dilaton range considered. The measured scalar amplitude then fixes $H_*\simeq1.5\times10^{13}\,$GeV. The isocurvature suppression, instead, grows steeply with $n$, allowing the exponent to satisfy the bound without appreciably changing the inflationary observables. At the reference coupling $\xi_\chi=5\times10^{-5}$, this requires $n\geq4$. The interaction scales estimated for the deformation remain well above $H_*$ in the regime examined.

This realisation sits in the anthropic window. The non-minimal coupling fixes $f_0=M_P/\sqrt{\xi_\chi}$, giving a super-Planckian present decay constant and, for QCD, an ultralight axion. At the reference coupling, its mass is $m_0\simeq1.7\times10^{-14}\,$eV and the observed abundance requires $\varphi_{\rm ini}\simeq2.4\times10^{-6}$. The construction thus provides the required hierarchy while retaining a predictive inflaton sector, but does not reach the classic-window benchmark of Sec.~\ref{sec:DM}: the same coupling fixes both $f_0$ and the inflationary kinetic pole. Its super-Planckian decay constants also leave the axion weak-gravity \cite{Arkani-Hamed:2006emk} and quality questions open. The latter is aggravated by identifying the radial PQ field with the dilaton, which removes the usual suppression of symmetry-breaking operators by their dimension. These features belong to the explicit embedding rather than to the geometric mechanism itself.

The main unresolved question concerns the evolution after inflation. Our suppression is evaluated at horizon crossing, whereas the observational bound applies to the late-time isocurvature mode. Oscillations of $\Theta$ about its minimum modulate $K_\varphi$ and can amplify axion fluctuations, as discussed in Sec.~\ref{sec:axion_sector}. Establishing the late-time viability therefore requires following the coupled dynamics through reheating \cite{Garcia-Bellido:2008ycs,Repond:2016sol,Sfakianakis:2018lzf}. The inflationary results derived here provide the initial conditions for that analysis.

The scenario is testable on several fronts. Its tensor-to-scalar ratio lies within reach of next-generation $B$-mode experiments, while its spectral index remains near the Starobinsky value. As noted in Sec.~\ref{subsec:inflation_obs}, data combinations including ACT DR6 already place that value at $N_*=55$ in up to $\sim3\sigma$ tension. The isocurvature amplitude supplies a complementary probe: it is sensitive to the deformation exponent even where the adiabatic observables are nearly independent of it. For the QCD realisation, black-hole superradiance further constrains the non-minimal coupling, leaving axion masses below the interval currently disfavoured by black-hole spins and motivating measurements of heavier systems.

Beyond the post-inflationary dynamics, a natural extension is a scale-invariant construction that separates the late-time decay constant from the coupling controlling the inflationary geometry, allowing an explicit test of the classic-window benchmark. Further questions concern the scale-invariant embedding of the axion coupling to gauge fields and the origin of the $\Theta^{-n}$ deformation in an ultraviolet completion, with Weyl-invariant Einstein-Cartan constructions providing a possible setting in which the axion originates from gravitational dynamics \cite{Karananas:2024xja,Karananas:2025ews,Karananas:2025uhy}. We leave these directions for future work. 

\section*{Acknowledgements}
MDA is supported by the Leverhulme Trust. J.~R. is supported by a Ram\'on y Cajal contract of the Spanish Ministry of Science and Innovation with Ref.~RYC2020-028870-I. This research was further supported by the project PID2022-139841NB-I00 of MICIU/AEI/10.13039/501100011033 and FEDER, UE, and the 2025 Leonardo Grant for Scientific Research and Cultural Creation from the BBVA Foundation. The BBVA Foundation is not responsible for the opinions, comments, and content included in the project and/or its resulting outcomes, which are the sole and exclusive responsibility of the authors. 

\appendix
\section{Covariant multifield perturbations}\label{sec:cov_multif}

Following closely the covariant treatment in Refs.~\cite{Gong:2011uw,Elliston:2012ab,Kaiser:2012ak},  let us 
introduce a covariant vector ${\cal Q}^I$ connecting $ \phi^I$ and $\phi^I + \delta \phi^I$,
\begin{equation}
{\cal D}_\lambda \phi^I \vert_{\lambda = 0} = {\cal Q}^I\,, 
\end{equation}
 with $\lambda$ parametrizing the geodesic in the field-space manifold. The expansion $\delta \phi^I$ in powers of ${\cal Q}^I$, $
\delta \phi^I = {\cal Q}^I - \frac{1}{2!} \Gamma^I_{\>\> JK} (\phi^L) { {\cal Q}^J\cal Q^K} + \ldots$,
allows us to identify $\delta \phi^I \leftrightarrow {\cal Q}^I$ at linear order. Defining the gauge-invariant Mukhanov-Sasaki variables,
\begin{equation}
Q^I \equiv {\cal Q}^I + \frac{\dot{\phi}^I}{H} \psi\,,
\end{equation}
with $H\equiv \dot a/a$ the Hubble rate and separating the field equations into background and first-order perturbations, we get \begin{eqnarray}
&&\displaystyle {\cal D}_t^2 \phi^I + 3 H {\cal D}_t{\phi}^I + {K}^{IK} U_{, K} = 0\,, \label{eq:Bphi}\\
&&\displaystyle {\cal D}_t^2 Q^I + 3 H {\cal D}_t Q^I + (\Omega^2_Q)^I_{\,J} \, Q^J = 0\,,\label{eq:Pphi}
\end{eqnarray}
with
\begin{align}
&(\Omega^2_Q)^I_{\,J}\equiv  \frac{k^2}{a^2} \delta^I_{\>\> J} + {\cal M}^I_{\>\> J} - \frac{1}{ M_{P}^2 a^3}{\cal D}_t \left( \frac{a^3}{H} \dot{\phi}^I \dot{\phi}_J \right) \,, \\
&{\cal M}^I_{\>\> J} \equiv {K}^{IK} \left( \nabla_J \nabla_K U\right) - {\cal R}^I_{\>\> LMJ} \dot{\phi}^L \dot{\phi}^M\,,\label{MIJ}
\end{align}
and ${\cal R}^I_{\>\> LMJ}$ the Riemann tensor for the field metric $K_{IJ}$. Since the background quantities $\phi^I$ are just coordinates on the field-space manifold $K_{IJ}$, the action of the directional derivative on them, 
\begin{equation}
 \mathcal{D}_t \phi^I = \dot \phi^I\,,\hspace{8mm} {\cal D}_t^2 \phi^I = \ddot \phi^I+\Gamma^I_{JK}\dot \phi^J\dot \phi^K\,,
\end{equation}
gives rise to a simple geodesic equation in field-space supplemented by a Hubble friction term $3 H \dot {\phi}^I$ and an external force $K^{IK}U_{,K}$. The effect of ${\cal D}_t$ on the field perturbations $Q^I$ is significantly more involved 
\begin{eqnarray}
 \mathcal{D}_t Q^I &=& \dot{ Q}^I  +\Gamma^I_{\;JK}\dot \phi^JQ^K\,,\label{eq:covariant_derivative}
\\
\mathcal{D}_t^2 Q^I &=& \ddot{ Q}^I + \left(\partial_L\Gamma^I_{\;JK}+\Gamma^I_{\;LM}\Gamma^M_{JK}\right)\dot \phi^L\dot \phi^J Q^K \nonumber 
\\ &&+  \Gamma^I_{\;JK}\ddot \phi^J Q^K + 2\Gamma^I_{\;JK}\dot \phi^J\dot{ Q}^K \,.
\end{eqnarray}
 In order to connect with observable quantities, it is convenient to define a background field
 \begin{equation}
 \dot{\sigma}^2 = {K}_{IJ} \dot{\phi}^I \dot{\phi}^J\,,
 \end{equation}
together with a set of projector operators 
\begin{equation}
\hat{\sigma}^I \equiv \frac{\dot{\phi}^I}{\dot{\sigma}}\,,
\hspace{10mm}
\hat{s}^{IJ} \equiv {K}^{IJ} - \hat{\sigma}^I \hat{\sigma}^J\,.\label{eq:projectors}
\end{equation}
Decomposing the perturbations $Q^I$ into components 
\begin{equation}
Q_\sigma \equiv \hat{\sigma}_I Q^I , \hspace{10mm}
Q_s^I \equiv \hat{s}^I_{\>\> J} Q^J\,,\label{eq:pert_decomp}
\end{equation}
parallel and perpendicular to the trajectory. In terms of these variables, Eqs.~\eqref{eq:Bphi} and \eqref{eq:Pphi} in Fourier space become
\begin{align}
&\ddot \sigma +3H \dot \sigma+ \hat \sigma^I U_{,I}=0 \,, \label{sigmaB} \\ 
& \ddot{Q}_\sigma + 3 H \dot{Q}_\sigma +\Omega^2_{\sigma} Q_\sigma = S_\sigma\,, \label{curvaturepert}  \\
&{\cal D}_t^2 Q_s^I + \left[ 3 H \delta^I_{\>\> J} + 2 \hat{\sigma}^I \omega_J \right] {\cal D}_t Q_s^J  +(\Omega^2_s)^I{}_J Q_s^J =S_s^I\,,  \label{isopert}
\end{align}
with 
\begin{align}
&\Omega_{\sigma}^2\equiv  \frac{k^2}{a^2} + {\cal M}_{\sigma\sigma} - \omega^2 - \frac{1}{M_{P}^2 a^3} \frac{d}{dt} \left( \frac{a^3 \dot{\sigma}^2}{H} \right)\,,  \\
&(\Omega_{s}^2)^I{}_J\equiv \frac{k^2}{a^2} \delta^I_{\>\> J} + {\cal M}^I_{\>\> J} - 2 \hat{\sigma}^I \left( {\cal M}_{\sigma J} + \frac{\ddot{\sigma}}{\dot{\sigma}} \omega_J \right) \,, \\
& S_\sigma\equiv  2 \left[ \frac{\partial}{\partial t} -  \frac{ U_{, \sigma}}{\dot{\sigma}} - \frac{\dot{H}}{H} \right] \left( \omega_J Q_s^J \right)\,, \label{Ssigma} \\
& S^I_s\equiv  - 2 \omega^I \left( \dot{Q}_\sigma + \frac{\dot{H}}{H} Q_\sigma - \frac{\ddot{\sigma}}{\dot{\sigma}} Q_\sigma \right)\,,
\end{align}
and\footnote{Note that for ${\cal M}_{\sigma\sigma}$ the curvature contribution in \eqref{MIJ} vanishes because the Riemann tensor is antisymmetric in its first index pair. This is unique to the adiabatic perturbations and does not occur for the isocurvature perturbations.}
\begin{eqnarray}
{\cal M}_{\sigma J} &\equiv& \hat{\sigma}_I  {\cal M}^I_{\>\> J}\,, \\
{\cal M}_{\sigma\sigma} &\equiv& \hat{\sigma}_I \hat{\sigma}^J {\cal M}^I_{\>\> J}
= \hat{\sigma}^K \hat{\sigma}^J \left( {\nabla}_K {\nabla}_J U \right)\,. 
\end{eqnarray}
The adiabatic and entropy perturbations in Eqs.~\eqref{curvaturepert}  and \eqref{isopert} are coupled through the so-called \textit{turning rate} $\omega^I$, which measures the change of the unit vector $\hat{\sigma}^I$,
\begin{equation} \label{omega_def}
\omega^I \equiv {\cal D}_t \hat{\sigma}^I=-\frac{1}{\dot \sigma}U_{,K}\hat{s}^{IK}\,,
\end{equation}
and is, by definition, orthogonal to $\hat\sigma^I$, i.e. $\omega_I\hat\sigma^I=0$.

The perturbations in the adiabatic direction are proportional to the gauge invariant curvature perturbation,
\begin{equation}
{\cal R}_c=\frac{H}{\dot \sigma}Q_\sigma\,.
\end{equation}
In terms of it, we have \cite{Kaiser:2012ak}
\begin{equation}\label{conservR}
\dot{\cal R}_c = \frac{H}{\dot{H}} \frac{k^2}{a^2} \Psi + \frac{2H}{\dot{\sigma}} \left( \omega_I Q_s^I \right)\,,
\end{equation}
with $\Psi$ the gauge-invariant Bardeen potential \cite{Malik:2008im}.

\subsection{Perturbations on the scale-invariant attractor}\label{appendix:multifield}

On the scale-invariant attractor of the undeformed class only $\Theta$ rolls, $\dot\phi^I=\dot\Theta\,\delta^I_{\Theta}$ (see Sec.~\ref{sec:currents}), so that $\hat{\sigma}^I\propto \delta_{\Theta}^I$. By \eqref{eq:projectors} and \eqref{eq:pert_decomp}, for $Q^I=(Q^\Theta,Q^\Phi,Q^\varphi)$ we have
\begin{equation}
Q_\sigma = M_P\sqrt{K_\Theta}\, Q^\Theta    , \hspace{5mm}
Q_s^\Phi =Q^\Phi\,, \hspace{5mm}  Q_s^\varphi =Q^\varphi\,.
\end{equation}
The turning rate \eqref{omega_def} then vanishes identically. Since $U$ depends on $\Theta$ alone, $U_{,K}=U_{,\Theta}\,\delta^\Theta_K$, while $\hat\sigma^\Theta=\dot\Theta/\dot\sigma=1/(M_P\sqrt{K_\Theta})$ for $\dot\Theta>0$, so that $\hat s^{I\Theta}=K^{I\Theta}-\hat\sigma^I\hat\sigma^\Theta=\delta^I_\Theta/(M_P^2K_\Theta)-\delta^I_\Theta/(M_P^2K_\Theta)=0$ and $\omega^I=-U_{,\Theta}\,\hat s^{I\Theta}/\dot\sigma=0$, which is \eqref{eq:turning_rate}. The sources vanish and Eqs.~\eqref{curvaturepert} and \eqref{isopert} take the explicit form
\begin{align}
& \ddot{Q}_\sigma + 3 H \dot{Q}_\sigma +\Omega^2_{\sigma} Q_\sigma = 0\,,  \\
&{\cal D}_t^2 Q_s^\Phi + 3 H   {\cal D}_t Q_s^\Phi  +(\Omega_{s}^2)^\Phi{}_\Phi Q_s^\Phi =0\,, \label{QPhi}\\ 
&{\cal D}_t^2 Q_s^\varphi + 3 H   {\cal D}_t Q_s^\varphi  +(\Omega_{s}^2)^\varphi{}_\varphi Q_s^\varphi =0\,, \label{Qvarphi} 
\end{align}
The entropic modes $Q_s^I$ carry a field-space index and are not canonically normalised, so the mode equations \eqref{QPhi} and \eqref{Qvarphi} involve the covariant derivative ${\cal D}_t$, with
\begin{align}
&\Omega_{\sigma}^2\equiv  \frac{k^2}{a^2} + {\cal M}_{\sigma\sigma} - \frac{1}{M_{P}^2 a^3} \frac{d}{dt} \left( \frac{a^3 \dot{\sigma}^2}{H} \right)\,,  \\
&(\Omega_{s}^2)^\Phi{}_\Phi\equiv \frac{k^2}{a^2} +{\cal M}_{\Phi\Phi}\,,  \\
&(\Omega_{s}^2)^\varphi{}_\varphi\equiv \frac{k^2}{a^2}+{\cal M}_{\varphi\varphi}\,,
\end{align}
and the effective masses ${\cal M}_{\sigma\sigma}=U_{,\sigma\sigma}$, ${\cal M}_{\Phi\Phi}$ and ${\cal M}_{\varphi\varphi}$. Using the relations
\begin{equation}
\dot \sigma = \sqrt{2\epsilon}M_{P}H    
\end{equation}
and
\begin{eqnarray}
\hspace{-4mm} 
&& U_{,\sigma} = -3H\dot\sigma-\ddot\sigma 
= -\sqrt{2\epsilon}\left(3-\epsilon+\frac{\epsilon_2}{2}\right)M_{P}H^2 \,, \\ 
\hspace{-4mm} 
&& U_{,\sigma\sigma} = \left(6\epsilon-\frac{3}{2}\epsilon_2+\frac{5}{2}\epsilon\,\epsilon_2-2\epsilon^2-\frac{1}{4}\epsilon_2^2-\frac{1}{2}\epsilon_2\epsilon_3\right)H^2\,,    
\end{eqnarray}
with 
\begin{equation}
\epsilon \equiv -\frac{\dot H}{H^2}\,, \hspace{5mm} \epsilon_2 = \frac{\dot \epsilon}{\epsilon H}\,,
\hspace{5mm} \epsilon_3\equiv \frac{\dot\epsilon_2}{\epsilon_2 H}\,,
\end{equation}
we obtain 
\begin{eqnarray}
\Omega_{\sigma}^2=\frac{k^2}{a^2}-\frac{1}{2}\epsilon_2  \left(3-\epsilon+\frac12\epsilon_2+\epsilon_3\right) H^2\,.
\label{eq:omega_sigma}
\end{eqnarray}
The conservation of ${\cal R}_c$ relies on the two shift symmetries. Indeed, replacing $U(\Theta)$ in \eqref{eq:sigmamodel} by a separable potential,  $U(\Theta)\to V(\Theta,\varphi)=U(\Theta)+W(\varphi)$ and using the fact that in the slow-roll approximation,
\begin{equation}
\hat{\sigma}^I\simeq -\frac{X^I}{\sqrt{X^K X_K}}\,, \hspace{5mm} \omega^I=\frac{M_P\sqrt{V}}{\sqrt{3}}\, X^J \nabla_J \left[\frac{X^I}{\sqrt{X^K X_K}}\right]\,,
\end{equation}
with $X^K=(\ln V)^{,K}$, where we used $\dot\sigma\simeq M_P\sqrt{V}\,\sqrt{X^KX_K}/\sqrt{3}$ in slow roll, we have
\begin{equation}
\omega^2\propto\left(\frac{\partial_\varphi \ln  V\, \partial_\Theta \ln P-\partial_\Theta \ln V\, \partial_\varphi \log P}{\sqrt{ K_\Theta K_\varphi}}\right)^2\,,
\end{equation}
with
\begin{equation}
P^2(\Theta,\varphi)\equiv \frac{K_\Theta K_\varphi}{K_\varphi V_{,\Theta}^2+K_\Theta V_{,\varphi}^2}\,.
\end{equation} 
This vanishes identically when $W_{,\varphi}=0$ but not otherwise: a potential for the phase makes the trajectory turn and reopens the transfer of isocurvature into the curvature perturbation.

The reduction of \eqref{QPhi} and \eqref{Qvarphi} to the ordinary-derivative equations \eqref{eq:ddotSvarphi} of the main text, and of the adiabatic equation to \eqref{eq:curvcons}, is carried out in App.~\ref{sec:cov_ord_der}.

\subsection{From the covariant to the ordinary-derivative entropic equations}\label{sec:cov_ord_der}

Here we show that, once the modes are canonically normalised on the same footing as the adiabatic one\footnote{For the adiabatic direction this is already built in, $Q_\sigma=M_P\sqrt{K_\Theta}\,Q^\Theta$, so that $M_P\sqrt{K_\Theta}\,\mathcal{D}_t Q^\Theta=\dot Q_\sigma$ and its equation carries ordinary derivatives from the outset.} $Q_\sigma$, the derivatives become ordinary and \eqref{eq:ddotSvarphi} follow. We work with the $\Phi$ mode. The $\varphi$ mode is identical with $\tilde{\gamma}$ in place of $\beta$.

Since only $\Theta$ rolls on the attractor, in the covariant derivative \eqref{eq:covariant_derivative}, only $J=\Theta$ contributes, and of the connection coefficients $\Gamma^I_{\Theta K}$ exactly one survives for each $I$,
\begin{align}
    \mathcal{D}_{t}Q^{\Phi}&=\dot Q^\Phi+\Gamma^{\Phi}_{\Theta \Phi}\dot{\Theta}Q^\Phi,\\
    \mathcal{D}_{t}Q^{\varphi}&=\dot Q^\varphi+\Gamma^{\varphi}_{\Theta \varphi}\dot{\Theta}Q^\varphi,
\end{align}
where $\Gamma^{\Phi}_{\Theta \Phi}\dot{\Theta}=\dot{\beta}$ and $\Gamma^{\varphi}_{\Theta \varphi}\dot{\Theta}=\dot{\tilde{\gamma}}$. On the attractor $Q_s^\Phi=Q^\Phi$ and $Q_s^\varphi=Q^\varphi$, so the two notations are used interchangeably. 
Applying ${\cal D}_t$ once more, we get
\begin{equation}
\mathcal{D}_t^2 Q_s^\Phi=\ddot Q_s^\Phi+2\dot\beta\,\dot Q_s^\Phi+\big(\ddot\beta+\dot\beta^2\big)Q_s^\Phi\,,
\end{equation}
so that \eqref{QPhi} reads, in ordinary derivatives,
\begin{equation}
\ddot Q_s^\Phi+\big(3H+2\dot\beta\big)\dot Q_s^\Phi+\Big(\ddot\beta+\dot\beta^2+3H\dot\beta+\frac{k^2}{a^2}+\mathcal{M}_{\Phi\Phi}\Big)Q_s^\Phi=0\,.
\label{eq:Qder_termsmore}
\end{equation}
The above equation carries an extra $2\dot\beta$ friction and extra connection-mass dependent terms. In order to cancel these terms, we need to canonically normalise the entropic field\footnote{This is the standard construction of \cite{GrootNibbelink:2001qt,Gordon:2000hv}, where the reduction is carried out in an orthonormal basis adapted to the trajectory. In general, such a basis rotates along the trajectory and the reduction leaves residual terms proportional to the turning rate. Here it is exact as $\omega^I=0$ by \eqref{eq:turning_rate}, and ${\cal M}^{I}{}_{J}$ is diagonal in the $(\Theta,\Phi,\varphi)$ basis, so the entropic modes neither source ${\cal R}_c$ nor mix with one another.} (as we did in \eqref{eq:normalised_pert})
\begin{equation}
\delta s^\Phi\equiv\sqrt{K_{\Phi\Phi}}\,Q_s^\Phi=e^{\beta}Q_s^\Phi\,,
\label{eq:deltas_canon}
\end{equation}
so that
\begin{align}
\dot Q_s^\Phi&=e^{-\beta}\big(\dot{\delta s}^\Phi-\dot\beta\,\delta s^\Phi\big)\,,\\
\ddot Q_s^\Phi&=e^{-\beta}\big(\ddot{\delta s}^\Phi-2\dot\beta\,\dot{\delta s}^\Phi+(\dot\beta^2-\ddot\beta)\,\delta s^\Phi\big)\,.
\end{align}
Substituting these into \eqref{eq:Qder_termsmore}, the friction coefficient reduces to the standard $3H$ one and the $\beta-$dependent mass term cancels identically leaving
\begin{equation}
\ddot{\delta s}^\Phi+3H\,\dot{\delta s}^\Phi +\Big(\frac{k^2}{a^2}+{\cal M}_{\Phi\Phi}\Big)\delta s^\Phi=0\,.
\label{eq:appA_reduced}
\end{equation}
The covariant derivatives have become ordinary and the mass is unchanged. 
Finally we pass to ${\cal S}^\Phi=(H/\dot\sigma)\,\delta s^\Phi$, i.e.
\begin{equation}
    \delta s^\Phi=\sqrt{2\epsilon}\,M_P\,{\cal S}^\Phi\,,
\end{equation}
where we used $\dot\sigma=\sqrt{2\epsilon}\,M_P H$. Differentiating the rescaling factor with respect to time gives
\begin{equation}
\frac{\dot{\big(\sqrt{2\epsilon}\big)}}{\sqrt{2\epsilon}}=\frac{\dot\epsilon}{2\epsilon}=\frac{\epsilon_2}{2}H\,,
\label{eq:appA_eps2}
\end{equation}
by the definition of $\epsilon_2$. Inserting this into \eqref{eq:appA_reduced} and dividing by $\sqrt{2\epsilon}\,M_P$, we find that the first derivative of the rescaling adds to the friction and its second derivative to the mass,
\begin{equation}
\ddot{\cal S}^\Phi+\big(3H+\epsilon_2 H\big)\dot{\cal S}^\Phi+\Big(\frac{k^2}{a^2}+{\cal M}_{\Phi\Phi}+\frac{3}{2}\epsilon_2H^2
+\frac{\ddot{\big(\sqrt{2\epsilon}\big)}}{\sqrt{2\epsilon}}\Big){\cal S}^\Phi=0\,.
\end{equation}
Differentiating \eqref{eq:appA_eps2} once more and using $\dot H=-\epsilon H^2$ together with $\dot\epsilon_2=\epsilon_2\epsilon_3 H$,
\begin{equation}
\frac{\ddot{\big(\sqrt{2\epsilon}\big)}}{\sqrt{2\epsilon}}=\frac{\epsilon_2}{2}\Big(\epsilon_3-\epsilon+\frac{\epsilon_2}{2}\Big)H^2\,,
\end{equation}
so that the mass terms combine into
\begin{equation}
\Delta_\epsilon\equiv\frac{\epsilon_2}{2}\Big(3-\epsilon+\frac{\epsilon_2}{2}+\epsilon_3\Big)H^2\,,
\end{equation}
and there is a shift in the friction term
\begin{equation}
\ddot{\cal S}^\Phi+(3+\epsilon_2)H\,\dot{\cal S}^\Phi+\Big(\frac{k^2}{a^2}+{\cal M}_{\Phi\Phi}+\Delta_\epsilon\Big){\cal S}^\Phi=0,
\label{eq:appA_final}
\end{equation}
which is \eqref{eq:ddotSvarphi} with $I=\Phi$. The $\varphi$ mode gives \eqref{eq:ddotSvarphi} with $\tilde{\gamma}\equiv\tfrac12\ln K_{\varphi\varphi}$.

The curvature perturbation follows the same steps with $Q_\sigma$ (already canonical) in place of $\delta s$. Here, since $\Omega_\sigma^2=k^2/a^2-\Delta_\epsilon$ (cf. \eqref{eq:omega_sigma}), the shift cancels the mass and one recovers the massless equation \eqref{eq:curvcons},
\begin{equation}
\ddot{\cal R}_c+(3+\epsilon_2)H\,\dot{\cal R}_c+\frac{k^2}{a^2}{\cal R}_c=0\,,
\end{equation}
consistent with the super-horizon conservation of ${\cal R}_c$ in the absence of sources.
\section{Evaluation of the isocurvature mass}\label{app:mass_der}

The effective isocurvature mass of the axion mode, derived in Sec.~\ref{se:ps} as \eqref{eq:m_phiphi}, reads
\begin{equation}
    \mathcal{M}_{\varphi\varphi}= \tilde{\gamma}_{,\sigma}\,U_{,\sigma}- \bigl(\tilde{\gamma}^2_{,\sigma} +\tilde{\gamma}_{,\sigma\sigma}\bigr)\dot\sigma^2\,,
    \quad
    \tilde{\gamma} \equiv \tfrac{1}{2}\ln K_\varphi^{\rm new}(\Theta)\,.
\end{equation}
Here we evaluate it explicitly in the slow-roll attractor. Throughout this appendix we set $M_P=U_0=1$. The ratio $\mathcal{M}_{\varphi\varphi}/H^2$ quoted in the main text is independent of both.
\begin{itemize}
    \item \textbf{Hubble rate and kinetic energy:}
    \begin{equation}
        H^2 = \frac{(1-\Theta)^2}{3}\,,
        \qquad
        \dot\sigma^2 = \frac{4}{3\,\widetilde{K}_\Theta(\Theta)}\,.
    \end{equation}
    The second relation follows from the slow-roll equation of motion $3H\,\widetilde{K}_\Theta\,\dot\Theta = -U'$, which gives $\dot\sigma^2 = \widetilde{K}_\Theta\,\dot\Theta^2 = (U')^2/(9H^2\widetilde{K}_\Theta)= 4/(3\widetilde{K}_\Theta)$.
 
    \item \textbf{Potential gradient along $\sigma$:}
    \begin{equation}
        U_{,\sigma} = \frac{U'}{\sqrt{\widetilde{K}_\Theta}} = \frac{-2(1-\Theta)}{\sqrt{\widetilde{K}_\Theta(\Theta)}}\,.
    \end{equation}
 
    \item \textbf{Analytical $\tilde{\gamma}$ derivatives:} From $K_\varphi^{\rm new}(\Theta) = \Theta^{-n}(\kappa\Theta+c)/(\kappa c)$,
    \begin{align}
        \tilde{\gamma}'  &= \frac{1}{2}\!\left(-\frac{n}{\Theta}+\frac{\kappa}{\kappa\Theta+c}\right),\\
        \tilde{\gamma}'' &= \frac{1}{2}\!\left(\frac{n}{\Theta^2}-\frac{\kappa^2}{(\kappa\Theta+c)^2}\right).
    \end{align}
    Note that $\tilde{\gamma}$ involves $K_\varphi^{\rm new}$, which is unaffected by the off-diagonal correction, and only the adiabatic quantities ($\dot\sigma^2$, $U_{,\sigma}$, $\sigma$-derivatives) use $\widetilde{K}_\Theta$. Moreover, the same function $\tilde{\gamma}$ controls the strong-coupling scales of App.~\ref{app:cutoff}.
 
    \item \textbf{$\sigma$-direction derivatives of $\tilde{\gamma}$:}
    \begin{align}
        \tilde{\gamma}_{,\sigma} &= \frac{\tilde{\gamma}'}{\sqrt{\widetilde{K}_\Theta}}\,,\\
        \tilde{\gamma}_{,\sigma\sigma}&= \frac{1}{\widetilde{K}_\Theta}\!\left(\tilde{\gamma}''- \frac{(\widetilde{K}_\Theta)'}{2\,\widetilde{K}_\Theta}\,\tilde{\gamma}'\right).
    \end{align}
 \end{itemize}
All quantities are evaluated at $\Theta_*(\xi_\chi)$, the field value $N_*=55$ $e$-folds before the end of inflation.

The same quantity also fixes the tilt \eqref{eq:n_iso} of the isocurvature spectrum. We read \eqref{eq:PS_axion} for a generic mode, with the subscript $*$ denoting evaluation at its own horizon exit, $k=aH$. Then
\begin{equation}
    \frac{{\rm d}\ln k}{{\rm d}N}=1+\frac{{\rm d}\ln H}{{\rm d}N}=1-\epsilon\,,
\end{equation}
so that ${\rm d}/{\rm d}\ln k={\rm d}/{\rm d}N$ to first order in slow roll. The axion fraction $F_A$ and the background misalignment $\varphi_{\rm ini}$ are properties of the homogeneous solution and carry no dependence on $k$, so the only mode dependence of \eqref{eq:PS_axion} resides in $H_*$ and $K_\varphi(\Theta_*)$, giving
\begin{align}
n_{\rm iso}-1=\frac{{\rm d}\ln\mathcal{P}_S^{(A)}}{{\rm d}\ln k}&=2\frac{{\rm d}\ln H}{{\rm d}N}-\frac{{\rm d}\ln K_\varphi}{{\rm d}N}\nonumber \\&=-2\epsilon_*-\left.\frac{{\rm d}\ln K_\varphi}{{\rm d}N}\right|_*\,.
    \label{eq:n_iso_app}
\end{align}
The second term is nothing but the leading contribution to $\mathcal{M}_{\varphi\varphi}$. Using the slow-roll equation in the form $U_{,\sigma}=-3H\dot\sigma$ and $\dot{\tilde{\gamma}}=\tilde{\gamma}_{,\sigma}\dot\sigma$, the first term of \eqref{eq:m_phiphi} reads
\begin{equation}
\tilde{\gamma}_{,\sigma}U_{,\sigma}=-3H\dot{\tilde{\gamma}}=-3H^2\frac{{\rm d}\tilde{\gamma}}{{\rm d}N}=-\frac{3}{2}H^2\frac{{\rm d}\ln K_\varphi}{{\rm d}N}\,,
    \label{eq:Mphiphi_slowroll}
\end{equation}
the remaining term $-(\tilde{\gamma}_{,\sigma}^2+\tilde{\gamma}_{,\sigma\sigma})\dot\sigma^2$ being of second order in slow roll. Substituting \eqref{eq:Mphiphi_slowroll} into \eqref{eq:n_iso_app} returns \eqref{eq:n_iso}.

\section{The classical Higgs-Dilaton model in the Einstein frame}

\label{app:HD}
In this appendix we collect the conformal-transformation steps used in Sec.~\ref{sec:HD} to pass to the Einstein frame, and review the standard formulation of the unmodified ($n=0$) Higgs-Dilaton model \cite{Wetterich:1987fm,Shaposhnikov:2008xi,Casas:2017wjh,Casas:2018fum,Kubo:2018kho,Rubio:2020zht,Kubo:2020fdd,Garces:2026cbj}, to which our construction reduces at late times.
 
Under the Weyl rescaling \eqref{eq:Weyl}, the Ricci scalar transforms as
\begin{equation}
    R= \Omega^2\left[\tilde{R}+ 6\,\tilde\Box\ln\Omega-6\,\tilde{g}^{\mu\nu}\partial_\mu\ln\Omega\,\partial_\nu\ln\Omega\right],
    \label{eq:Ricci_transform}
\end{equation}
while $\sqrt{-g} = \sqrt{-\tilde{g}}\,\Omega^{-4}$. Substituting into the non-minimal coupling term and using
$\Omega^{-4}\cdot(B/2)\cdot\Omega^2 = M_P^2/2$, one obtains
\begin{align}
    \sqrt{-g}\,\frac{B}{2}\,R=&\sqrt{-\tilde{g}}\,\frac{M_P^2}{2}\,\tilde{R}+ \sqrt{-\tilde{g}}\,3M_P^2\,\tilde\Box\ln\Omega \\&
    - \sqrt{-\tilde{g}}\,3M_P^2\,\tilde{g}^{\mu\nu}\partial_\mu\ln\Omega\,\partial_\nu\ln\Omega\,.
    \label{eq:Weyl_NMC}
\end{align}
The second term on the right-hand side of \eqref{eq:Weyl_NMC} is a total derivative,
\begin{equation}
    \sqrt{-\tilde{g}}\,\tilde\Box\ln\Omega=\partial_\mu\!\left(\sqrt{-\tilde{g}}\,\tilde{g}^{\mu\nu}\partial_\nu\ln\Omega\right),
\end{equation}
and is dropped from the action. Using
\begin{equation}
    \partial_\mu\ln\Omega= \frac{1}{2}\,\partial_\mu\ln B= \frac{\xi_h h\,\partial_\mu h + \xi_\chi\chi\,\partial_\mu\chi}{B}\,,
\end{equation}
the remaining gradient term gives
\begin{equation}
    -3M_P^2\,\tilde{g}^{\mu\nu}\partial_\mu\ln\Omega\,\partial_\nu\ln\Omega=-\frac{3M_P^2}{B^2}\left(\xi_h h\,\partial_\mu h + \xi_\chi\chi\,\partial_\mu\chi\right)^2\,.
    \label{eq:conformal_term}
\end{equation}
Since only the non-minimal coupling $(B/2)\,R$ is involved, these steps are independent of the deformation $\Theta^{-n}$ and hold both for the modified model of Sec.~\ref{sec:HD} and for the classical Higgs-Dilaton model, to which we now turn.
 
Setting $n=0$, combining \eqref{eq:conformal_term} with the rescaled kinetic terms \eqref{eq:hkin_rescaled}--\eqref{eq:phikin_rescaled} and performing, on top of the Weyl rescaling \eqref{eq:Weyl}, the field
redefinition
\begin{equation}
\exp\left[\gamma\, \Phi\right] \equiv \sqrt{\frac{\kappa_c}{\kappa}} \frac{\bar\rho}{M_P}\,, \hspace{10mm}
\gamma^{-2} \Theta \equiv \frac{\bar\rho^2}{\Omega^2 M_P^2}\,,\label{eq:Thetadef}
\end{equation}
with $\bar \rho^2(h,\chi)\equiv  (1+6\xi_h)h^2+ (1+6\xi_\chi)\chi^2$ and
\begin{equation}\label{kappadef}
\kappa_c \equiv-\frac{\xi_h}{1+6 \xi_h}\,, \hspace{5mm}
\kappa \equiv\kappa_c\left(1-\frac{\xi_\chi}{\xi_h}\right)\,,\hspace{5mm}
\gamma^2 \equiv \frac{\xi_\chi}{1+6  \xi_\chi}\,,
\end{equation}
we obtain an Einstein-frame action of the form \eqref{eq:sigmamodel} with
\begin{align}\label{Ktheta}
&K_\Theta=-\frac{1}{4\, \Theta}\left(\frac{1}{\kappa\,\Theta+c}+\frac{a_\kappa}{1-\Theta}\right)\simeq -\frac{1}{4\, \Theta}\left(\frac{1}{\kappa\,\Theta+c}\right)\,, \\
& K_\Phi(\Theta)=\Theta\,, \label{eq:KphiHD}\\
& K_\varphi(\Theta)=\frac{\kappa\,\Theta+c }{\kappa\,c} \,,\label{eq:Kvarphi}\\
& U(\Theta)= U_0(1-\Theta)^2\,.\label{eq:Utheta}
\end{align}
Here
\begin{equation}\label{cdef}
U_0\equiv\frac{\lambda \, a_\kappa^2 M_P^4}{4}\,,\,\hspace{5mm}
a_\kappa \equiv \frac{1+6  \kappa}{\kappa}\,, \, \hspace{5mm}
c\equiv\frac{\kappa}{\kappa_c}\gamma^2\,.
\end{equation}
Specialising the entropic frequencies entering \eqref{QPhi} and \eqref{Qvarphi} to these kinetic functions gives
\begin{align}
&(\Omega_{s}^2)^\Phi{}_\Phi\equiv \frac{k^2}{a^2} + \frac{\kappa}{M_P^2}\,\dot \sigma^2 + \beta_{,\sigma} U_{,\sigma}\,,  \\
&(\Omega_{s}^2)^\varphi{}_\varphi\equiv \frac{k^2}{a^2}+ \frac{\kappa}{M_P^2} \,\dot \sigma^2 +  \tilde{\gamma}_{,\sigma} U_{,\sigma} \,.
\end{align}
Note that during inflation this three-field kinetic manifold is maximally symmetric, with constant sectional curvature
\begin{equation}
{\cal K}\simeq\frac{\kappa}{M_P^2}\,,
\end{equation}
i.e.\ a hyperbolic space with field-space Ricci scalar
\begin{equation}
R=\frac{6\kappa}{M_P^2}\simeq-\frac{1}{M_P^2}.
\end{equation}

\section{Derivation of the kinetic coefficients} \label{app:kinetic} 

In this appendix we derive the kinetic coefficients of the $(\Theta,\Phi,\varphi)$ basis mentioned in Sec.~\ref{sec:HD}. We start from the Einstein-frame field-space metric generated by the Weyl rescaling \eqref{eq:Weyl}. The explicit kinetic terms in \eqref{eq:Jmod} transform as
\begin{align}
(\partial h)^2&\;\longrightarrow\;\frac{M_P^2}{B}\,(\partial h)^2\,,
\label{eq:hkin_rescaled}\\
\Theta^{-n}(\partial\chi)^2&\;\longrightarrow\;\frac{\Theta^{-n}M_P^2}{B}\,(\partial\chi)^2\,,
\label{eq:chixin_rescaled}\\
\Theta^{-n}\chi^2(\partial\varphi)^2&\;\longrightarrow\;\frac{\Theta^{-n}\chi^2 M_P^2}{B}\,(\partial\varphi)^2\,.
\label{eq:phikin_rescaled}
\end{align}
Collecting \eqref{eq:conformal_term}--\eqref{eq:phikin_rescaled}, the Einstein-frame Lagrangian takes the form
\begin{equation}
    \frac{\mathcal{L}_E}{\sqrt{-\tilde{g}}}=\frac{M_P^2}{2}\tilde{R}- \frac{1}{2}\mathcal{G}_{AB}\,\tilde{g}^{\mu\nu}\partial_\mu\phi^A\partial_\nu\phi^B- U(h,\chi)\,,
    \label{eq:LE}
\end{equation}
where $\phi^A=(h,\chi,\varphi)$ and the field-space metric $\mathcal{G}_{AB}$ has components 
\begin{align}
\mathcal{G}_{hh}&= \frac{M_P^2}{B}+\frac{6\xi_h^2 h^2 M_P^2}{B^2}\,,
&\mathcal{G}_{h\chi}&= \frac{6\xi_h\xi_\chi h\chi\,M_P^2}{B^2}\,,
\label{eq:Ghh_Ghchi}\\
\mathcal{G}_{\chi\chi}&= \frac{\Theta^{-n}M_P^2}{B}+ \frac{6\xi_\chi^2\chi^2 M_P^2}{B^2}\,,&
\mathcal{G}_{\varphi\varphi}&= \frac{\Theta^{-n}\chi^2 M_P^2}{B}\,.
\label{eq:Gchichi_Gphiphi}
\end{align}
The standard HD kinetic matrix is recovered for $n=0$, and at late times, $\Theta_0=1$, for any $n$. Splitting $\mathcal{G}_{AB}$ into the HD piece and the modification $\delta \mathcal{G}_{\rm AB}$ gives
\begin{equation}
\mathcal{G}_{AB}=\frac{M_P^2}{B}\,\delta_{AB}+\frac{6M_P^2}{B^2}\,v_A v_B+\bigl(\Theta^{-n}-1\bigr)\frac{M_P^2}{B}\,
    \delta_A^\chi\delta_B^\chi,
\label{eq:Gsplit}
\end{equation}
with $v_A\equiv(\xi_h h,\;\xi_\chi\chi)= \tfrac{1}{2}\,\partial_A B$. Eq.~\eqref{eq:Gsplit} refers to the $(h,\chi)$ block, the axion entry $\mathcal{G}_{\varphi\varphi}$ is given in \eqref{eq:Gchichi_Gphiphi} and carries the same $\Theta^{-n}$ factor.
The HD metric (the first two terms) decomposes into a direction-independent piece and a rank-one term along the non-minimal-coupling gradient $v_A$, generated by the Weyl rescaling of \eqref{eq:Weyl}; the modification adds weight only to the $\chi\chi$ entry.

We now pass to the $(\Theta,\Phi,\varphi)$ basis. The change of variables $(h,\chi)\to(\Theta,\Phi)$ defines
two tangent vectors in the $(h,\chi)$ plane,
\begin{align}
e_\Phi^A&\equiv\Bigl(  \frac{\partial h}{\partial\Phi}\Big|_\Theta,\frac{\partial\chi}{\partial\Phi}\Big|_\Theta\Bigr)= \gamma(h,\chi)\,,\label{eq:phi_base}\\
e_\Theta^A&\equiv\Bigl(\frac{\partial h}{\partial\Theta}\Big|_\Phi,\frac{\partial\chi}{\partial\Theta}\Big|_\Phi\Bigr)=\bar\rho\,\bigl(f_1'(\Theta),\,f_2'(\Theta)\bigr)\,,
\label{eq:theta_base}
\end{align}
with $\bar\rho$ from \eqref{eq:A}. The first follows from $\gamma\,d\Phi=d\ln\bar\rho$ (namely \eqref{eq:Thetadef}): moving along $\Phi$ at fixed $\Theta$ is a uniform rescaling $h\to\lambda h$, $\chi\to\lambda\chi$ (the dilaton direction). Moving along $\Theta$ at fixed $\Phi$ instead changes the ratio $r=h/\chi$ at constant scale $\bar\rho$ (the inflaton direction); writing $h=\bar\rho\,f_1(\Theta)$ and $\chi=\bar\rho\,f_2(\Theta)$,
the functions $f_1,f_2$ are determined by
\begin{align}
(1+6\xi_h)f_1^2+(1+6\xi_\chi)f_2^2 &= 1\,,
\label{eq:f_constraint1}\\
\xi_h f_1^2+\xi_\chi f_2^2 &= \frac{\gamma^2}{\Theta}\,,
\label{eq:f_constraint2}
\end{align}
enforcing \eqref{eq:Thetadef}. Differentiating with respect to $\Theta$
\begin{align}
(1+6\xi_h)f_1 f_1'+(1+6\xi_\chi)f_2 f_2'&=0\,,\\
\xi_h f_1 f_1'+\xi_\chi f_2 f_2'&= -\frac{\gamma^2}{2\Theta^2}\,,
\label{eq:f_deriv_constraints}
\end{align}
and solving for $\xi_h\neq\xi_\chi$ we obtain
\begin{equation}
f_1 f_1'= \frac{(1+6\xi_\chi)\,\gamma^2}{2\Theta^2(\xi_\chi-\xi_h)}\,,\quad f_2 f_2'= \frac{(1+6\xi_h)\,\gamma^2}{2\Theta^2(\xi_h-\xi_\chi)}.
\label{eq:fi_fprimei}
\end{equation}
The full change of variables $\partial_\mu h = e_\Phi^h\,\partial_\mu\Phi+e_\Theta^h\,\partial_\mu\Theta$
(and similarly for $\chi$; there is no $\partial_\mu\varphi$ contribution since $h,\chi$ do not depend on the axion phase) leads, after expansion, to three terms \begin{align}
M_P^2 K_\Phi^{\rm new}&= \mathcal{G}_{AB}\,e_\Phi^A\,e_\Phi^B\,,
\label{eq:KPhi_gen}\\[2pt]
 M_P^2 K_\Theta^{\rm new}&= \mathcal{G}_{AB}\,e_\Theta^A\,e_\Theta^B\,,
\label{eq:KTheta_gen}\\[2pt]
 M_P^2 K_{\Theta\Phi}&= \mathcal{G}_{AB}\,e_\Theta^A\,e_\Phi^B\,.
\label{eq:KTP_gen}
\end{align}

We compute the three projections explicitly, exhibiting in each the split into the HD result and the $\Theta^{-n}$ correction.
We start by calculating the kinetic coefficient $K_{\Phi}$ as
\begin{widetext}
\begin{align}
M_P^2\,K_\Phi^{\rm new} &= \gamma^2\!\left[\mathcal{G}_{hh}\,h^2+2\,\mathcal{G}_{h\chi}\,h\chi+\mathcal{G}_{\chi\chi}\,\chi^2\right]\nonumber\\
&= \gamma^2\!\left[\left(\frac{M_P^2}{B}+\frac{6\xi_h^2 h^2 M_P^2}{B^2}\right)h^2+ 2\,\frac{6\xi_h\xi_\chi h\chi\,M_P^2}{B^2}\,h\chi+ \left(\frac{6\xi_\chi^2\chi^2 M_P^2}{B^2}+\frac{\Theta^{-n}M_P^2}{B}\right)\chi^2\right]\nonumber\\
&= \gamma^2 M_P^2\!\left[\frac{h^2+\Theta^{-n}\chi^2}{B}+ \frac{6\!\left(\xi_h h^2+\xi_\chi\chi^2\right)^2}{B^2}
   \right]\nonumber\\
&= \gamma^2 M_P^2\!\left[\frac{h^2+\Theta^{-n}\chi^2}{B} + 6\right].
\end{align}
\end{widetext}
Writing $h^2+\Theta^{-n}\chi^2 = (h^2+\chi^2)+(\Theta^{-n}-1)\chi^2$ and using $(h^2+\chi^2)/B = A/B - 6$, one obtains
\begin{equation}
    K_\Phi^{\rm new}(\Theta)= \gamma^2\frac{A}{B}+ \left(\Theta^{-n}-1\right)\frac{\gamma^2\chi^2}{B}\,.
\end{equation}
The first term is the standard HD result $K_\Phi^{\rm HD}=\Theta$, recovered exactly for $n=0$.
The correction is the projection of the modified radial $\chi$ kinetic term onto the $\Phi$ direction.
Expressing $\chi^2/B = 1/(\xi_h r^2(\Theta)+\xi_\chi)$ as a function of $\Theta$ via the ratio $r(\Theta)\equiv h/\chi$ determined by \eqref{eq:Thetadef}, one has \eqref{eq:Kphi_new} in terms of the couplings,
\begin{equation}
    K_\Phi^{\rm new}(\Theta) = K_{\Phi}^{\rm HD}(\Theta)+ \frac{(\Theta^{-n}-1)\,\gamma^2}{\xi_h r^2(\Theta)+\xi_\chi}\,.
\end{equation}
The inflaton coefficient follows from the analogous projection onto $e_\Theta=\bar\rho\,(f_1',f_2')$. Using $h=\bar\rho f_1$, $\chi=\bar\rho f_2$ and the second constraint of \eqref{eq:f_deriv_constraints},
$\xi_h f_1f_1'+\xi_\chi f_2f_2'=-\gamma^2/(2\Theta^2)$, together with $\bar\rho^2/B=\Theta/\gamma^2$,
\begin{align}
M_P^2\,K_\Theta^{\rm new}&= \bar\rho^2\!\left[\mathcal{G}_{hh}\,f_1'^{\,2}+ 2\,\mathcal{G}_{h\chi}\,f_1'f_2'+\mathcal{G}_{\chi\chi}\,f_2'^{\,2}\right]\nonumber\\
 &= \bar\rho^2 M_P^2\!\left[\frac{f_1'^{\,2}+\Theta^{-n}f_2'^{\,2}}{B}+ \frac{6\bigl(\xi_h h f_1'+\xi_\chi\chi f_2'\bigr)^2}{B^2}\right]\nonumber\\
&= M_P^2\!\left[\frac{\Theta}{\gamma^2}\left(f_1'^{\,2}+\Theta^{-n}f_2'^{\,2}\right)+ \frac{3}{2\Theta^2}\right].
\label{eq:KTheta_calc_app}
\end{align}
Splitting $\Theta^{-n}=1+(\Theta^{-n}-1)$, we get
\begin{equation}
    K_\Theta^{\rm new}(\Theta)= K_\Theta^{\rm HD}(\Theta)+ \left(\Theta^{-n}-1\right)\frac{\Theta}{\gamma^2}\,f_2'^{\,2}\,,
\end{equation}
where the $n$-independent piece, $K_\Theta^{\rm HD}=(\Theta/\gamma^2)(f_1'^{\,2}+f_2'^{\,2})+3/(2\Theta^2)$, coincides with \eqref{Ktheta} of App.~\ref{app:HD} upon using \eqref{eq:fi_fprimei} and the explicit
$f_1^2$, $f_2^2$; this is \eqref{eq:KTheta_new} of the main text.

Finally, the off-diagonal coefficient carries exactly one factor of each tangent vector,
\begin{align}
M_P^2\,K_{\Theta\Phi}&= \gamma\bar\rho\!\left[\mathcal{G}_{hh}\,h f_1'+\mathcal{G}_{h\chi}\bigl(h f_2'+\chi f_1'\bigr)+ \mathcal{G}_{\chi\chi}\,\chi f_2'\right]\nonumber\\
 &= \frac{\gamma\bar\rho^2 M_P^2}{B}\!\left[f_1f_1'+\Theta^{-n}f_2f_2'+ 6\bigl(\xi_h f_1f_1'+\xi_\chi f_2f_2'\bigr)
   \right]\nonumber\\
& = \left(\Theta^{-n}-1\right)\frac{\Theta}{\gamma}\,f_2f_2'\;M_P^2\,.
\label{eq:KTP_calc_app}
\end{align}
At $n=0$ the prefactor vanishes, so the standard Higgs-Dilaton model is exactly diagonal in the $(\Theta,\Phi)$ variables, $K_{\Theta\Phi}^{\rm HD}=0$.

\subsection{Near-pole structure of the effective kinetic function}\label{app:pole}

Here we derive the two regimes of the effective inflaton kinetic function \eqref{eq:Keff_closed} near the field-space boundary $\Theta_{\rm bd}$, used in Sec.~\ref{subsec:inflation_obs}. With $\delta\equiv\Theta-\Theta_{\rm bd}$ and using the near-pole expansion 
\begin{align}
f_2^2&\simeq2A_0\,\delta,\\ f_2'^{\,2}&\simeq \frac{A_0}{2\delta},\\ 
A_0&=\frac{(1+6\xi_h)\gamma^2}{2\Theta_{\rm bd}^2(\xi_h-\xi_\chi)}>0,
\label{eq:nearpole_exp}
\end{align}
the HD term develops a simple pole,
\begin{equation}
    K_\Theta^{\rm HD}\simeq\frac{\mathcal{R}^{\rm HD}}{\delta}\,,
    \qquad
    \mathcal{R}^{\rm HD}\simeq\frac{1+6\xi_h}{4\,\xi_h\,\Theta_{\rm bd}}\,.
    \label{eq:RHD}
\end{equation}
The off-diagonal coefficient~\eqref{eq:KTP_box} stays finite at the pole, so it does not introduce a new singularity. In the immediate neighbourhood of the pole ($f_2^2\to0$) the subtraction in~\eqref{eq:Keff_def} is negligible and one recovers the simple pole with a rescaled residue,
\begin{equation}
    \widetilde K_\Theta \simeq \frac{\mathcal{R}^{\rm new}}{\delta}\,,\qquad \mathcal{R}^{\rm new}=\Theta_{\rm bd}^{-n}\,\mathcal{R}^{\rm HD}\,,
    \label{eq:Rnew}
\end{equation}
so the simple-pole regime is confined to a thin shell. The subtraction in~\eqref{eq:Keff_closed} becomes important for $(\Theta^{-n}-1)\,f_2^2\gtrsim1$, i.e.\ $\delta\gtrsim\delta_c$ with\footnote{The crossover is fixed by $(\Theta^{-n}-1)\,f_2^2=1$. Using the near-pole expansions \eqref{eq:nearpole_exp} and $(\Theta^{-n}-1)\simeq\Theta_{\rm bd}^{-n}\gg1$, this reads $\Theta_{\rm bd}^{-n}\,2A_0\,\delta_c=1$, giving~\eqref{eq:crossover}.}
\begin{equation}
    \delta_c\;\equiv\;\frac{\Theta_{\rm bd}^{\,n}}{2A_0}\,.
    \label{eq:crossover}
\end{equation}
Outside this shell, using
\begin{equation}
    \frac{f_2'}{f_2}=\frac{\Theta_{\rm bd}}{2\,\Theta\,(\Theta-\Theta_{\rm bd})}\,,
\end{equation}
the correction in~\eqref{eq:Keff_closed} saturates to \eqref{eq:saturated}.
The simple pole \eqref{eq:Rnew}, with residue $\mathcal{R}^{\rm new}$, and the second-order pole \eqref{eq:Ktil_pole}, with residue $a$, describe opposite sides of the crossover \eqref{eq:crossover}. They match there, $a/\delta_c=\mathcal{R}^{\rm new}$ up to the factor $\xi_h/(\xi_h-\xi_\chi)$.

\section{Background-dependent cutoff of the effective theory} \label{app:cutoff}

In this appendix we estimate the cutoff associated with the field-dependent axion kinetic term, and show that the effective description remains weakly coupled throughout inflation. Since the scalar sector defines a nonlinear sigma model, the cutoff is generally background dependent. The relevant part of the action is
\begin{equation}
S\supset \int d^4x\sqrt{-g} \left[-\frac12(\partial\sigma)^2 -\frac12M_P^2K_\varphi(\sigma)(\partial\varphi)^2 -U(\sigma)
\right].
\end{equation}
It is useful to define
\begin{equation}
K_\varphi(\sigma)\equiv e^{2\tilde{\gamma}(\sigma)} .
\end{equation}
Expanding around an arbitrary background value, $ \sigma=\bar\sigma+s$, we define the locally canonically normalised angular fluctuation as 
\begin{equation}
q=M_P\sqrt{K_\varphi(\bar\sigma)}\,\varphi .
\end{equation}
The angular kinetic term becomes
\begin{equation}
-\frac12M_P^2 e^{2\tilde{\gamma}(\sigma)}(\partial\varphi)^2 =-\frac12e^{2[\tilde{\gamma}(\bar\sigma+s)-\tilde{\gamma}(\bar\sigma)]}(\partial q)^2 .
\end{equation}
Expanding in powers of \(s\), we find
\begin{equation}
e^{2[\tilde{\gamma}(\bar\sigma+s)-\tilde{\gamma}(\bar\sigma)]}=1+2\tilde{\gamma}' s+\left(\tilde{\gamma}''+2\tilde{\gamma}'^2\right)s^2 +\cdots ,
\end{equation}
where primes denote derivatives with respect to \(\sigma\), evaluated at \(\bar\sigma\). Hence
\begin{equation}
\mathcal L_\varphi =-\frac12(\partial q)^2-\tilde{\gamma}' s(\partial q)^2-\frac12\left(\tilde{\gamma}''+2\tilde{\gamma}'^2\right)s^2(\partial q)^2 +\cdots .
\end{equation}
The cubic operator $s(\partial q)^2$ has mass dimension five, so its coefficient $\tilde{\gamma}'$ sets the corresponding strong-coupling scale. Including the usual $4\pi$ loop factor,
\begin{equation}
\Lambda_3\sim\frac{4\pi}{|\tilde{\gamma}'|}.
\end{equation}
Similarly, the quartic operator \(s^2(\partial q)^2\) has mass dimension six and gives
\begin{equation}
\Lambda_4\sim\frac{4\pi}{\sqrt{|\tilde{\gamma}''+2\tilde{\gamma}'^2|}}.
\end{equation}
We now apply this estimate to the class of angular metric deformations considered in the main text. For the deformation $K_\varphi(\Theta)=\Theta^{-n}K_\varphi^{\rm HD}(\Theta)$, with $n>0$ and \(K_\varphi^{\rm HD}\) the angular metric of the undeformed model,  we have
\begin{equation}
\tilde{\gamma}(\Theta)=\tilde{\gamma}_0(\Theta)-\frac{n}{2}\ln\Theta,
\qquad
\tilde{\gamma}_0(\Theta)\equiv\frac12\ln K_\varphi^{\rm HD}(\Theta).
\end{equation}
The canonical adiabatic field is related to \(\Theta\) by $d\sigma=M_P\sqrt{\widetilde K_\Theta(\Theta)}\,d\Theta$. Therefore, one obtains
\begin{equation}
\tilde{\gamma}'=\frac{1}{M_P\sqrt{\widetilde K_\Theta(\Theta)}}\left[\frac{d\tilde{\gamma}_0}{d\Theta}-\frac{n}{2\Theta}\right].
\end{equation}
Since $2\tilde\gamma=\ln K_\varphi^{\rm new}$, the cubic cutoff $\Lambda_3\sim4\pi/|\tilde\gamma'|$ is controlled by the rate of change of the axion normalisation,
\begin{equation}
\Lambda_3\sim\frac{8\pi M_P\sqrt{\widetilde K_\Theta(\Theta)}}{\left|\,{\rm d}\ln K_\varphi^{\rm new}/{\rm d}\Theta\,\right|}\,,\qquad\frac{{\rm d}\ln K_\varphi^{\rm new}}{{\rm d}\Theta}=-\frac{n}{\Theta}+\frac{\kappa}{\kappa\,\Theta+c}\,,
\label{eq:lambda3_exact}
\end{equation}
the second equality following from \eqref{eq:Kphi_new} and \eqref{eq:Kvarphi}.
At the pivot the first term dominates, so that
\begin{equation}
|\tilde{\gamma}'|\simeq\frac{n}{2M_P\Theta\sqrt{\widetilde K_\Theta(\Theta)}}.
\end{equation}
This approximation is conservative, since $d\tilde{\gamma}_0/d\Theta=1/[2(\Theta-\Theta_{\rm bd})]>0$ partially cancels the deformation term, so the true cutoff is higher than the estimate that follows. The cubic cutoff is therefore
\begin{equation}\label{eq:lambda3}
\Lambda_3\sim\frac{8\pi M_P\Theta\sqrt{\widetilde K_\Theta(\Theta)}}{n}.
\end{equation}
For the same deformation the quartic coefficient is dominated by $\tilde{\gamma}'^2$, and in the regime $|\tilde{\gamma}''|\ll2\tilde{\gamma}'^2$, we get
\begin{equation}
    \Lambda_4\simeq\frac{4\pi}{\sqrt{2}\,|\tilde{\gamma}'|}=\frac{\Lambda_3}{\sqrt{2}}\,.
\end{equation}
Increasing $n$ lowers $\Lambda_3$, imposing an upper bound complementary to the isocurvature lower bound. Increasing $\xi_\chi$ raises $\Theta_*$ at essentially fixed $\widetilde K_\Theta(\Theta_*)$, since the pivot lies at a fixed distance from the pole (Sec.~\ref{subsec:inflation_obs}), and hence raises the cutoff. The smallest cutoff therefore occurs at the largest exponent and smallest coupling. At the reference coupling $\xi_\chi=5\times10^{-5}$, where $\Theta_*\simeq1.3\times10^{-2}$, \eqref{eq:lambda3} reads $\Lambda_3\simeq32\,M_P/n$, the minimal exponent $n=4$ of \eqref{eq:n_bound} gives $\Lambda_3\simeq7.9\,M_P$ and $\Lambda_4\simeq5.6\,M_P$, while the largest exponent considered, $n=8$, gives $\Lambda_3\simeq3.9\,M_P$ and $\Lambda_4\simeq2.8\,M_P$, so that with $H_*$ as in \eqref{eq:Hstar_fixed} the expansion parameter is $H_*/\Lambda_3\lesssim1.6\times10^{-6}$. The complementary upper bound is never reached in practice: $\Lambda_3$ would fall to $M_P$ only near $n\simeq32$, far above any exponent the isocurvature bound selects. The $\Theta^{-n}$ deformation thus introduces no new scale below $M_P$, and the effective description remains weakly coupled throughout inflation for every exponent admitted.

\bibliography{references}

@article{Gordon:2000hv,
    author = "Gordon, Christopher and Wands, David and Bassett, Bruce A. and Maartens, Roy",
    title = "{Adiabatic and entropy perturbations from inflation}",
    eprint = "astro-ph/0009131",
    archivePrefix = "arXiv",
    doi = "10.1103/PhysRevD.63.023506",
    journal = "Phys. Rev. D",
    volume = "63",
    pages = "023506",
    year = "2000"
}

@article{GrootNibbelink:2001qt,
    author = "Groot Nibbelink, S. and van Tent, B. J. W.",
    title = "{Scalar perturbations during multiple field slow-roll inflation}",
    eprint = "hep-ph/0107272",
    archivePrefix = "arXiv",
    reportNumber = "SPIN-2001-19, ITP-UU-01-27",
    doi = "10.1088/0264-9381/19/4/302",
    journal = "Class. Quant. Grav.",
    volume = "19",
    pages = "613--640",
    year = "2002"
}

@article{Barman:2025lvk,
    author = "Barman, Basabendu and Bernal, Nicol{\'a}s and Rubio, Javier",
    title = "{Two or three things particle physicists (mis)understand about (pre)heating}",
    eprint = "2503.19980",
    archivePrefix = "arXiv",
    primaryClass = "hep-ph",
    reportNumber = "IPARCOS-UCM-25-019",
    doi = "10.1016/j.nuclphysb.2025.116996",
    journal = "Nucl. Phys. B",
    volume = "1018",
    pages = "116996",
    year = "2025"
}

@article{Garcia-Bellido:2012npk,
    author = "Garcia-Bellido, Juan and Rubio, Javier and Shaposhnikov, Mikhail",
    title = "{Higgs-Dilaton cosmology: Are there extra relativistic species?}",
    eprint = "1209.2119",
    archivePrefix = "arXiv",
    primaryClass = "hep-ph",
    reportNumber = "PREPRINT-IFT-UAM-CSIC-12-87",
    doi = "10.1016/j.physletb.2012.10.075",
    journal = "Phys. Lett. B",
    volume = "718",
    pages = "507--511",
    year = "2012"
}

@article{Piani:2022gon,
    author = "Piani, Matteo and Rubio, Javier",
    title = "{Higgs-Dilaton inflation in Einstein-Cartan gravity}",
    eprint = "2202.04665",
    archivePrefix = "arXiv",
    primaryClass = "gr-qc",
    doi = "10.1088/1475-7516/2022/05/009",
    journal = "JCAP",
    volume = "05",
    number = "05",
    pages = "009",
    year = "2022"
}

@article{Karananas:2020qkp,
    author = "Karananas, Georgios K. and Michel, Marco and Rubio, Javier",
    title = "{One residue to rule them all: Electroweak symmetry breaking, inflation and field-space geometry}",
    eprint = "2006.11290",
    archivePrefix = "arXiv",
    primaryClass = "hep-th",
    reportNumber = "LMU-ASC 29/20, MPP-2020-102, HIP-2020-19/TH",
    doi = "10.1016/j.physletb.2020.135876",
    journal = "Phys. Lett. B",
    volume = "811",
    pages = "135876",
    year = "2020"
}

@article{Rubio:2020zht,
    author = "Rubio, Javier",
    title = "{Scale symmetry, the Higgs and the Cosmos}",
    eprint = "2004.00039",
    archivePrefix = "arXiv",
    primaryClass = "gr-qc",
    reportNumber = "HIP-2020-8/TH",
    doi = "10.22323/1.376.0074",
    journal = "PoS",
    volume = "CORFU2019",
    pages = "074",
    year = "2020"
}

@article{Planck:2018jri,
    author = "Akrami, Y. and others",
    collaboration = "Planck",
    title = "{Planck 2018 results. X. Constraints on inflation}",
    eprint = "1807.06211",
    archivePrefix = "arXiv",
    primaryClass = "astro-ph.CO",
    doi = "10.1051/0004-6361/201833887",
    journal = "Astron. Astrophys.",
    volume = "641",
    pages = "A10",
    year = "2020"
}

@article{AtacamaCosmologyTelescope:2025nti,
    author = "Calabrese, Erminia and others",
    collaboration = "Atacama Cosmology Telescope",
    title = "{The Atacama Cosmology Telescope: DR6 constraints on extended cosmological models}",
    eprint = "2503.14454",
    archivePrefix = "arXiv",
    primaryClass = "astro-ph.CO",
    reportNumber = "FERMILAB-PUB-25-0157-PPD",
    doi = "10.1088/1475-7516/2025/11/063",
    journal = "JCAP",
    volume = "11",
    pages = "063",
    year = "2025"
}

@article{Galante:2014ifa,
    author = "Galante, Mario and Kallosh, Renata and Linde, Andrei and Roest, Diederik",
    title = "{Unity of Cosmological Inflation Attractors}",
    eprint = "1412.3797",
    archivePrefix = "arXiv",
    primaryClass = "hep-th",
    doi = "10.1103/PhysRevLett.114.141302",
    journal = "Phys. Rev. Lett.",
    volume = "114",
    number = "14",
    pages = "141302",
    year = "2015"
}

@article{Rubio:2018ogq,
    author = "Rubio, Javier",
    title = "{Higgs inflation}",
    eprint = "1807.02376",
    archivePrefix = "arXiv",
    primaryClass = "hep-ph",
    doi = "10.3389/fspas.2018.00050",
    journal = "Front. Astron. Space Sci.",
    volume = "5",
    pages = "50",
    year = "2019"
}

@article{Elliston:2012ab,
    author = "Elliston, Joseph and Seery, David and Tavakol, Reza",
    title = "{The inflationary bispectrum with curved field-space}",
    eprint = "1208.6011",
    archivePrefix = "arXiv",
    primaryClass = "astro-ph.CO",
    doi = "10.1088/1475-7516/2012/11/060",
    journal = "JCAP",
    volume = "11",
    pages = "060",
    year = "2012"
}

@article{Gong:2011uw,
    author = "Gong, Jinn-Ouk and Tanaka, Takahiro",
    title = "{A covariant approach to general field space metric in multi-field inflation}",
    eprint = "1101.4809",
    archivePrefix = "arXiv",
    primaryClass = "astro-ph.CO",
    reportNumber = "CERN-PH-TH-2011-012, YITP-11-9",
    doi = "10.1088/1475-7516/2012/02/E01",
    journal = "JCAP",
    volume = "03",
    pages = "015",
    year = "2011",
    note = "[Erratum: JCAP 02, E01 (2012)]"
}

@article{Malik:2008im,
    author = "Malik, Karim A. and Wands, David",
    title = "{Cosmological perturbations}",
    eprint = "0809.4944",
    archivePrefix = "arXiv",
    primaryClass = "astro-ph",
    doi = "10.1016/j.physrep.2009.03.001",
    journal = "Phys. Rept.",
    volume = "475",
    pages = "1--51",
    year = "2009"
}

@article{Kaiser:2012ak,
    author = "Kaiser, David I. and Mazenc, Edward A. and Sfakianakis, Evangelos I.",
    title = "{Primordial Bispectrum from Multifield Inflation with Nonminimal Couplings}",
    eprint = "1210.7487",
    archivePrefix = "arXiv",
    primaryClass = "astro-ph.CO",
    reportNumber = "MIT-CTP-4410",
    doi = "10.1103/PhysRevD.87.064004",
    journal = "Phys. Rev. D",
    volume = "87",
    pages = "064004",
    year = "2013"
}

@article{Weinberg:1977ma,
    author = "Weinberg, Steven",
    title = "{A New Light Boson?}",
    reportNumber = "HUTP-77/A074",
    doi = "10.1103/PhysRevLett.40.223",
    journal = "Phys. Rev. Lett.",
    volume = "40",
    pages = "223--226",
    year = "1978"
}

@article{Wilczek:1977pj,
    author = "Wilczek, Frank",
    title = "{Problem of Strong  $P$  and  $T$  Invariance in the Presence of Instantons}",
    reportNumber = "Print-77-0939 (COLUMBIA)",
    doi = "10.1103/PhysRevLett.40.279",
    journal = "Phys. Rev. Lett.",
    volume = "40",
    pages = "279--282",
    year = "1978"
}

@article{Wantz:2009it,
    author = "Wantz, Olivier and Shellard, E.P.S.",
    title = "{Axion Cosmology Revisited}",
    eprint = "0910.1066",
    archivePrefix = "arXiv",
    primaryClass = "astro-ph.CO",
    doi = "10.1103/PhysRevD.82.123508",
    journal = "Phys. Rev. D",
    volume = "82",
    pages = "123508",
    year = "2010"
}

@article{Sikivie:2006ni,
    author = "Sikivie, Pierre",
    editor = "Kuster, Markus and Raffelt, Georg and Beltran, Berta",
    title = "{Axion Cosmology}",
    eprint = "astro-ph/0610440",
    archivePrefix = "arXiv",
    reportNumber = "UFIFT-HEP-06-16",
    doi = "10.1007/978-3-540-73518-2_2",
    journal = "Lect. Notes Phys.",
    volume = "741",
    pages = "19--50",
    year = "2008"
}

@article{Peccei:1977hh,
    author = "Peccei, R.D. and Quinn, Helen R.",
    title = "{CP Conservation in the Presence of Instantons}",
    reportNumber = "ITP-568-STANFORD",
    doi = "10.1103/PhysRevLett.38.1440",
    journal = "Phys. Rev. Lett.",
    volume = "38",
    pages = "1440--1443",
    year = "1977"
}

@article{Peccei:2006as,
    author = "Peccei, R.D.",
    editor = "Kuster, Markus and Raffelt, Georg and Beltran, Berta",
    title = "{The Strong CP problem and axions}",
    eprint = "hep-ph/0607268",
    archivePrefix = "arXiv",
    doi = "10.1007/978-3-540-73518-2_1",
    journal = "Lect. Notes Phys.",
    volume = "741",
    pages = "3--17",
    year = "2008"
}

@article{Preskill:1982cy,
    author = "Preskill, John and Wise, Mark B. and Wilczek, Frank",
    editor = "Srednicki, M.A.",
    title = "{Cosmology of the Invisible Axion}",
    reportNumber = "HUTP-82-A048, NSF-ITP-82-103",
    doi = "10.1016/0370-2693(83)90637-8",
    journal = "Phys. Lett. B",
    volume = "120",
    pages = "127--132",
    year = "1983"
}

@article{Abbott:1982af,
    author = "Abbott, L.F. and Sikivie, P.",
    editor = "Srednicki, M.A.",
    title = "{A Cosmological Bound on the Invisible Axion}",
    reportNumber = "PRINT-82-0695 (BRANDEIS)",
    doi = "10.1016/0370-2693(83)90638-X",
    journal = "Phys. Lett. B",
    volume = "120",
    pages = "133--136",
    year = "1983"
}

@article{Axenides:1983hj,
    author  = "Axenides, Minos and Brandenberger, Robert H. and Turner, Michael S.",
    title   = "{Development of Axion Perturbations in an Axion Dominated Universe}",
    journal = "Phys. Lett. B", volume = "126", pages = "178--182", year = "1983",
    doi     = "10.1016/0370-2693(83)90586-5"
}

@article{Dine:1982ah,
    author = "Dine, Michael and Fischler, Willy",
    editor = "Srednicki, M.A.",
    title = "{The Not So Harmless Axion}",
    reportNumber = "UPR-0201T",
    doi = "10.1016/0370-2693(83)90639-1",
    journal = "Phys. Lett. B",
    volume = "120",
    pages = "137--141",
    year = "1983"
}

@article{Linde:1985yf,
    author  = "Linde, Andrei D.",
    title   = "{Generation of Isothermal Density Perturbations in the Inflationary Universe}",
    journal = "Phys. Lett. B", volume = "158", pages = "375--380", year = "1985",
    doi     = "10.1016/0370-2693(85)90436-8"
}

@article{Lyth:1989pb,
    author  = "Lyth, David H.",
    title   = "{A Limit on the Inflationary Energy Density From Axion Isocurvature Fluctuations}",
    journal = "Phys. Lett. B", volume = "236", pages = "408--410", year = "1990",
    doi     = "10.1016/0370-2693(90)90374-F"
}

@article{Turner:1990uz,
    author  = "Turner, Michael S. and Wilczek, Frank",
    title   = "{Inflationary Axion Cosmology}",
    journal = "Phys. Rev. Lett.", volume = "66", pages = "5--8", year = "1991",
    doi     = "10.1103/PhysRevLett.66.5"
}

@article{Tadepalli:2026mdc,
    author = "Tadepalli, Sai Chaitanya",
    title = "{Geometric suppression of CMB isocurvature and a blue-tilted spectrum}",
    eprint = "2601.22221",
    archivePrefix = "arXiv",
    primaryClass = "hep-ph",
    doi = "10.1103/fpd7-xdhl",
    journal = "Phys. Rev. D",
    volume = "114",
    number = "2",
    pages = "023550",
    year = "2026"
}

@article{Fox:2004kb,
    author        = "Fox, Patrick and Pierce, Aaron and Thomas, Scott D.",
    title         = "{Probing a QCD string axion with precision cosmological measurements}",
    eprint        = "hep-th/0409059",
    archivePrefix = "arXiv",
    year          = "2004"
}

@article{Beltran:2006sq,
    author        = "Beltran, Maria and Garcia-Bellido, Juan and Lesgourgues, Julien",
    title         = "{Isocurvature bounds on axions revisited}",
    eprint        = "hep-ph/0606107", archivePrefix = "arXiv",
    journal = "Phys. Rev. D", volume = "75", pages = "103507", year = "2007",
    doi     = "10.1103/PhysRevD.75.103507"
}

@article{Hertzberg:2008wr,
    author        = "Hertzberg, Mark P. and Tegmark, Max and Wilczek, Frank",
    title         = "{Axion Cosmology and the Energy Scale of Inflation}",
    eprint        = "0807.1726", archivePrefix = "arXiv",
    journal = "Phys. Rev. D", volume = "78", pages = "083507", year = "2008",
    doi     = "10.1103/PhysRevD.78.083507"
}

@article{Linde:1987bx,
    author  = "Linde, Andrei D.",
    title   = "{Inflation and Axion Cosmology}",
    journal = "Phys. Lett. B", volume = "201", pages = "437--439", year = "1988",
    doi     = "10.1016/0370-2693(88)90597-7"
}

@article{Tegmark:2005dy,
    author        = "Tegmark, Max and Aguirre, Anthony and Rees, Martin and Wilczek, Frank",
    title         = "{Dimensionless constants, cosmology and other dark matters}",
    eprint        = "astro-ph/0511774", archivePrefix = "arXiv",
    journal = "Phys. Rev. D", volume = "73", pages = "023505", year = "2006",
    doi     = "10.1103/PhysRevD.73.023505"
}

@article{Chakraborty:2025lyp,
    author        = "Chakraborty, Priyesh and Cheng, Junyi and Reece, Matthew and Wang, Zekai",
    title         = "{A Step in Flux to Suppress Axion Isocurvature}",
    eprint        = "2507.12519", archivePrefix = "arXiv",
    journal = "JHEP", volume = "03", pages = "046", year = "2026",
    doi     = "10.1007/JHEP03(2026)046"
}

@article{Kawasaki:2026jen,
    author        = "Kawasaki, Masahiro and Sheng, Jie and Yanagida, Tsutomu T.",
    title         = "{High-Quality Axion Dark Matter without Isocurvature Problem}",
    eprint        = "2605.29370", archivePrefix = "arXiv",
    year          = "2026"
}

@article{BICEP:2021xfz,
    author = "Ade, P. A. R. and others",
    collaboration = "BICEP, Keck",
    title = "{Improved Constraints on Primordial Gravitational Waves using Planck, WMAP, and BICEP/Keck Observations through the 2018 Observing Season}",
    eprint = "2110.00483",
    archivePrefix = "arXiv",
    primaryClass = "astro-ph.CO",
    doi = "10.1103/PhysRevLett.127.151301",
    journal = "Phys. Rev. Lett.",
    volume = "127",
    number = "15",
    pages = "151301",
    year = "2021"
}

@article{Borsanyi:2016ksw,
    author = "Borsanyi, Sz. and others",
    title = "{Calculation of the axion mass based on high-temperature lattice quantum chromodynamics}",
    eprint = "1606.07494",
    archivePrefix = "arXiv",
    primaryClass = "hep-lat",
    reportNumber = "DESY-16-105",
    doi = "10.1038/nature20115",
    journal = "Nature",
    volume = "539",
    number = "7627",
    pages = "69--71",
    year = "2016"
}

@article{GrillidiCortona:2015jxo,
    author = "Grilli di Cortona, Giovanni and Hardy, Edward and Pardo Vega, Javier and Villadoro, Giovanni",
    title = "{The QCD axion, precisely}",
    eprint = "1511.02867",
    archivePrefix = "arXiv",
    primaryClass = "hep-ph",
    doi = "10.1007/JHEP01(2016)034",
    journal = "JHEP",
    volume = "01",
    pages = "034",
    year = "2016"
}

@article{Ce:2015qha,
    author  = "C\`e, Marco and Consonni, Cristian and Engel, Georg P. and Giusti, Leonardo",
    title   = "{Non-Gaussianities in the topological charge distribution of the SU(3) Yang--Mills theory}",
    eprint  = "1506.06052", archivePrefix = "arXiv", primaryClass = "hep-lat",
    doi     = "10.1103/PhysRevD.92.074502",
    journal = "Phys. Rev. D", volume = "92", number = "7", pages = "074502", year = "2015"
}

@article{Boyd:1996bx,
    author  = "Boyd, G. and Engels, J. and Karsch, F. and Laermann, E. and Legeland, C. and Lutgemeier, M. and Petersson, B.",
    title   = "{Thermodynamics of SU(3) lattice gauge theory}",
    eprint  = "hep-lat/9602007", archivePrefix = "arXiv", primaryClass = "hep-lat",
    doi     = "10.1016/0550-3213(96)00170-8",
    journal = "Nucl. Phys. B", volume = "469", pages = "419--444", year = "1996"
}

@article{Visinelli:2009zm,
    author = "Visinelli, Luca and Gondolo, Paolo",
    title = "{Dark Matter Axions Revisited}",
    eprint = "0903.4377",
    archivePrefix = "arXiv",
    primaryClass = "astro-ph.CO",
    doi = "10.1103/PhysRevD.80.035024",
    journal = "Phys. Rev. D",
    volume = "80",
    pages = "035024",
    year = "2009"
}

@article{Maldacena:2002vr,
    author = "Maldacena, Juan Martin",
    title = "{Non-Gaussian features of primordial fluctuations in single field inflationary models}",
    eprint = "astro-ph/0210603",
    archivePrefix = "arXiv",
    doi = "10.1088/1126-6708/2003/05/013",
    journal = "JHEP",
    volume = "05",
    pages = "013",
    year = "2003"
}

@article{Kawasaki:2008pa,
    author = "Kawasaki, Masahiro and Nakayama, Kazunori and Sekiguchi, Toyokazu and Suyama, Teruaki and Takahashi, Fuminobu",
    title = "{A General Analysis of Non-Gaussianity from Isocurvature Perturbations}",
    eprint = "0810.0208",
    archivePrefix = "arXiv",
    primaryClass = "astro-ph",
    reportNumber = "IPMU-08-0069, ICRR-REPORT-529",
    doi = "10.1088/1475-7516/2009/01/042",
    journal = "JCAP",
    volume = "01",
    pages = "042",
    year = "2009"
}

@article{Kubo:2018kho,
    author = "Kubo, Jisuke and Lindner, Manfred and Schmitz, Kai and Yamada, Masatoshi",
    title = "{Planck mass and inflation as consequences of dynamically broken scale invariance}",
    eprint = "1811.05950", archivePrefix = "arXiv", primaryClass = "hep-ph",
    doi = "10.1103/PhysRevD.100.015037",
    journal = "Phys. Rev. D", volume = "100", number = "1", pages = "015037", year = "2019"
}

@article{Kubo:2020fdd,
    author = "Kubo, Jisuke and Kuntz, Jeffrey and Lindner, Manfred and Rezacek, Jonas and Saake, Philipp and Trautner, Andreas",
    title = "{Unified Emergence of Energy Scales and Cosmic Inflation}",
    eprint = "2012.09706", archivePrefix = "arXiv", primaryClass = "hep-ph",
    doi = "10.1007/JHEP08(2021)016",
    journal = "JHEP", volume = "08", pages = "016", year = "2021"
}

@article{Garces:2026cbj,
    author = "Garc\'es, Juan P. and Kubo, Jisuke and Lindner, Manfred and Reinig, Markus",
    title = "{Dark matter in scale-invariant gravity with hidden-sector condensation}",
    eprint = "2608.12456", archivePrefix = "arXiv", primaryClass = "hep-ph", year = "2026"
}

@article{Tokareva:2017nng,
    author = "Tokareva, Anna",
    title = "{A minimal scale invariant axion solution to the strong CP-problem}",
    eprint = "1705.10836", archivePrefix = "arXiv", primaryClass = "hep-ph",
    doi = "10.1140/epjc/s10052-018-5883-0",
    journal = "Eur. Phys. J. C", volume = "78", number = "5", pages = "423", year = "2018"
}

@article{Lyth:2005fi,
    author = "Lyth, David H. and Rodriguez, Yeinzon",
    title = "{The Inflationary prediction for primordial non-Gaussianity}",
    eprint = "astro-ph/0504045",
    archivePrefix = "arXiv",
    doi = "10.1103/PhysRevLett.95.121302",
    journal = "Phys. Rev. Lett.",
    volume = "95",
    pages = "121302",
    year = "2005"
}

@article{Sasaki:1995aw,
    author = "Sasaki, Misao and Stewart, Ewan D.",
    title = "{A General analytic formula for the spectral index of the density perturbations produced during inflation}",
    eprint = "astro-ph/9507001",
    archivePrefix = "arXiv",
    reportNumber = "LANCS-TH-9504, OU-TAP-22",
    doi = "10.1143/PTP.95.71",
    journal = "Prog. Theor. Phys.",
    volume = "95",
    pages = "71--78",
    year = "1996"
}

@article{Kawasaki:2008sn,
    author        = "Kawasaki, Masahiro and Nakayama, Kazunori and
                     Sekiguchi, Toyokazu and Suyama, Teruaki and
                     Takahashi, Fuminobu",
    title         = "{Non-Gaussianity from isocurvature perturbations}",
    eprint        = "0808.0009",
    archivePrefix = "arXiv",
    primaryClass  = "astro-ph",
    doi           = "10.1088/1475-7516/2008/11/019",
    journal       = "JCAP",
    volume        = "11",
    pages         = "019",
    year          = "2008"
}

@article{Blas:2011ac,
      author         = "Blas, Diego and Shaposhnikov, Mikhail and Zenhausern,
                        Daniel",
      title          = "{Scale-invariant alternatives to general relativity}",
      journal        = "Phys. Rev.",
      volume         = "D84",
      year           = "2011",
      pages          = "044001",
      doi            = "10.1103/PhysRevD.84.044001",
      eprint         = "1104.1392",
      archivePrefix  = "arXiv",
      primaryClass   = "hep-th",
      SLACcitation   = "%%CITATION = ARXIV:1104.1392;%%"
}

@article{Shaposhnikov:2008xi,
      author         = "Shaposhnikov, Mikhail and Zenhausern, Daniel",
      title          = "{Quantum scale invariance, cosmological constant and
                        hierarchy problem}",
      journal        = "Phys. Lett.",
      volume         = "B671",
      year           = "2009",
      pages          = "162-166",
      doi            = "10.1016/j.physletb.2008.11.041",
      eprint         = "0809.3406",
      archivePrefix  = "arXiv",
      primaryClass   = "hep-th",
      SLACcitation   = "%%CITATION = ARXIV:0809.3406;%%"
}

@article{Bezrukov:2007ep,
      author         = "Bezrukov, Fedor L. and Shaposhnikov, Mikhail",
      title          = "{The Standard Model Higgs boson as the inflaton}",
      journal        = "Phys. Lett.",
      volume         = "B659",
      year           = "2008",
      pages          = "703-706",
      doi            = "10.1016/j.physletb.2007.11.072",
      eprint         = "0710.3755",
      archivePrefix  = "arXiv",
      primaryClass   = "hep-th",
      SLACcitation   = "%%CITATION = ARXIV:0710.3755;%%"
}

@article{GarciaBellido:2011de,
      author         = "Garcia-Bellido, Juan and Rubio, Javier and Shaposhnikov,
                        Mikhail and Zenhausern, Daniel",
      title          = "{Higgs-Dilaton Cosmology: From the Early to the Late
                        Universe}",
      journal        = "Phys. Rev.",
      volume         = "D84",
      year           = "2011",
      pages          = "123504",
      doi            = "10.1103/PhysRevD.84.123504",
      eprint         = "1107.2163",
      archivePrefix  = "arXiv",
      primaryClass   = "hep-ph",
      reportNumber   = "IFT-UAM-CSIC-11-49",
      SLACcitation   = "%%CITATION = ARXIV:1107.2163;%%"
}

@article{Rubio:2019ypq,
      author         = "Rubio, Javier and Tomberg, Eemeli S.",
      title          = "{Preheating in Palatini Higgs inflation}",
      journal        = "JCAP",
      volume         = "1904",
      year           = "2019",
      number         = "04",
      pages          = "021",
      doi            = "10.1088/1475-7516/2019/04/021",
      eprint         = "1902.10148",
      archivePrefix  = "arXiv",
      primaryClass   = "hep-ph",
      SLACcitation   = "%%CITATION = ARXIV:1902.10148;%%"
}

@article{Bezrukov:2012hx,
      author         = "Bezrukov, Fedor and Karananas, Georgios K. and Rubio,
                        Javier and Shaposhnikov, Mikhail",
      title          = "{Higgs-Dilaton Cosmology: an effective field theory
                        approach}",
      journal        = "Phys. Rev.",
      volume         = "D87",
      year           = "2013",
      number         = "9",
      pages          = "096001",
      doi            = "10.1103/PhysRevD.87.096001",
      eprint         = "1212.4148",
      archivePrefix  = "arXiv",
      primaryClass   = "hep-ph",
      reportNumber   = "RBRC-1007",
      SLACcitation   = "%%CITATION = ARXIV:1212.4148;%%"
}

@article{Rubio:2014wta,
      author         = "Rubio, Javier and Shaposhnikov, Mikhail",
      title          = "{Higgs-Dilaton cosmology: Universality versus
                        criticality}",
      journal        = "Phys. Rev.",
      volume         = "D90",
      year           = "2014",
      pages          = "027307",
      doi            = "10.1103/PhysRevD.90.027307",
      eprint         = "1406.5182",
      archivePrefix  = "arXiv",
      primaryClass   = "hep-ph",
      SLACcitation   = "%%CITATION = ARXIV:1406.5182;%%"
}

@article{Karananas:2016kyt,
      author         = "Karananas, Georgios K. and Rubio, Javier",
      title          = "{On the geometrical interpretation of scale-invariant
                        models of inflation}",
      journal        = "Phys. Lett.",
      volume         = "B761",
      year           = "2016",
      pages          = "223-228",
      doi            = "10.1016/j.physletb.2016.08.037",
      eprint         = "1606.08848",
      archivePrefix  = "arXiv",
      primaryClass   = "hep-ph",
      SLACcitation   = "%%CITATION = ARXIV:1606.08848;%%"
}

@article{Aswathi:2025nxa,
    author = "Aswathi, P. S. and East, William E. and Siemonsen, Nils and Sun, Ling and Jones, Dana",
    title = "{Ultralight boson constraints from gravitational wave observations of spinning binary black holes}",
    eprint = "2507.20979",
    archivePrefix = "arXiv",
    primaryClass = "gr-qc",
    doi = "10.1103/n5hr-zljn",
    journal = "Phys. Rev. D",
    volume = "112",
    number = "12",
    pages = "123048",
    year = "2025"
}

@article{LIGOScientific:2025rsn,
    author = "Abac, A. G. and others",
    collaboration = "LIGO Scientific, VIRGO, KAGRA",
    title = "{GW231123: A Binary Black Hole Merger with Total Mass 190{\textendash}265 M$_\odot$}",
    eprint = "2507.08219",
    archivePrefix = "arXiv",
    primaryClass = "astro-ph.HE",
    reportNumber = "DCC: P2500026-v6, DCC: P2500026-v8",
    doi = "10.3847/2041-8213/ae0c9c",
    journal = "Astrophys. J. Lett.",
    volume = "993",
    number = "1",
    pages = "L25",
    year = "2025"
}

@article{LIGOScientific:2021usb,
    author = "Abbott, R. and others",
    collaboration = "LIGO Scientific, VIRGO",
    title = "{GWTC-2.1: Deep extended catalog of compact binary coalescences observed by LIGO and Virgo during the first half of the third observing run}",
    eprint = "2108.01045",
    archivePrefix = "arXiv",
    primaryClass = "gr-qc",
    reportNumber = "LIGO-P2100063",
    doi = "10.1103/PhysRevD.109.022001",
    journal = "Phys. Rev. D",
    volume = "109",
    number = "2",
    pages = "022001",
    year = "2024"
}

@article{Choi:2026tup,
    author = "Choi, Gongjun and Gherghetta, Tony",
    title = "{Suppressing Extra-Dimensional Axion Isocurvature Dynamically}",
    eprint = "2607.09969",
    archivePrefix = "arXiv",
    primaryClass = "hep-ph",
    reportNumber = "UMN-TH-4536/26, FTPI-MINN-26-16",
    month = "7",
    year = "2026"
}

@article{Nakayama:2015pba,
    author = "Nakayama, Kazunori and Takimoto, Masahiro",
    title = "{Higgs inflation and suppression of axion isocurvature perturbation}",
    eprint = "1505.02119",
    archivePrefix = "arXiv",
    primaryClass = "hep-ph",
    reportNumber = "UT-15-18",
    doi = "10.1016/j.physletb.2015.07.001",
    journal = "Phys. Lett. B",
    volume = "748",
    pages = "108--112",
    year = "2015"
}

@article{Chun:2014xva,
    author = "Chun, Eung Jin",
    title = "{Axion Dark Matter with High-Scale Inflation}",
    eprint = "1404.4284",
    archivePrefix = "arXiv",
    primaryClass = "hep-ph",
    doi = "10.1016/j.physletb.2014.06.017",
    journal = "Phys. Lett. B",
    volume = "735",
    pages = "164--167",
    year = "2014"
}

@article{Fairbairn:2014zta,
    author = "Fairbairn, Malcolm and Hogan, Robert and Marsh, David J. E.",
    title = "{Unifying inflation and dark matter with the Peccei-Quinn field: observable axions and observable tensors}",
    eprint = "1410.1752",
    archivePrefix = "arXiv",
    primaryClass = "hep-ph",
    doi = "10.1103/PhysRevD.91.023509",
    journal = "Phys. Rev. D",
    volume = "91",
    number = "2",
    pages = "023509",
    year = "2015"
}

@article{Sikivie:1982qv,
  title = {Axions, Domain Walls, and the Early Universe},
  author = {Sikivie, P.},
  journal = {Phys. Rev. Lett.},
  volume = {48},
  issue = {17},
  pages = {1156--1159},
  numpages = {0},
  year = {1982},
  month = {Apr},
  publisher = {American Physical Society},
  doi = {10.1103/PhysRevLett.48.1156},
  url = {https://link.aps.org/doi/10.1103/PhysRevLett.48.1156}
}

@article{Vilenkin:1982ks,
    author = "Vilenkin, A. and Everett, A. E.",
    title = "{Cosmic Strings and Domain Walls in Models with Goldstone and PseudoGoldstone Bosons}",
    doi = "10.1103/PhysRevLett.48.1867",
    journal = "Phys. Rev. Lett.",
    volume = "48",
    pages = "1867--1870",
    year = "1982"
}

@article{Hiramatsu:2012sc,
    author = "Hiramatsu, Takashi and Kawasaki, Masahiro and Saikawa, Ken'ichi and Sekiguchi, Toyokazu",
    title = "{Axion cosmology with long-lived domain walls}",
    eprint = "1207.3166",
    archivePrefix = "arXiv",
    primaryClass = "hep-ph",
    reportNumber = "ICRR-REPORT-620-2012-9, IPMU12-0140, YITP-12-58",
    doi = "10.1088/1475-7516/2013/01/001",
    journal = "JCAP",
    volume = "01",
    pages = "001",
    year = "2013"
}

@article{Hiramatsu:2012gg,
    author = "Hiramatsu, Takashi and Kawasaki, Masahiro and Saikawa, Ken'ichi and Sekiguchi, Toyokazu",
    title = "{Production of dark matter axions from collapse of string-wall systems}",
    eprint = "1202.5851",
    archivePrefix = "arXiv",
    primaryClass = "hep-ph",
    reportNumber = "ICRR-REPORT-608-2011-25, IPMU12-0025, YITP-12-9",
    doi = "10.1103/PhysRevD.85.105020",
    journal = "Phys. Rev. D",
    volume = "85",
    pages = "105020",
    year = "2012",
    note = "[Erratum: Phys.Rev.D 86, 089902 (2012)]"
}

@article{Kawasaki:2014sqa,
    author = "Kawasaki, Masahiro and Saikawa, Ken'ichi and Sekiguchi, Toyokazu",
    title = "{Axion dark matter from topological defects}",
    eprint = "1412.0789",
    archivePrefix = "arXiv",
    primaryClass = "hep-ph",
    reportNumber = "ICRR-REPORT-696-2014-22, IPMU14-0348",
    doi = "10.1103/PhysRevD.91.065014",
    journal = "Phys. Rev. D",
    volume = "91",
    number = "6",
    pages = "065014",
    year = "2015"
}

@article{Gorghetto:2020qws,
    author = "Gorghetto, Marco and Hardy, Edward and Villadoro, Giovanni",
    title = "{More axions from strings}",
    eprint = "2007.04990",
    archivePrefix = "arXiv",
    primaryClass = "hep-ph",
    doi = "10.21468/SciPostPhys.10.2.050",
    journal = "SciPost Phys.",
    volume = "10",
    number = "2",
    pages = "050",
    year = "2021"
}

@article{Seckel:1985tj,
    author  = "Seckel, D. and Turner, Michael S.",
    title   = "{Isothermal Density Perturbations in an Axion Dominated Inflationary Universe}",
    journal = "Phys. Rev. D", volume = "32", pages = "3178", year = "1985",
    doi     = "10.1103/PhysRevD.32.3178"
}

@article{Buschmann:2021sdq,
    author = "Buschmann, Malte and Foster, Joshua W. and Hook, Anson and Peterson, Adam and Willcox, Don E. and Zhang, Weiqun and Safdi, Benjamin R.",
    title = "{Dark matter from axion strings with adaptive mesh refinement}",
    eprint = "2108.05368",
    archivePrefix = "arXiv",
    primaryClass = "hep-ph",
    doi = "10.1038/s41467-022-28669-y",
    journal = "Nature Commun.",
    volume = "13",
    number = "1",
    pages = "1049",
    year = "2022"
}

@article{Linde:1991km,
    author = "Linde, Andrei D.",
    title = "{Axions in inflationary cosmology}",
    reportNumber = "SU-ITP-883",
    doi = "10.1016/0370-2693(91)90130-I",
    journal = "Phys. Lett. B",
    volume = "259",
    pages = "38--47",
    year = "1991"
}

@article{Allali:2022yvx,
    author = "Allali, Itamar J. and Hertzberg, Mark P. and Lyu, Yi",
    title = "{Altered axion abundance from a dynamical Peccei-Quinn scale}",
    eprint = "2203.15817",
    archivePrefix = "arXiv",
    primaryClass = "hep-ph",
    doi = "10.1103/PhysRevD.105.123517",
    journal = "Phys. Rev. D",
    volume = "105",
    number = "12",
    pages = "123517",
    year = "2022"
}

@article{Folkerts:2013tua,
    author = "Folkerts, Sarah and Germani, Cristiano and Redondo, Javier",
    title = "{Axion Dark Matter and Planck favor non-minimal couplings to gravity}",
    eprint = "1304.7270",
    archivePrefix = "arXiv",
    primaryClass = "hep-ph",
    reportNumber = "LMU-ASC-26-13",
    doi = "10.1016/j.physletb.2013.12.026",
    journal = "Phys. Lett. B",
    volume = "728",
    pages = "532--536",
    year = "2014"
}

@article{Lyu:2026rwo,
    author = "Lyu, Kun-Feng and Sinha, Kuver",
    title = "{A Stepping Stone Solution to the QCD Axion Isocurvature Problem}",
    eprint = "2609.28840",
    archivePrefix = "arXiv",
    primaryClass = "hep-ph",
    month = "9",
    year = "2026"
}

@article{Ng:2020ruv,
    author = "Ng, Ken K. Y. and Vitale, Salvatore and Hannuksela, Otto A. and Li, Tjonnie G. F.",
    title = "{Constraints on Ultralight Scalar Bosons within Black Hole Spin Measurements from the LIGO-Virgo GWTC-2}",
    eprint = "2011.06010",
    archivePrefix = "arXiv",
    primaryClass = "gr-qc",
    doi = "10.1103/PhysRevLett.126.151102",
    journal = "Phys. Rev. Lett.",
    volume = "126",
    number = "15",
    pages = "151102",
    year = "2021"
}

@article{Ferreira:2016wem,
      author         = "Ferreira, Pedro G. and Hill, Christopher T. and Ross,
                        Graham G.",
      title          = "{Weyl Current, Scale-Invariant Inflation and Planck Scale
                        Generation}",
      journal        = "Phys. Rev.",
      volume         = "D95",
      year           = "2017",
      number         = "4",
      pages          = "043507",
      doi            = "10.1103/PhysRevD.95.043507",
      eprint         = "1610.09243",
      archivePrefix  = "arXiv",
      primaryClass   = "hep-th",
      reportNumber   = "FERMILAB-PUB-16-458-T, FERMILAB-PUB--16-458-T",
      SLACcitation   = "%%CITATION = ARXIV:1610.09243;%%"
}

@article{Wetterich:1987fm,
      author         = "Wetterich, C.",
      title          = "{Cosmology and the Fate of Dilatation Symmetry}",
      journal        = "Nucl. Phys.",
      volume         = "B302",
      year           = "1988",
      pages          = "668-696",
      doi            = "10.1016/0550-3213(88)90193-9",
      eprint         = "1711.03844",
      archivePrefix  = "arXiv",
      primaryClass   = "hep-th",
      reportNumber   = "PRINT-87-0756, DESY-87-123",
      SLACcitation   = "%%CITATION = ARXIV:1711.03844;%%"
}

@article{Kallosh:2013yoa,
      author         = "Kallosh, Renata and Linde, Andrei and Roest, Diederik",
      title          = "{Superconformal Inflationary $\alpha$-Attractors}",
      journal        = "JHEP",
      volume         = "11",
      year           = "2013",
      pages          = "198",
      doi            = "10.1007/JHEP11(2013)198",
      eprint         = "1311.0472",
      archivePrefix  = "arXiv",
      primaryClass   = "hep-th",
      SLACcitation   = "%%CITATION = ARXIV:1311.0472;%%"
}

@article{Karananas:2025uhy,
    author = "Karananas, Georgios K. and Shaposhnikov, Mikhail and Zell, Sebastian",
    title = "{A non-compact QCD axion}",
    eprint = "2512.20290",
    archivePrefix = "arXiv",
    primaryClass = "hep-ph",
    month = "12",
    year = "2025"
}

@article{LiteBIRD:2022cnt,
    author = "Allys, E. and others",
    collaboration = "LiteBIRD",
    title = "{Probing Cosmic Inflation with the LiteBIRD Cosmic Microwave Background Polarization Survey}",
    eprint = "2202.02773",
    archivePrefix = "arXiv",
    primaryClass = "astro-ph.IM",
    doi = "10.1093/ptep/ptac150",
    journal = "PTEP",
    volume = "2023",
    number = "4",
    pages = "042F01",
    year = "2023"
}

@article{Chung:2017uzc,
    author = "Chung, Daniel J. H. and Upadhye, Amol",
    title = "{Search for strongly blue axion isocurvature}",
    eprint = "1711.06736",
    archivePrefix = "arXiv",
    primaryClass = "astro-ph.CO",
    doi = "10.1103/PhysRevD.98.023525",
    journal = "Phys. Rev. D",
    volume = "98",
    number = "2",
    pages = "023525",
    year = "2018"
}

@article{Kasuya:2009up,
    author = "Kasuya, Shinta and Kawasaki, Masahiro",
    title = "{Axion isocurvature fluctuations with extremely blue spectrum}",
    eprint = "0904.3800",
    archivePrefix = "arXiv",
    primaryClass = "astro-ph.CO",
    doi = "10.1103/PhysRevD.80.023516",
    journal = "Phys. Rev. D",
    volume = "80",
    pages = "023516",
    year = "2009"
}

@article{Rigouzzo:2025ycb,
    author = "Rigouzzo, Cristian and Zell, Sebastian",
    title = "{Nonminimal couplings to gravity and axion isocurvature bounds}",
    eprint = "2512.16754",
    archivePrefix = "arXiv",
    primaryClass = "hep-ph",
    doi = "10.1103/PhysRevD.113.123502",
    journal = "Phys. Rev. D",
    volume = "113",
    number = "12",
    pages = "123502",
    year = "2026"
}

@article{Bezrukov:2010jz,
    author = "Bezrukov, F. and Magnin, A. and Shaposhnikov, M. and Sibiryakov, S.",
    title = "{Higgs inflation: consistency and generalisations}",
    eprint = "1008.5157",
    archivePrefix = "arXiv",
    primaryClass = "hep-ph",
    doi = "10.1007/JHEP01(2011)016",
    journal = "JHEP",
    volume = "01",
    pages = "016",
    year = "2011"
}

@article{Rigouzzo:2025hza,
    author = "Rigouzzo, Claire and Zell, Sebastian",
    title = "{No dark matter axion during minimal Higgs inflation}",
    eprint = "2504.02952",
    archivePrefix = "arXiv",
    primaryClass = "hep-ph",
    doi = "10.1103/369q-dxtt",
    journal = "Phys. Rev. D",
    volume = "113",
    number = "12",
    pages = "L121301",
    year = "2026"
}

@article{Kobayashi:2016qld,
    author = "Kobayashi, Takeshi and Takahashi, Fuminobu",
    title = "{Cosmological Perturbations of Axion with a Dynamical Decay Constant}",
    eprint = "1607.04294",
    archivePrefix = "arXiv",
    primaryClass = "hep-ph",
    reportNumber = "SISSA-39-2016-FISI, TU-1026, IPMU16-0118",
    doi = "10.1088/1475-7516/2016/08/056",
    journal = "JCAP",
    volume = "08",
    pages = "056",
    year = "2016"
}

@article{Kearney:2016vqw,
    author = "Kearney, John and Orlofsky, Nicholas and Pierce, Aaron",
    title = "{High-Scale Axions without Isocurvature from Inflationary Dynamics}",
    eprint = "1601.03049",
    archivePrefix = "arXiv",
    primaryClass = "hep-ph",
    reportNumber = "FERMILAB-PUB-16-005-T, MCTP-16-02",
    doi = "10.1103/PhysRevD.93.095026",
    journal = "Phys. Rev. D",
    volume = "93",
    number = "9",
    pages = "095026",
    year = "2016"
}

@article{Kawasaki:2013iha,
    author = "Kawasaki, Masahiro and Yanagida, Tsutomu T. and Yoshino, Kazuyoshi",
    title = "{Domain wall and isocurvature perturbation problems in axion models}",
    eprint = "1305.5338",
    archivePrefix = "arXiv",
    primaryClass = "hep-ph",
    reportNumber = "ICRR-REPORT-653-2013-2, IPMU-13-0104",
    doi = "10.1088/1475-7516/2013/11/030",
    journal = "JCAP",
    volume = "11",
    pages = "030",
    year = "2013"
}

@article{Harigaya:2015hha,
    author = "Harigaya, Keisuke and Ibe, Masahiro and Kawasaki, Masahiro and Yanagida, Tsutomu T.",
    title = "{Dynamics of Peccei-Quinn Breaking Field after Inflation and Axion Isocurvature Perturbations}",
    eprint = "1507.00119",
    archivePrefix = "arXiv",
    primaryClass = "hep-ph",
    reportNumber = "IPMU-15-0099",
    doi = "10.1088/1475-7516/2015/11/003",
    journal = "JCAP",
    volume = "11",
    pages = "003",
    year = "2015"
}

@article{Ballesteros:2021bee,
    author = "Ballesteros, Guillermo and Ringwald, Andreas and Tamarit, Carlos and Welling, Yvette",
    title = "{Revisiting isocurvature bounds in models unifying the axion with the inflaton}",
    eprint = "2104.13847",
    archivePrefix = "arXiv",
    primaryClass = "hep-ph",
    reportNumber = "IFT-UAM/CSIC-21-45, DESY 21-048, DESY-21-048, TUM-HEP-1334-21",
    doi = "10.1088/1475-7516/2021/09/036",
    journal = "JCAP",
    volume = "09",
    pages = "036",
    year = "2021"
}

@article{Graham:2025iwx,
    author = "Graham, Peter W. and Racco, Davide",
    title = "{Revisiting isocurvature bounds on the minimal QCD axion}",
    eprint = "2506.03348",
    archivePrefix = "arXiv",
    primaryClass = "hep-ph",
    reportNumber = "ZU-TH 38/25",
    doi = "10.1007/JHEP12(2025)028",
    journal = "JHEP",
    volume = "12",
    pages = "028",
    year = "2025"
}

@article{Karananas:2024xja,
    author = "Karananas, Georgios K. and Shaposhnikov, Mikhail and Zell, Sebastian",
    title = "{Weyl-invariant Einstein-Cartan gravity: unifying the strong CP and hierarchy puzzles}",
    eprint = "2406.11956",
    archivePrefix = "arXiv",
    primaryClass = "hep-th",
    doi = "10.1007/JHEP11(2024)146",
    journal = "JHEP",
    volume = "11",
    pages = "146",
    year = "2024"
}

@article{Karananas:2025ews,
    author = "Karananas, Georgios K. and Shaposhnikov, Mikhail and Zell, Sebastian",
    title = "{Gravitational Origin of the QCD Axion}",
    eprint = "2506.11836",
    archivePrefix = "arXiv",
    primaryClass = "hep-th",
    doi = "10.1103/d7mp-sjvc",
    journal = "Phys. Rev. Lett.",
    volume = "135",
    number = "24",
    pages = "241001",
    year = "2025"
}

@article{Artymowski:2016pjz,
      author         = "Artymowski, Michal and Rubio, Javier",
      title          = "{Endlessly flat scalar potentials and
                        $\alpha$-attractors}",
      journal        = "Phys. Lett.",
      volume         = "B761",
      year           = "2016",
      pages          = "111-114",
      doi            = "10.1016/j.physletb.2016.08.024",
      eprint         = "1607.00398",
      archivePrefix  = "arXiv",
      primaryClass   = "astro-ph.CO",
      SLACcitation   = "%%CITATION = ARXIV:1607.00398;%%"
}

@article{Bezrukov:2014ipa,
      author         = "Bezrukov, Fedor and Rubio, Javier and Shaposhnikov,
                        Mikhail",
      title          = "{Living beyond the edge: Higgs inflation and vacuum
                        metastability}",
      journal        = "Phys. Rev.",
      volume         = "D92",
      year           = "2015",
      number         = "8",
      pages          = "083512",
      doi            = "10.1103/PhysRevD.92.083512",
      eprint         = "1412.3811",
      archivePrefix  = "arXiv",
      primaryClass   = "hep-ph",
      SLACcitation   = "%%CITATION = ARXIV:1412.3811;%%"
}

@article{Piani:2023aof,
    author = "Piani, Matteo and Rubio, Javier",
    title = "{Preheating in Einstein-Cartan Higgs Inflation: oscillon formation}",
    eprint = "2304.13056",
    archivePrefix = "arXiv",
    primaryClass = "hep-ph",
    reportNumber = "IPARCOS-UCM-23-031",
    doi = "10.1088/1475-7516/2023/12/002",
    journal = "JCAP",
    volume = "12",
    pages = "002",
    year = "2023"
}

@article{Bezrukov:2017dyv,
    author = "Bezrukov, Fedor and Pauly, Martin and Rubio, Javier",
    title = "{On the robustness of the primordial power spectrum in renormalized Higgs inflation}",
    eprint = "1706.05007",
    archivePrefix = "arXiv",
    primaryClass = "hep-ph",
    doi = "10.1088/1475-7516/2018/02/040",
    journal = "JCAP",
    volume = "02",
    pages = "040",
    year = "2018"
}

@article{DeAngelis:2026mwx,
    author = "De Angelis, Mariaveronica and Rubio, Javier",
    title = "{The Goldstone Awakens: Unimodular dark energy in scale-invariant $R^2$ gravity}",
    eprint = "2607.16405",
    archivePrefix = "arXiv",
    primaryClass = "astro-ph.CO",
    reportNumber = "IPARCOS-UCM-26-037",
    month = "7",
    year = "2026"
}

@article{Repond:2016sol,
      author         = "Repond, Joel and Rubio, Javier",
      title          = "{Combined Preheating on the lattice with applications to
                        Higgs inflation}",
      journal        = "JCAP",
      volume         = "1607",
      year           = "2016",
      number         = "07",
      pages          = "043",
      doi            = "10.1088/1475-7516/2016/07/043",
      eprint         = "1604.08238",
      archivePrefix  = "arXiv",
      primaryClass   = "astro-ph.CO",
      SLACcitation   = "%%CITATION = ARXIV:1604.08238;%%"
}

@article{Casas:2017wjh,
      author         = "Casas, Santiago and Pauly, Martin and Rubio, Javier",
      title          = "{Higgs-dilaton cosmology: An inflation–dark-energy
                        connection and forecasts for future galaxy surveys}",
      journal        = "Phys. Rev.",
      volume         = "D97",
      year           = "2018",
      number         = "4",
      pages          = "043520",
      doi            = "10.1103/PhysRevD.97.043520",
      eprint         = "1712.04956",
      archivePrefix  = "arXiv",
      primaryClass   = "astro-ph.CO",
      SLACcitation   = "%%CITATION = ARXIV:1712.04956;%%"
}

@article{Casas:2018fum,
      author         = "Casas, Santiago and Karananas, Georgios K. and Pauly,
                        Martin and Rubio, Javier",
      title          = "{Scale-invariant alternatives to general relativity. III.
                        The inflation-dark energy connection}",
      journal        = "Phys. Rev.",
      volume         = "D99",
      year           = "2019",
      number         = "6",
      pages          = "063512",
      doi            = "10.1103/PhysRevD.99.063512",
      eprint         = "1811.05984",
      archivePrefix  = "arXiv",
      primaryClass   = "astro-ph.CO",
      SLACcitation   = "%%CITATION = ARXIV:1811.05984;%%"
}

@article{Cembranos:2026cat,
    author = "Cembranos, Jose A. R. and Luque, Jes{\'u}s and Rubio, Javier",
    title = "{Tunnelling out of Starobinsky inflation: Raising the spectral tilt}",
    eprint = "2607.15354",
    archivePrefix = "arXiv",
    primaryClass = "astro-ph.CO",
    month = "7",
    year = "2026"
}

@article{Starobinsky:1980te,
    author  = "Starobinsky, Alexei A.",
    title   = "{A New Type of Isotropic Cosmological Models Without Singularity}",
    journal = "Phys. Lett. B", volume = "91", pages = "99--102", year = "1980",
    doi     = "10.1016/0370-2693(80)90670-X"
}

@article{Kallosh:2013hoa,
    author        = "Kallosh, Renata and Linde, Andrei",
    title         = "{Universality Class in Conformal Inflation}",
    eprint        = "1306.5220", archivePrefix = "arXiv",
    journal = "JCAP", volume = "07", pages = "002", year = "2013",
    doi     = "10.1088/1475-7516/2013/07/002"
}

@article{AtacamaCosmologyTelescope:2025blo,
    author        = "Louis, Thibaut and others",
    collaboration = "Atacama Cosmology Telescope",
    title         = "{The Atacama Cosmology Telescope: DR6 power spectra, likelihoods and $\Lambda$CDM parameters}",
    eprint        = "2503.14452", archivePrefix = "arXiv",
    journal = "JCAP", volume = "11", pages = "062", year = "2025",
    doi     = "10.1088/1475-7516/2025/11/062"
}

@article{Haque:2025uri,
    author        = "Haque, Md Riajul and Pal, Sourav and Paul, Debarun",
    title         = "{ACT DR6 insights on $\alpha$-attractor inflationary models and reheating}",
    eprint        = "2505.01517", archivePrefix = "arXiv",
    journal = "Phys. Rev. D", volume = "114", pages = "023521", year = "2026",
    doi     = "10.1103/1kj4-s4ts"
}

@article{Kamionkowski:1992mf,
    author        = "Kamionkowski, Marc and March-Russell, John",
    title         = "{Planck scale physics and the Peccei-Quinn mechanism}",
    eprint        = "hep-th/9202003", archivePrefix = "arXiv",
    journal = "Phys. Lett. B", volume = "282", pages = "137--141", year = "1992",
    doi     = "10.1016/0370-2693(92)90492-M"
}

@article{Kofman:1995fi,
    author = "Kofman, Lev and Linde, Andrei D. and Starobinsky, Alexei A.",
    title = "{Nonthermal phase transitions after inflation}",
    eprint = "hep-th/9510119",
    archivePrefix = "arXiv",
    reportNumber = "SU-ITP-95-21, UH-IFA-95-51, SU-ITP-95-21",
    doi = "10.1103/PhysRevLett.76.1011",
    journal = "Phys. Rev. Lett.",
    volume = "76",
    pages = "1011--1014",
    year = "1996"
}

@article{Krajewski:2018moi,
    author = "Krajewski, Tomasz and Turzy{\'n}ski, Krzysztof and Wieczorek, Micha{\l}",
    title = "{On preheating in $\alpha$-attractor models of inflation}",
    eprint = "1801.01786",
    archivePrefix = "arXiv",
    primaryClass = "astro-ph.CO",
    doi = "10.1140/epjc/s10052-019-7155-z",
    journal = "Eur. Phys. J. C",
    volume = "79",
    number = "8",
    pages = "654",
    year = "2019"
}

@article{Iarygina:2018kee,
    author = "Iarygina, Oksana and Sfakianakis, Evangelos I. and Wang, Dong-Gang and Achucarro, Ana",
    title = "{Universality and scaling in multi-field $\alpha$-attractor preheating}",
    eprint = "1810.02804",
    archivePrefix = "arXiv",
    primaryClass = "astro-ph.CO",
    reportNumber = "Nikhef 2018-050",
    doi = "10.1088/1475-7516/2019/06/027",
    journal = "JCAP",
    volume = "06",
    pages = "027",
    year = "2019"
}

@article{Adshead:2023nhk,
    author = "Adshead, Peter and Giblin, Jr., John T. and Pfaltzgraff-Carlson, Reid",
    title = "{Kinetic preheating after {\ensuremath{\alpha}}-attractor inflation}",
    eprint = "2311.17237",
    archivePrefix = "arXiv",
    primaryClass = "astro-ph.CO",
    doi = "10.1016/j.physletb.2024.138928",
    journal = "Phys. Lett. B",
    volume = "856",
    pages = "138928",
    year = "2024"
}

@article{Ema:2016dny,
    author = "Ema, Yohei and Jinno, Ryusuke and Mukaida, Kyohei and Nakayama, Kazunori",
    title = "{Violent Preheating in Inflation with Nonminimal Coupling}",
    eprint = "1609.05209",
    archivePrefix = "arXiv",
    primaryClass = "hep-ph",
    doi = "10.1088/1475-7516/2017/02/045",
    journal = "JCAP",
    volume = "02",
    pages = "045",
    year = "2017"
}

@article{DeCross:2015uza,
    author = "DeCross, Matthew P. and Kaiser, David I. and Prabhu, Anirudh and Prescod-Weinstein, C. and Sfakianakis, Evangelos I.",
    title = "{Preheating after Multifield Inflation with Nonminimal Couplings, I: Covariant Formalism and Attractor Behavior}",
    eprint = "1510.08553",
    archivePrefix = "arXiv",
    primaryClass = "astro-ph.CO",
    reportNumber = "PREPRINT-MIT-CTP-4716",
    doi = "10.1103/PhysRevD.97.023526",
    journal = "Phys. Rev. D",
    volume = "97",
    number = "2",
    pages = "023526",
    year = "2018"
}

@article{Garcia-Bellido:2008ycs,
    author = "Garcia-Bellido, Juan and Figueroa, Daniel G. and Rubio, Javier",
    title = "{Preheating in the Standard Model with the Higgs-Inflaton coupled to gravity}",
    eprint = "0812.4624",
    archivePrefix = "arXiv",
    primaryClass = "hep-ph",
    reportNumber = "IFT-UAM-CSIC-08-93",
    doi = "10.1103/PhysRevD.79.063531",
    journal = "Phys. Rev. D",
    volume = "79",
    pages = "063531",
    year = "2009"
}

@article{Sfakianakis:2018lzf,
    author = "Sfakianakis, Evangelos I. and van de Vis, Jorinde",
    title = "{Preheating after Higgs Inflation: Self-Resonance and Gauge boson production}",
    eprint = "1810.01304",
    archivePrefix = "arXiv",
    primaryClass = "hep-ph",
    reportNumber = "Nikhef-2018-044",
    doi = "10.1103/PhysRevD.99.083519",
    journal = "Phys. Rev. D",
    volume = "99",
    number = "8",
    pages = "083519",
    year = "2019"
}

@article{Tkachev:1995md,
    author = "Tkachev, I. I.",
    title = "{Phase transitions at preheating}",
    eprint = "hep-th/9510146",
    archivePrefix = "arXiv",
    reportNumber = "OSU-TA-21-95",
    doi = "10.1016/0370-2693(96)00297-3",
    journal = "Phys. Lett. B",
    volume = "376",
    pages = "35--40",
    year = "1996"
}

@article{Holman:1992us,
    author        = "Holman, Richard and Hsu, Stephen D. H. and Kephart, Thomas W. and Kolb, Edward W. and Watkins, Richard and Widrow, Lawrence M.",
    title         = "{Solutions to the strong CP problem in a world with gravity}",
    eprint        = "hep-ph/9203206", archivePrefix = "arXiv",
    journal = "Phys. Lett. B", volume = "282", pages = "132--136", year = "1992",
    doi     = "10.1016/0370-2693(92)90491-L"
}

@article{Arkani-Hamed:2006emk,
    author        = "Arkani-Hamed, Nima and Motl, Lubos and Nicolis, Alberto and Vafa, Cumrun",
    title         = "{The String landscape, black holes and gravity as the weakest force}",
    eprint        = "hep-th/0601001", archivePrefix = "arXiv",
    journal = "JHEP", volume = "06", pages = "060", year = "2007",
    doi     = "10.1088/1126-6708/2007/06/060"
}

@article{PhysRevD.33.889,
  title = {Cosmic and local mass density of ``invisible'' axions},
  author = {Turner, Michael S.},
  journal = {Phys. Rev. D},
  volume = {33},
  issue = {4},
  pages = {889--896},
  numpages = {0},
  year = {1986},
  month = {Feb},
  publisher = {American Physical Society},
  doi = {10.1103/PhysRevD.33.889},
  url = {https://link.aps.org/doi/10.1103/PhysRevD.33.889}
}

@article{Pisarski:1980md,
    author = "Pisarski, R. D. and Yaffe, L. G.",
    title = "{THE DENSITY OF INSTANTONS AT FINITE TEMPERATURE}",
    doi = "10.1016/0370-2693(80)90559-6",
    journal = "Phys. Lett. B",
    volume = "97",
    pages = "110--112",
    year = "1980"
}

@article{Visinelli:2009kt,
    author = "Visinelli, Luca and Gondolo, Paolo",
    title = "{Axion cold dark matter in non-standard cosmologies}",
    eprint = "0912.0015",
    archivePrefix = "arXiv",
    primaryClass = "astro-ph.CO",
    doi = "10.1103/PhysRevD.81.063508",
    journal = "Phys. Rev. D",
    volume = "81",
    pages = "063508",
    year = "2010"
}

@article{Marsh:2015xka,
    author = "Marsh, David J. E.",
    title = "{Axion Cosmology}",
    eprint = "1510.07633",
    archivePrefix = "arXiv",
    primaryClass = "astro-ph.CO",
    reportNumber = "KCL-PH-TH-2015-50",
    doi = "10.1016/j.physrep.2016.06.005",
    journal = "Phys. Rept.",
    volume = "643",
    pages = "1--79",
    year = "2016"
}

\end{document}